\documentclass[useAMS,usenatbib]{mn2e}
\usepackage{graphicx}
\usepackage{rotating}
\usepackage{longtable}
\usepackage{lscape}
\usepackage{amssymb,amsmath}

\def\lsim{\mathrel{\rlap{\lower 3pt \hbox{$\sim$}} \raise 2.0pt \hbox{$<$}}}
\def\gsim{\mathrel{\rlap{\lower 3pt \hbox{$\sim$}} \raise 2.0pt \hbox{$>$}}}

\def\ad{\Delta\theta}
\def\dv{\Delta{\rm V}}
\def\kms{km\,s$^{-1}$}
\def\lbol{{\rm L}_{\rm bol}}
\def\msun{{\rm M}_\odot}
\def\mbh{{\rm M}_{\rm BH}}
\def\pd{{\rm pd}}
\def\qsof{QSO$_{\rm F}$}
\def\qsob{QSO$_{\rm B}$}
\def\z{{\rm z}}
\def\zf{{\rm z}_{\rm F}}
\def\zb{{\rm z}_{\rm B}}
\def\fc{f_{\rm C}}
\def\fchi{f_{\rm C}\left({\rm H\,I}\right)}
\def\fcfeii{f_{\rm C}\left({\rm Fe\,II}\right)}
\def\fcmgii{f_{\rm C}}

\def\ew{{\rm W}}
\def\ewr{{\rm W}_{\rm r}}

\def\civ{{\rm C\,{\sc IV}}}
\def\feii{{\rm Fe\,{\sc II}}}
\def\hi{{\rm H\,{\sc I}}}
\def\lya{{\rm Ly\,{\sc $\alpha$}}}
\def\mgii{{\rm Mg\,{\sc II}}}
\def\mgi{{\rm Mg\,{\sc I}}}

\def\nev{{\rm Ne\,{\sc V}}}

\def\siiv{{\rm Si\,{\sc IV}}}

\def\mhost{{\rm M}_{{\rm{host}}}}

\title[The extent of the Mg\,II absorbing CGM of quasars]
{The extent of the Mg\,II absorbing circum--galactic medium of quasars\thanks{
Based on observations made with the Gran Telescopio Canarias (GTC), installed 
in the Spanish Observatorio del Roque de los Muchachos of the Instituto de 
Astrof{\'i}sica de Canarias, in the island of La Palma.}}
\author[Farina et al.]{
	\parbox{\textwidth}{
	E.~P.~Farina$^{1,2,3}$\thanks{E--mail: {\texttt emanuele.paolo.farina@gmail.com}}, 
        R.~Falomo$^{4}$,
	R.~Scarpa$^{5}$,
	R.~Decarli$^{3}$,
	A.~Treves$^{1,2}$\thanks{Associated to INAF}, and
	J.~K.~Kotilainen$^{6}$}\vspace{0.6cm}\\
       	\parbox{\textwidth}{
	$^{1}$ Universit\`a degli Studi dell'Insubria --- via Valleggio 11, I-22100 Como, Italy\\
	$^{2}$ INFN Milano--Bicocca --- Universit\`a degli Studi di Milano--Bicocca, Piazza della Scienza 3, I-20126 Milano, Italy\\
	$^{3}$ Max--Planck--Institut f{\"u}r Astronomie --- K{\"o}nigstuhl 17, D-69117 Heidelberg, Germany\\
       	$^{4}$ INAF --- Osservatorio Astronomico di Padova, Vicolo dell'Osservatorio 5, I-35122 Padova, Italy\\
	$^{5}$ Instituto de astrof{\`i}sica de Canarias --- c/via Lactea s/n, E-38205 San Cristobal de la Laguna, Spain\\
       	$^{6}$ Finnish Centre for Astronomy with ESO (FINCA) --- University of Turku, V{\"a}is{\"a}l{\"a}ntie 20, FI-21500 Piikki{\"o}, Finland
	}}
\begin{document}

\date{\today}

\pagerange{\pageref{firstpage}--\pageref{lastpage}} \pubyear{2014}

\maketitle

\label{firstpage}

\begin{abstract}
	We investigate the extent and the properties of the \mgii\ 
	cool, low--density absorbing gas located in the halo and in 
	the circum--galactic environment of quasars, using a sample 	
	of 31 projected quasar pairs with impact parameter $\pd<200
	$\,kpc in the redshift range $0.5\lsim\z\lsim1.6$.
	In the transverse direction, we detect 18 \mgii\ absorbers
	associated with the foreground quasars, while no absorption 
	system originated by the gas surrounding the quasar itself 
	is found along the line--of--sight.
	This suggests that the quasar emission induces an anisotropy 
	in the absorbing gas distribution.
	Our observations indicate that the covering fraction ($\fc$) 
	of \mgii\ absorption systems with rest frame equivalent width 
	$\ewr(\lambda2796)>0.3$\,\AA\ ranges from $\fc\sim1.0$ at 
	$\pd\lsim65$\,kpc to $\fc\sim0.2$ at \mbox{$\pd\gsim150$\,kpc},
	and appears to be higher than for galaxies. 
	Our findings support a scenario where the luminosity/mass of 
	the host galaxies affect the extent and the richness of the 
	absorbing \mgii\ circum--galactic medium.
\end{abstract}



\section{Introduction}\label{sec:1}

Major mergers between galaxies are believed to be responsible for 
intense starbursts in galaxies and for channelling large amount of 
gas down to the circum--nuclear regions that triggers the activity 
of the central supermassive black hole \citep[e.g.][]{Hernquist1989, 
Kauffmann2000, Canalizo2001, Dimatteo2005, Hennawi2006bin, Bennert2008}.
In this scenario, the circum--galactic medium (CGM) of quasars is 
expected to harbour a large amount of enriched gas \citep[e.g.][]{
Prochaska2013a} and its study offers an opportunity to investigate 
the link between the nuclear activity of quasars and their immediate 
environment. 

The study of absorption lines represents a powerful way to probe the quasar 
CGM \citep[e.g.][]{Shaver1982, Shaver1983, Shaver1985, Hennawi2006, Bowen2006, 
Hennawi2007, Prochaska2009, Tytler2009, Decarli2009, Farina2013}, since 
it is typically too dim to be detected directly (\citealp{Chelouche2008,
Hennawi2013}).
In particular the \mgii\,$\lambda\lambda2796,2803$ doublet is well--suited 
for this aim. 
It falls within the optical wavelength range at intermediate redshift, 
probes regions of metal enriched, photoionised gas at temperatures around 
${\rm T}\sim10^4$\,K \citep{Bergeron1986, Charlton2003}, and traces a 
wide range of neutral hydrogen column densities \citep[i.e., from 
$N_\textrm{HI}\sim10^{16.5}$\,cm$^{-2}$ to greater than 
$N_\textrm{HI}\sim10^{21.5}$\,cm$^{-2}$, e.g. ][]{Bergeron1986, Rao2000, 
Rao2006}.

Observations of low redshift galaxies ($\z\lsim1$) showed the presence of 
a large halo of cool \mgii\ absorbing gas, extending up to $\sim200$\,kpc 
\citep[e.g.][]{Bahcall1969, Boksenberg1978, Steidel1994, Steidel1997, 
Churchill2005, Chen2010a, Chen2010b, Bowen2011, Nielsen2013a, Nielsen2013b}, 
that is strongly linked to galaxy properties such as: 
luminosity \citep[e.g.][]{Chen2008, Chen2010a};
mass \citep[e.g.][]{Bouche2006, Chen2010b, Churchill2013a};
colour \citep[e.g.][]{Zibetti2007};
star formation rate \citep[e.g.][]{Prochter2006, Menard2011, Nestor2011};
morphology \citep[e.g.][]{Kacprzak2007}; 
and galactic environment \citep[e.g.][]{Bordoloi2011}.
These findings have motivated the search for possible mechanism responsible 
for the enrichment of the CGM at such large scale. 
Detailed studies of the kinematics of the absorption systems have shown that
both outflows due to galactic wind \citep[e.g.][]{Tremonti2007, Weiner2009,
Rubin2010} and inflows onto the galaxies \citep[e.g.][]{Ribaudo2011, Rubin2012}
could supply the CGM of cool enriched gas. 

In this Paper we extend these studies to quasar host galaxies: if two 
quasars are angularly close but have discordant redshifts, one can probe the 
CGM of the foreground target (\qsof) through the study of absorption features 
imprinted on spectra of the background source (\qsob).
This technique has allowed a detailed study of the distribution of neutral 
hydrogen around quasars.  
For instance \citet{Hennawi2006}, starting from a sample of 149 projected 
quasar pairs (projected distance: $30$\,kpc$\lsim \pd \lsim 2.5$\,Mpc; 
redshift: $1.8<\z<4.0$), found that the probability to have an absorber with 
$N_{\rm HI}>10^{19}$\,cm$^{-2}$ coincident within 200\,kpc with a \qsof\ is 
high ($\sim50\%$), and that the distribution of these absorbers is highly 
anisotropic \citep{Hennawi2007}.
Considering a larger sample of pairs \citet{Prochaska2013a} and 
\citet{Prochaska2013b} recently confirm the presence of a large
number of absorbers with $N_{\rm HI}>10^{17.3}$\,cm$^{-2}$ in the
proximity of quasars.

It is worth noting that the quasar host galaxies are typically more massive 
than normal galaxies and may trace group/cluster environments \citep[e.g.][]{
Wold2001, Serber2006, Hutchings2009}, hence their CGM is expected to be 
richer and could exhibit different physical characteristics. 
In addition, the presence of the central SMBH may have a substantial effect. 
In fact its emission may photoionise gas from tens to several hundreds of kpc 
and photo--evaporate cool clumps \citep[e.g.][]{Hennawi2007, Wild2008, 
Chelouche2008, Farina2013}.

The first attempt to explore the \mgii\ absorbing CGM of quasars was
performed by \citet{Bowen2006}, who detected absorption lines in all 
the 4~close projected quasar pairs investigated.
In \citet{Farina2013} we have studied 10 additional systems, exploring 
projected separations between 60~and~120\,kpc.
Here we aim to further expand this work and to extend it up to separations 
of 200\,kpc, in order to determine the gas covering fraction and the size 
of the haloes hosting quasars. 
The new sample of projected quasar pairs is described in \S\ref{sec:2} and 
the analysis of the collected spectra in \S\ref{sec:3}. 
In \S\ref{sec:4} we investigate the properties of the detected \mgii\ 
absorption systems. 
We compare and contrast our results with those found in inactive galaxies in 
\S\ref{sec:5} and we summarise our conclusions in 
\S\ref{sec:6}.
 
Throughout this paper we assume a concordance cosmology with 
H$_0$=70\,km\,s$^{-1}$Mpc$^{-1}$, $\Omega_{\rm m}$=0.3, 
and \mbox{$\Omega_\Lambda$=0.7}. 	


\section[]{The sample}\label{sec:2}

In order to study the CGM of quasars in absorption, we searched in the 
\citet{Schneider2010} catalogue \citep[based on the 7th data release of
the Sloan Digital Sky Survey, SDSS,][]{Abazajian2009} for quasar pairs 
that have physical projected separations $\pd<200$\,kpc (calculated 
in the frame of the foreground targets) and line--of--sight velocity differences 
$\dv>5000$\,\kms. 
17 of the 85 retrieved systems were observed with the intermediate resolution 
grisms R2500V and R2500R of the Optical System for Imaging and low 
Resolution Integrated Spectroscopy \citep[OSIRIS,][]{Cepa2000, Cepa2003} 
mounted on the 10.4\,m Gran Telescopio de Canarias (GTC). 
We have selected these targets to have the \mgii\ doublet lines at 
$\z=\zf$ well within the spectral range covered by the two considered 
grisms (nominally from 4470\,\AA\ to 5950\,\AA\ for R2500V, and from 
5630\,\AA\ to 7540\,\AA\ for R2500R). 
This constrains the \qsof\ redshifts in the range: $0.6\lsim\zf\lsim1.6$.
Data for an additional pair (QQ01, see Table~\ref{tab:sample}) were 
collected with the grism 1400V of the FOcal Reducer and low dispersion 
Spectrograph \citep[FORS2,][]{Appenzeller1998} installed on the Antu Very 
Large Telescope (VLT) of the European Southern Observatory (ESO). 
This system is part of the sample of south quasar pairs we have investigated 
in \citet{Farina2013}.

\begin{figure}
\centering
\includegraphics[width=1.0\columnwidth]{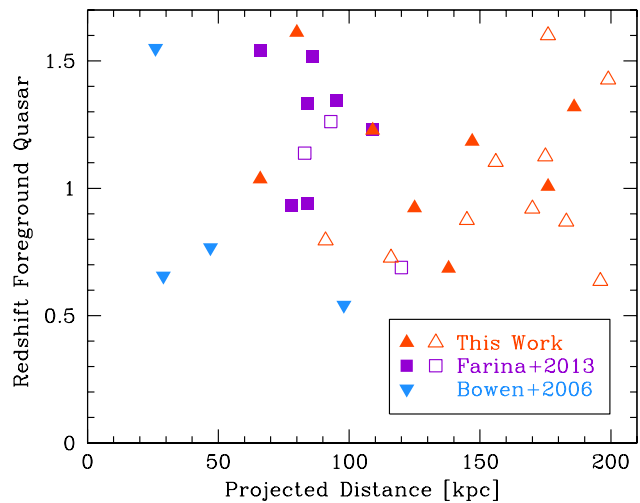}
\caption{Distribution of observed projected quasar pairs in the 
redshift--impact parameter plane. 
Orange and cyan triangles, and violet squares are targets from 
this work, \citet{Bowen2006}, and \citet{Farina2013}, respectively. 
Systems showing an absorption feature associated to the foreground 
quasar are highlighted with filled points.
}\label{fig:sample}
\end{figure}

In Table~\ref{tab:sample} and in Figure~\ref{fig:sample} we present the 
general properties of the 18~observed pairs. 
These are radio quiet quasars, with an average angular separation $\langle
\ad\rangle\sim18\farcs6$ that corresponds to an average projected distance 
$\langle\pd\rangle\sim146$\,kpc.
These data represent a substantial increase in the number of pairs 
investigated so far, especially at separations larger than 100\,kpc
(see Figure~\ref{fig:sample}).
Combined with data from \citet{Bowen2006} and \citet{Farina2013}, our 
sample yields constraints on the physical properties of the \mgii\ 
absorbing gas surrounding quasars on scales between 26\,kpc and 200\,kpc
at an average redshift $\langle\z\rangle\sim1.1$.
We stress that in the selection of the targets we did not take into 
account for the possible presence of absorbers {\it a priori}, hence our 
sample seems well--suited to estimate the unbiased frequency of \mgii\ 
absorption systems associated to quasars.

\begin{table*}
\centering
\scriptsize
\caption{
Properties of the observed projected quasar pairs: our identification label of 
the system (ID), position of the foreground target (RA,DEC), redshift derived 
from broad emission lines ($\z$), SDSS BEST~AB r--band magnitude \citep[r,][]{
Abazajian2009}, angular ($\ad$) and projected ($\pd$) separation between the two
quasars, bolometric luminosity ($\lbol$, see Appendix~\ref{app:1} for 
details), black hole mass ($\mbh$, see \citealt{Vestergaard2006} and 
Appendix~\ref{app:1} for details) and host galaxy mass ($\mhost$ derived from
$\mbh$, see \citealt{Decarli2010b} and text for details) of the foreground quasar, 
seeing during the observations (See.), and signal--to--noise ratio per pixel on 
the continuum close to the expected position of the \mgii\ absorption lines (S/N). 
The labels ${\rm F}$ and ${\rm B}$ refer to the foreground and background 
quasar, respectively.
}\label{tab:sample}
\begin{tabular}{lcccccccccccccc}
\hline
  ID                     & RA$_{\rm F}$ & DEC$_{\rm F}$ & $\zf$ & $\zb$ & r$_{\rm F}$               & r$_{\rm B}$               & $\ad$          & $\pd$	 & ${\rm L}_{\rm bol,F}$ & ${\rm M}_{\rm BH,F}$ & ${\rm M}_{\rm host,F}$ &  See.     & S/N$_{\rm F}$ & S/N$_{\rm B}$ \\
                         & (J2000)      & (J2000)       & 	&       & (mag) 		    & (mag)		        & (arcsec)	 & (kpc)	 & ($10^{46}$\,erg/s)    &	($10^8\,\msun$) &     ($10^{11}\,\msun$) &  (arcsec) &  	     &  	     \\     
\hline
QQ01$^{\rm a}$	         & 00:38:23.8   & --29:13:11    & 0.793 & 2.699 & 19.54$^{\rm b}$	    & 19.87$^{\rm b}$ 	        &	    12.1 & \phantom{1}91 &	   0.09$\pm$0.01 &	 \phantom{1}1.9 &		     0.9 &	 1.2 &  	  11 &  	  11 \\
QQ02\phantom{$^{\rm a}$} & 00:47:57.2	&  +14:47:42	& 1.620 & 2.746 & 19.17\phantom{$^{\rm b}$} & 19.07\phantom{$^{\rm b}$} & \phantom{1}9.4 & \phantom{1}80 &	   1.95$\pm$0.14 &       \phantom{1}3.3 &		     0.9 &	 1.0 &  	  15 &  	  12 \\ 
QQ03\phantom{$^{\rm a}$} & 00:49:51.9	&  +00:03:44	& 1.121 & 1.332 & 19.13\phantom{$^{\rm b}$} & 19.71\phantom{$^{\rm b}$} &	    21.4 &	     175 &	   1.04$\pm$0.08 &       \phantom{1}7.5 &		     2.9 &	 0.9 &  	  32 &  	  22 \\
QQ04\phantom{$^{\rm a}$} & 01:03:48.1	&  +15:01:57	& 1.319 & 1.744 & 18.89\phantom{$^{\rm b}$} & 17.47\phantom{$^{\rm b}$} &	    22.2 &	     186 &	   2.14$\pm$0.09 &      	   17.1 &		     5.9 &	 1.1 &  	  33 &  	  43 \\
QQ05\phantom{$^{\rm a}$} & 01:35:00.8	& --00:40:54	& 1.003 & 1.259 & 19.55\phantom{$^{\rm b}$} & 19.10\phantom{$^{\rm b}$} &	    22.0 &	     176 &	   0.23$\pm$0.04 &       \phantom{1}2.9 &		     1.2 &	 1.4 &  	  10 &  	   9 \\
QQ06\phantom{$^{\rm a}$} & 01:46:30.1	&  +00:15:21	& 0.923 & 1.019 & 20.12\phantom{$^{\rm b}$} & 18.67\phantom{$^{\rm b}$} &	    15.9 &	     125 &	   0.30$\pm$0.06 &       \phantom{1}3.3 &		     1.5 &	 0.8 &  	  11 &  	  17 \\
QQ07\phantom{$^{\rm a}$} & 02:15:52.5	&  +01:10:00	& 0.875 & 2.215 & 19.60\phantom{$^{\rm b}$} & 20.01\phantom{$^{\rm b}$} &	    18.8 &	     145 &	   0.35$\pm$0.03 &       \phantom{1}1.3 &		     0.6 &	 0.8 &  	  22 &  	  13 \\
QQ08\phantom{$^{\rm a}$} & 02:19:53.0	& --00:44:34	& 0.685 & 1.035 & 19.22\phantom{$^{\rm b}$} & 19.22\phantom{$^{\rm b}$} &	    19.5 &	     138 &	   0.18$\pm$0.02 &       \phantom{1}2.4 &		     1.3 &	 0.8 &  	  14 &  	  10 \\
QQ09\phantom{$^{\rm a}$} & 02:21:58.7	& --00:10:44	& 1.036 & 3.213 & 19.87\phantom{$^{\rm b}$} & 20.03\phantom{$^{\rm b}$} & \phantom{1}8.2 & \phantom{1}66 &	   0.31$\pm$0.03 &       \phantom{1}4.1 &		     1.7 &	 0.9 &  	  18 &  	  17 \\
QQ10\phantom{$^{\rm a}$} & 02:29:47.1	& --00:53:32	& 0.727 & 2.307 & 20.55\phantom{$^{\rm b}$} & 20.09\phantom{$^{\rm b}$} &	    16.0 &	     116 &	   0.10$\pm$0.01 &       \phantom{1}0.8 &		     0.4 &	 0.9 &  	  11 &  	   8 \\
QQ11\phantom{$^{\rm a}$} & 02:46:03.6	& --00:32:11	& 1.600 & 2.161 & 18.47\phantom{$^{\rm b}$} & 20.13\phantom{$^{\rm b}$} &	    22.0 &	     176 &	   2.14$\pm$0.07 &      	   10.1 &		     2.9 &	 1.1 &  	  22 &  	   6 \\
QQ12\phantom{$^{\rm a}$} & 03:38:54.2   & --00:23:17    & 1.189 & 2.664 & 20.16\phantom{$^{\rm b}$} & 19.99\phantom{$^{\rm b}$} & 	    17.8 &	     147 &         0.25$\pm$0.03 &       \phantom{1}6.3 &		     2.4 &	 1.4 &  	  12 &  	  10 \\
QQ13\phantom{$^{\rm a}$} & 16:24:49.6   &  +49:30:32    & 0.634 & 2.275 & 19.33\phantom{$^{\rm b}$} & 19.74\phantom{$^{\rm b}$} &	    28.6 &	     196 &	   0.14$\pm$0.01 &       \phantom{1}6.1 &		     3.3 &	 0.8 &  	  25 &  	  17 \\
QQ14\phantom{$^{\rm a}$} & 16:26:07.1   &  +14:00:27    & 1.103 & 2.341 & 19.50\phantom{$^{\rm b}$} & 19.59\phantom{$^{\rm b}$} &	    19.0 &	     156 &	   0.58$\pm$0.02 &       \phantom{1}1.7 &		     0.7 &	 0.8 &  	  18 &  	  22 \\
QQ15\phantom{$^{\rm a}$} & 16:41:45.2   &  +27:36:22    & 0.920 & 1.649 & 19.18\phantom{$^{\rm b}$} & 18.61\phantom{$^{\rm b}$} &	    21.7 &	     170 &	   0.51$\pm$0.04 &       \phantom{1}1.5 &		     0.7 &	 0.9 &  	  15 &  	  20 \\
QQ16\phantom{$^{\rm a}$} & 22:06:44.4   & --00:39:06    & 1.228 & 1.514 & 19.54\phantom{$^{\rm b}$} & 19.72\phantom{$^{\rm b}$} &	    13.1 &	     109 &	   0.71$\pm$0.03 &       \phantom{1}7.1 &		     2.6 &	 0.8 &  	  28 &  	  19 \\
QQ17\phantom{$^{\rm a}$} & 22:59:02.9   &  +00:32:44    & 0.868 & 1.465 & 20.04\phantom{$^{\rm b}$} & 19.66\phantom{$^{\rm b}$} &	    23.7 &	     183 &	   0.26$\pm$0.03 &       \phantom{1}3.8 &		     1.8 &	 0.8 &  	  13 &  	   9 \\
QQ18\phantom{$^{\rm a}$} & 23:53:15.0   & --00:50:34    & 1.427 & 1.946 & 19.14\phantom{$^{\rm b}$} & 19.50\phantom{$^{\rm b}$} &	    23.6 &	     199 &	   0.79$\pm$0.07 &       \phantom{1}5.4 &		     1.8 &	 0.9 &  	  34 &  	  32 \\
\hline																											  
\multicolumn{14}{l}{$^{\rm a}$ Quasar pair observed with FORS2 at ESO--VLT.} \\
\multicolumn{14}{l}{$^{\rm b}$ R--band magnitude from the USNO--B catalogue, first epoch \citep{Monet2003}.}                           \\
\end{tabular}
\end{table*}

\section[]{Observations and Data Analysis}\label{sec:3}

Observations were carried out with the OSIRIS R2500R and R2500V grisms, 
yielding a spectral resolution of ${\rm R}=\lambda/\Delta\lambda\sim2500$ 
(with the 1\arcsec\ slit). 
This corresponds to a resolution element of ${\rm FWHM}\sim120$\,\kms\ that
allows one to separate the doublet components but that is larger than the 
typical width of \mgii\ absorbers \citep[$\lsim100$\,\kms, see e.g.][]{Charlton1998}.
Thus the internal dynamics of the absorption systems can not be resolved. 
The position angle of the slit was oriented so that the spectrum of both the 
sources could be acquired simultaneously. 
The integration times range from 900\,s to 4800\,s, depending on the magnitude 
of the background quasar. 
In order to correct for cosmic rays and for CCD cosmetic defects, for each 
target we have taken a series of~3~consecutive exposures, respectively shifted 
by $\sim5\arcsec$. 
Details about the spectra gathered at ESO--VLT are given in \citet{Farina2013}.
 
Standard \texttt{IRAF}\footnote{\texttt{IRAF} \citep{Tody1986} is distributed 
by the National Optical Astronomy Observatories, which are operated by the 
Association of Universities for Research in Astronomy, Inc., under cooperative 
agreement with the National Science Foundation.} tools were used for the data 
reduction. The \texttt{ccdred} package was employed to perform bias subtraction,
flat field correction, image alignment and combination. 
The spectra extraction, the background subtraction and the calibrations both 
in wavelength and in flux were performed with \texttt{twod} and \texttt{oned} 
packages. 
Residuals of wavelength calibration are around $0.03$\,\AA\ (sub--pixel).
Standard stars' spectra were collected during the same nights of the targets. 
The absolute calibration of the spectra was obtained through the photometry 
of field stars present in r--band short exposures of the targets. This 
procedure yields to a photometric accuracy of $\sim0.1$\,mag \citep[see][for 
details]{Decarli2008}.
The Galactic extinction was taken into account considering the estimates of 
\citet{Schlegel1998} and assuming a standard interstellar extinction curve 
\citep[R$_{\rm V}=3.1$, e.g.][]{Cardelli1989}. 
The spectra of the quasar pairs are reproduced in Figure~\ref{fig:allspec}.

\begin{figure*}
\centering
\includegraphics[width=1.99\columnwidth]{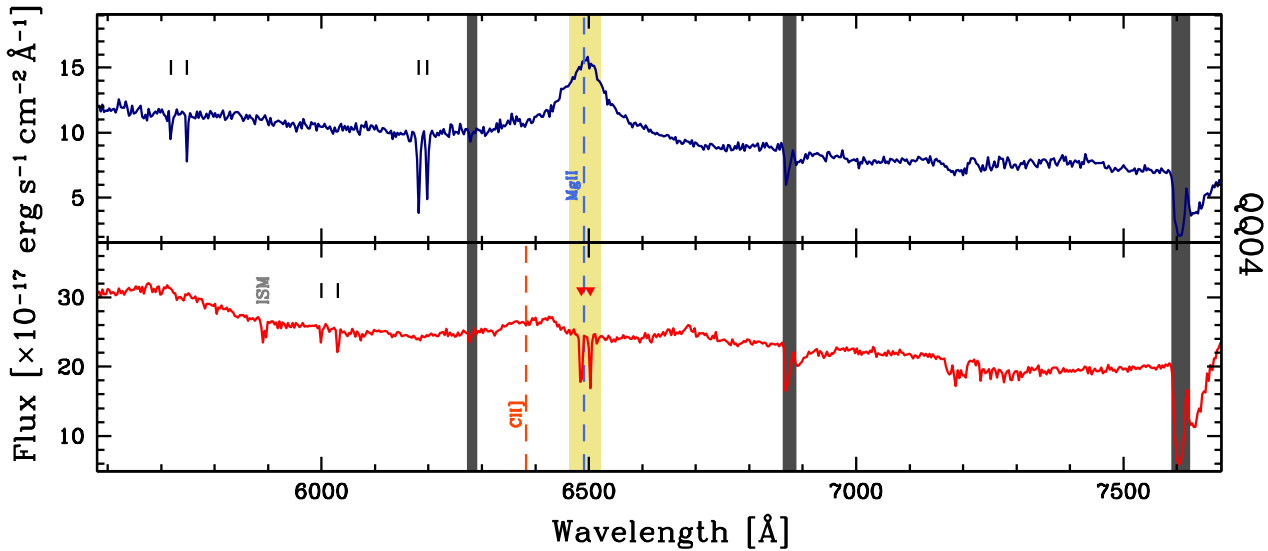}
\caption{
Spectra of the projected quasar pair QQ04 corrected for Galactic extinction
and binned by 3 pixels. 
Blue and red solid lines refer to \qsof\ and to \qsob, respectively.
Main quasar emission lines and Galactic absorption are labelled. 
The shaded yellow region marks the wavelength range considered to associate a 
\mgii\ doublet to the \qsof\ (see~\S\ref{sec:4}) and dark gray bars 
cover regions affected by prominent telluric absorptions.
Black thicks point to absorption lines detected over a $3\sigma$ threshold (see
Appendix~\ref{app:2}), and red triangles highlight the \mgii\ absorption 
doublet associated to the \qsof\ (see Table~\ref{tab:abs}).
Figures of the other quasar pairs are available in the electronic edition of 
the Journal.
}\label{fig:allspec}
\end{figure*}

\addtocounter{figure}{-1}
\begin{figure*}
\centering
\includegraphics[width=1.99\columnwidth]{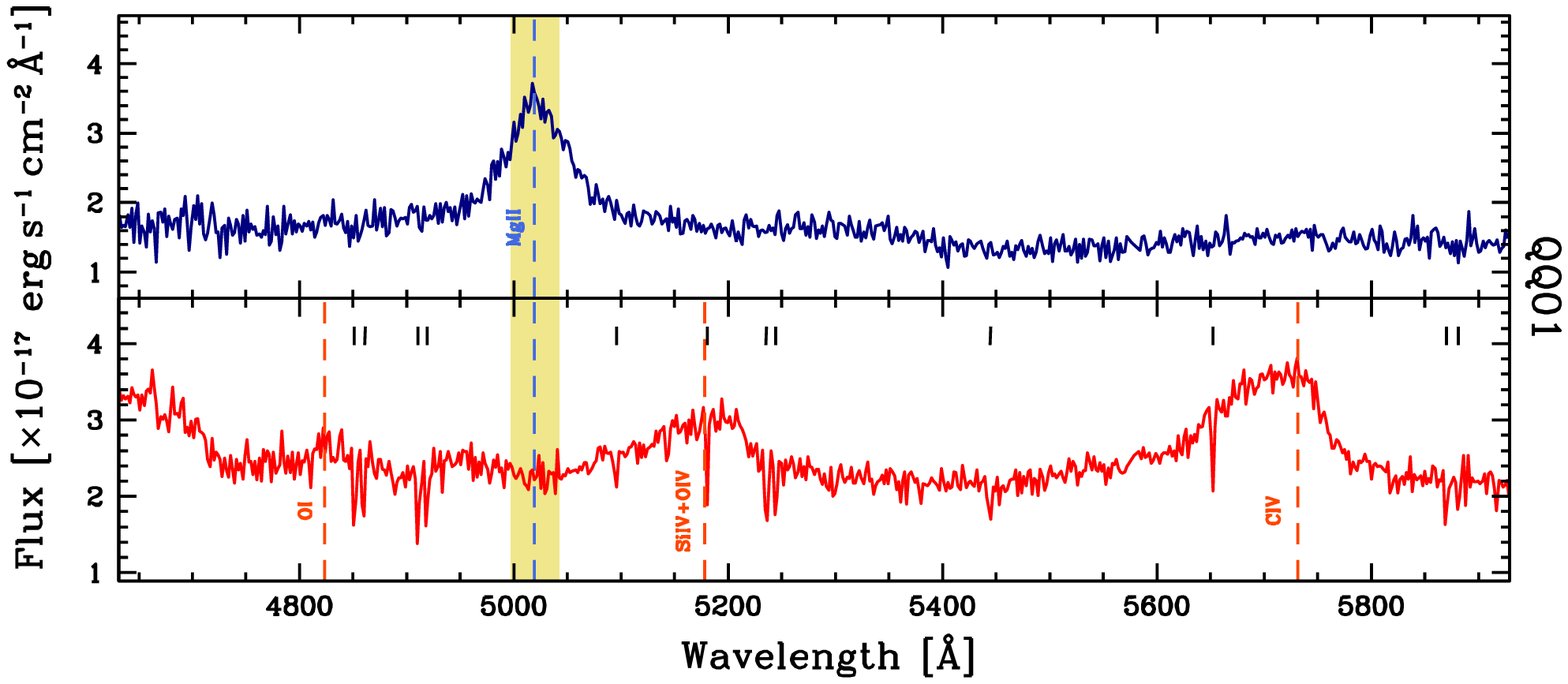}
\includegraphics[width=1.99\columnwidth]{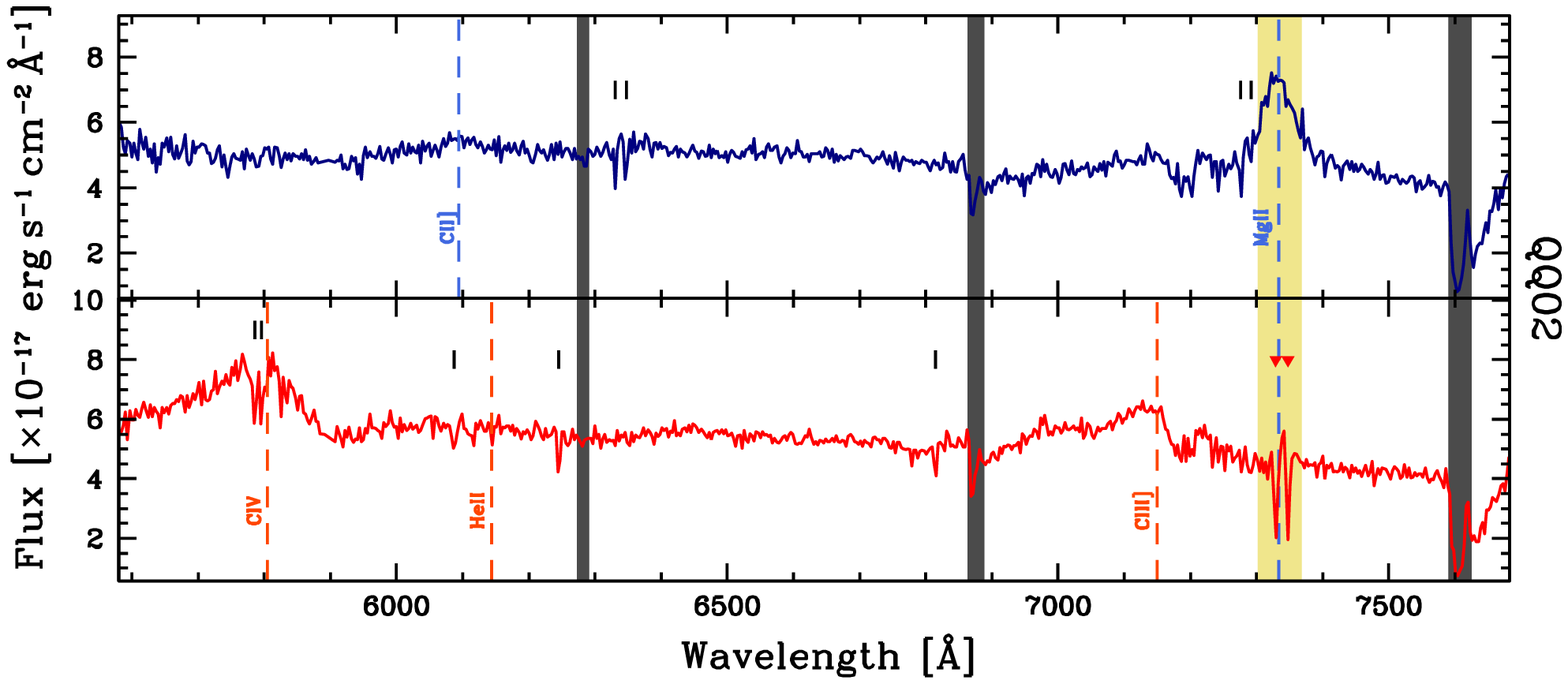}
\includegraphics[width=1.99\columnwidth]{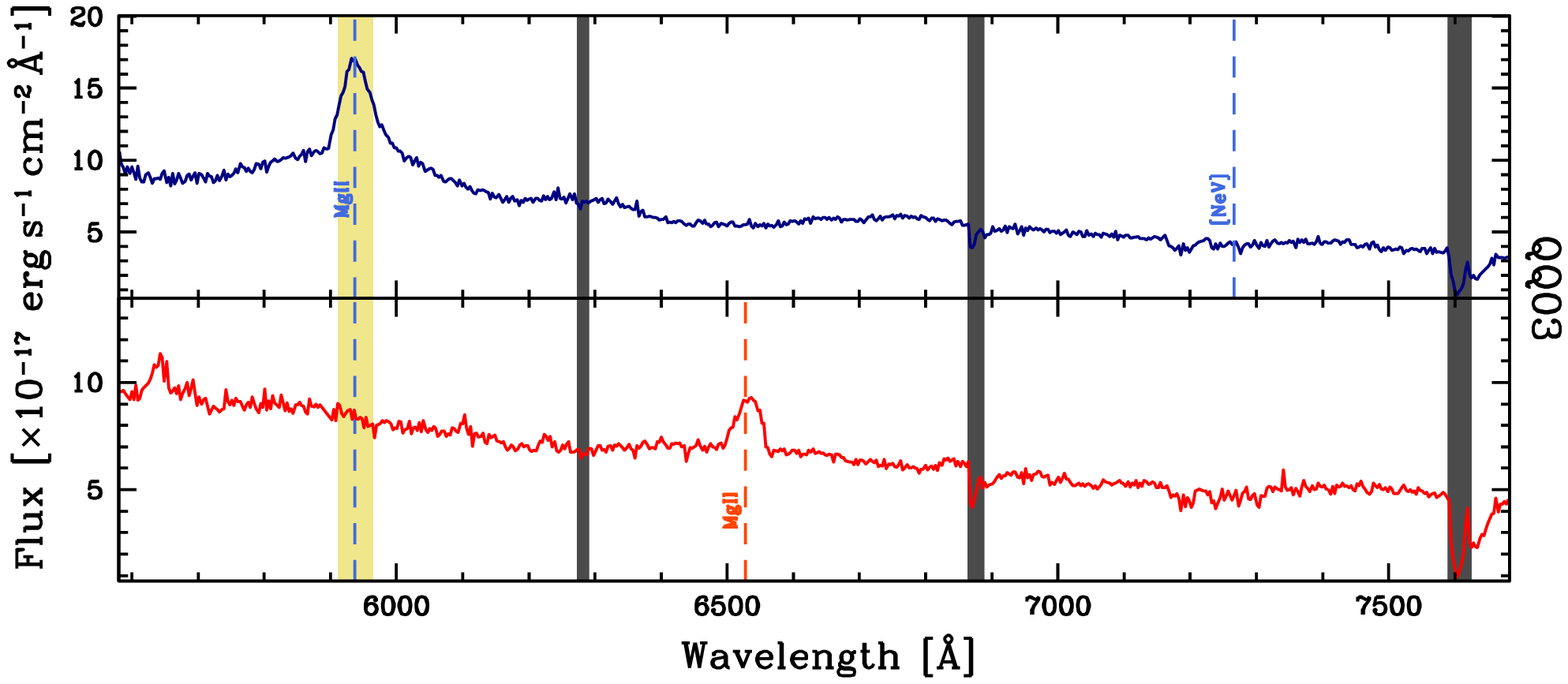}
\caption{ continued.}
\end{figure*}

\addtocounter{figure}{-1}
\begin{figure*}
\centering
\includegraphics[width=1.99\columnwidth]{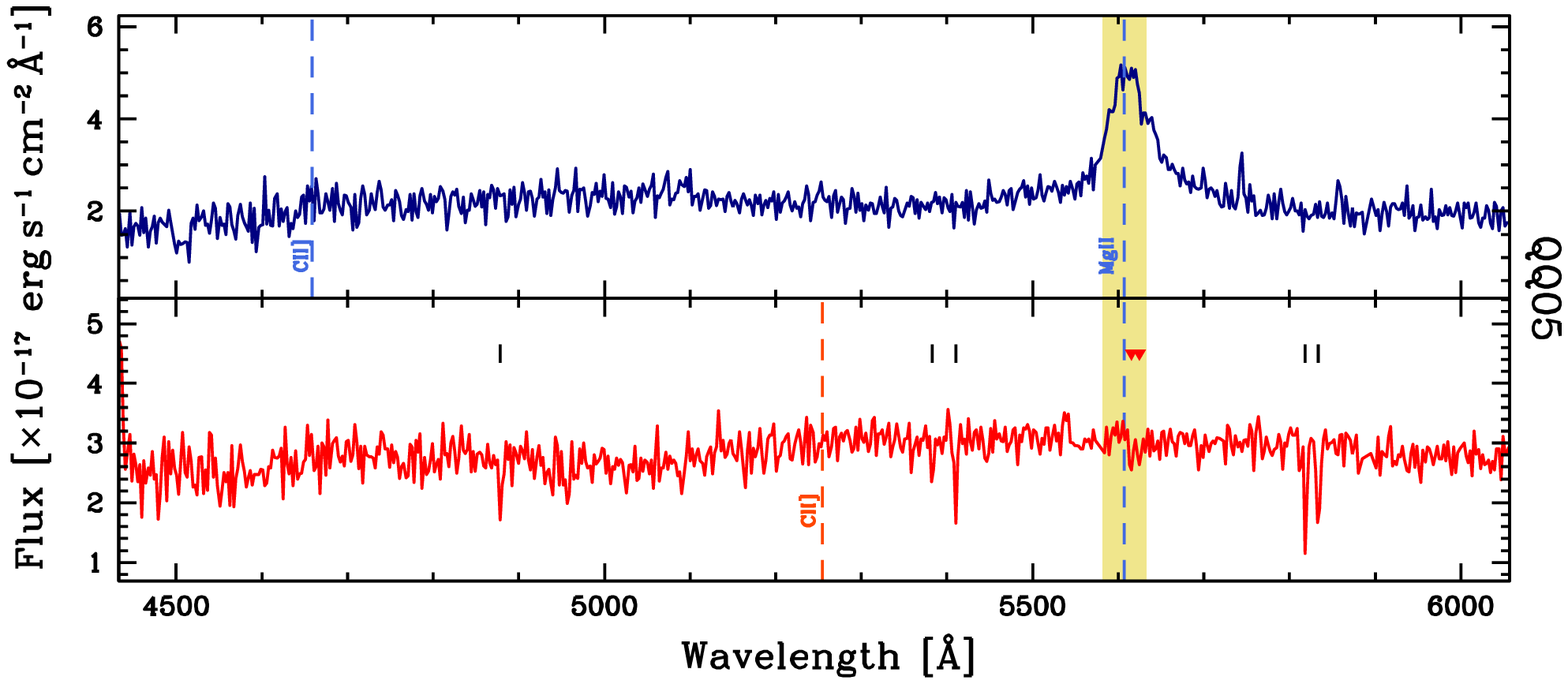}
\includegraphics[width=1.99\columnwidth]{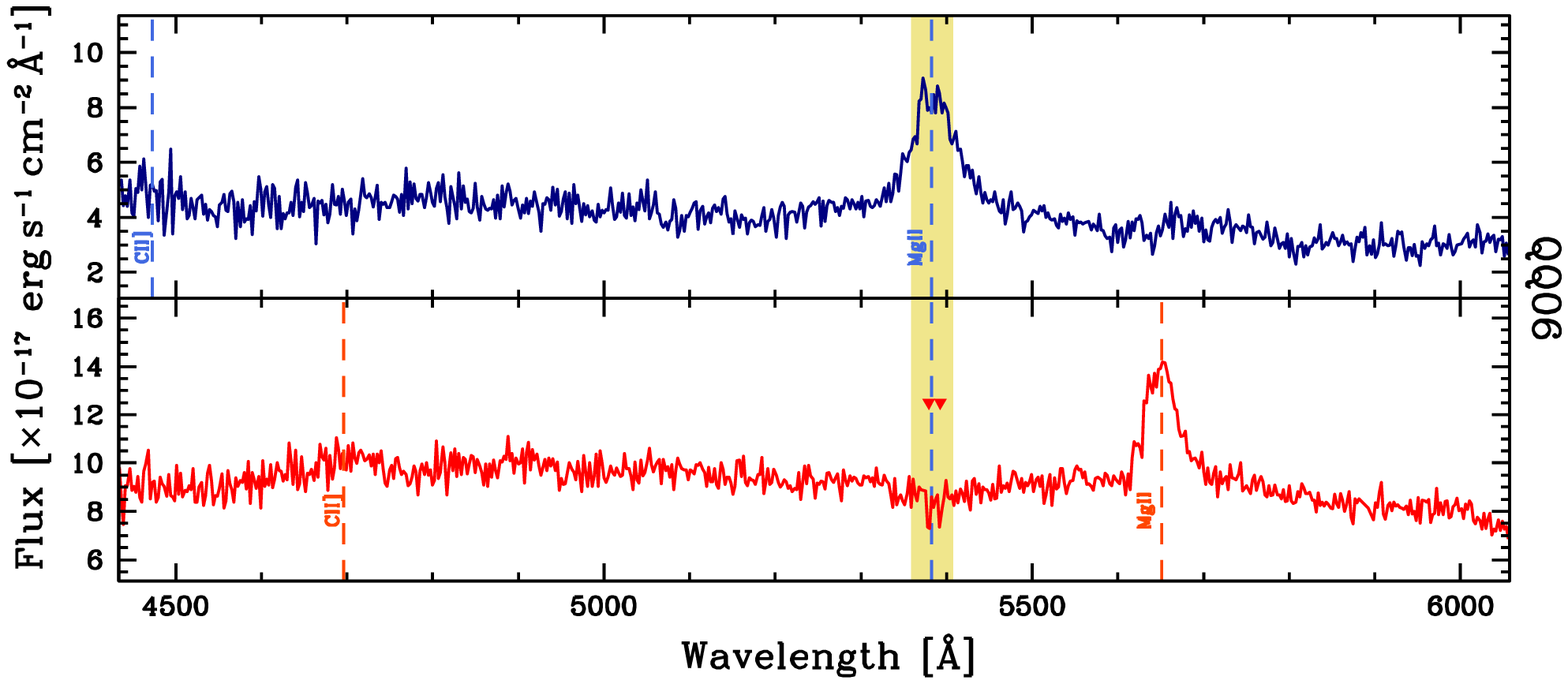}
\includegraphics[width=1.99\columnwidth]{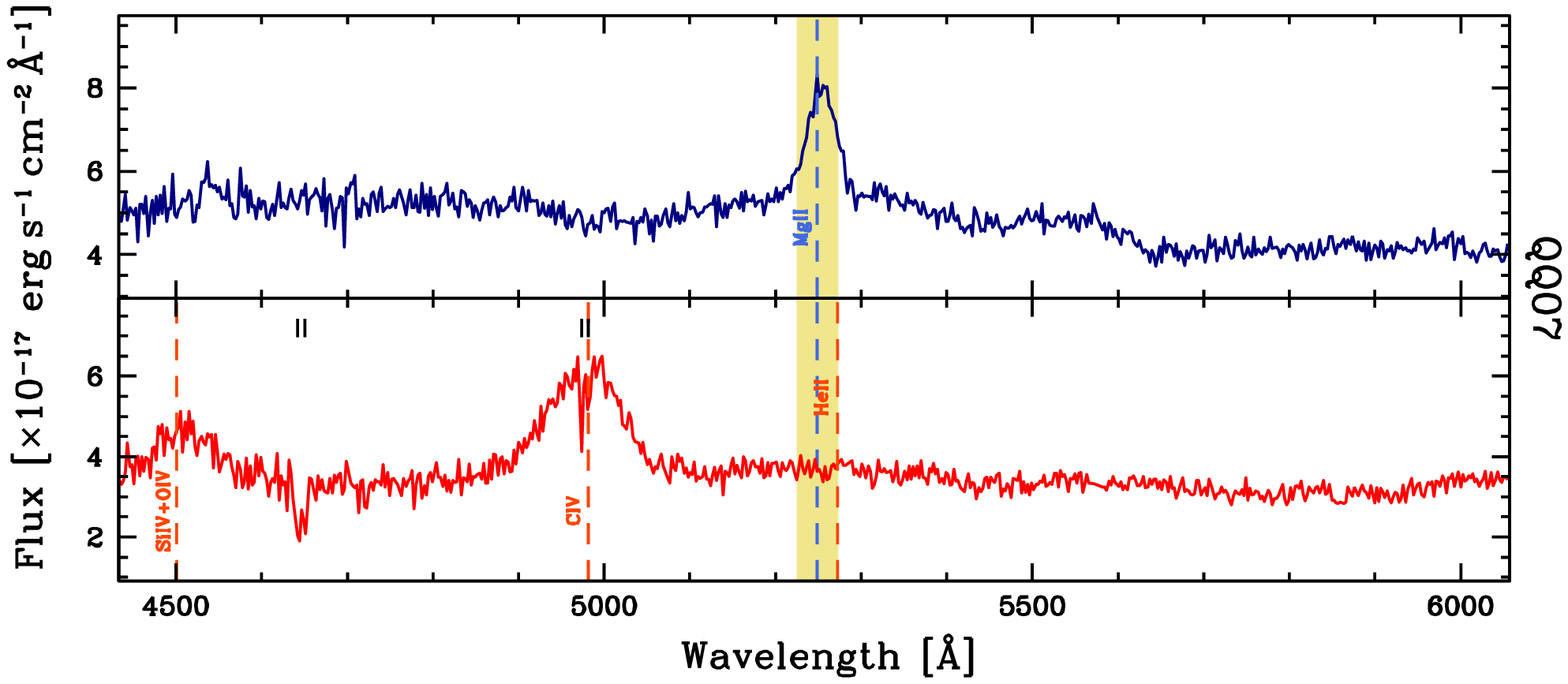}
\caption{ continued.}
\end{figure*}

\addtocounter{figure}{-1}
\begin{figure*}
\centering
\includegraphics[width=1.99\columnwidth]{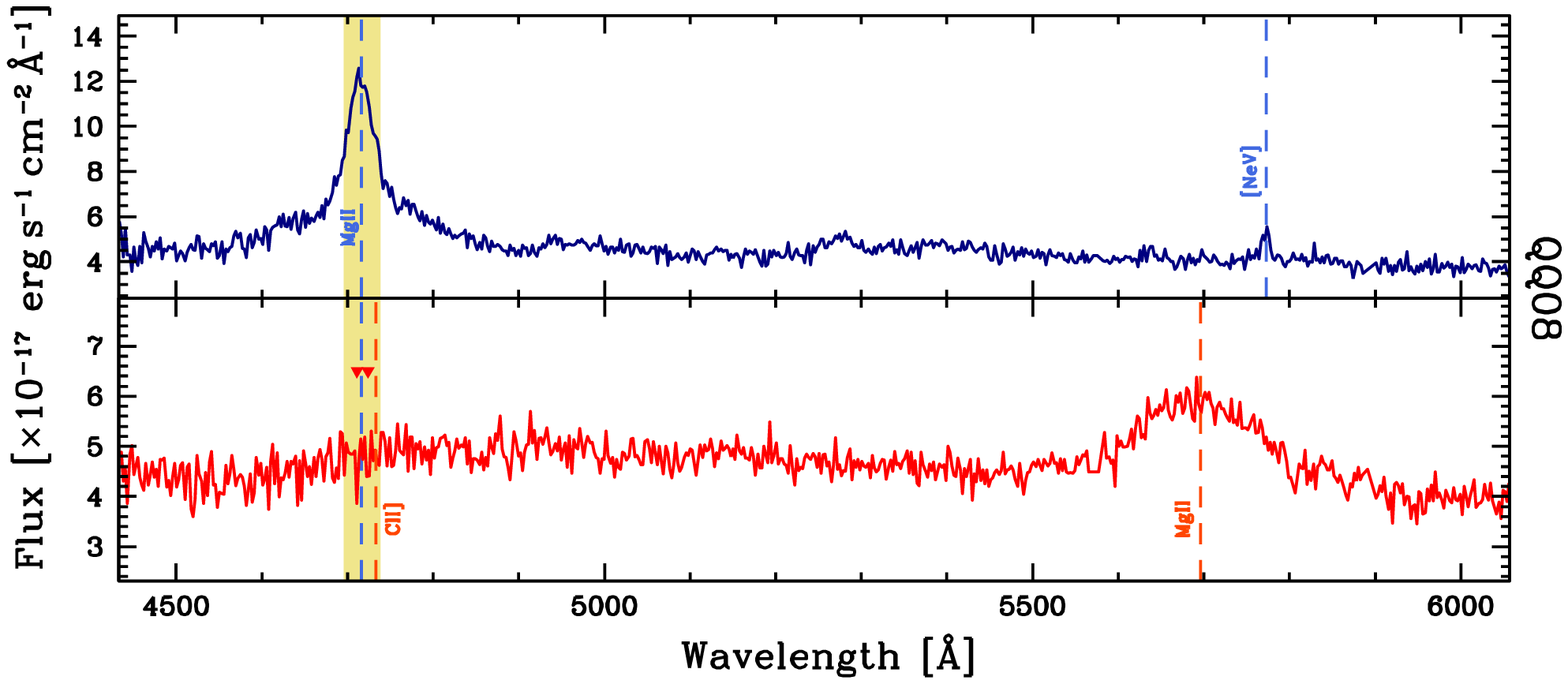}
\includegraphics[width=1.99\columnwidth]{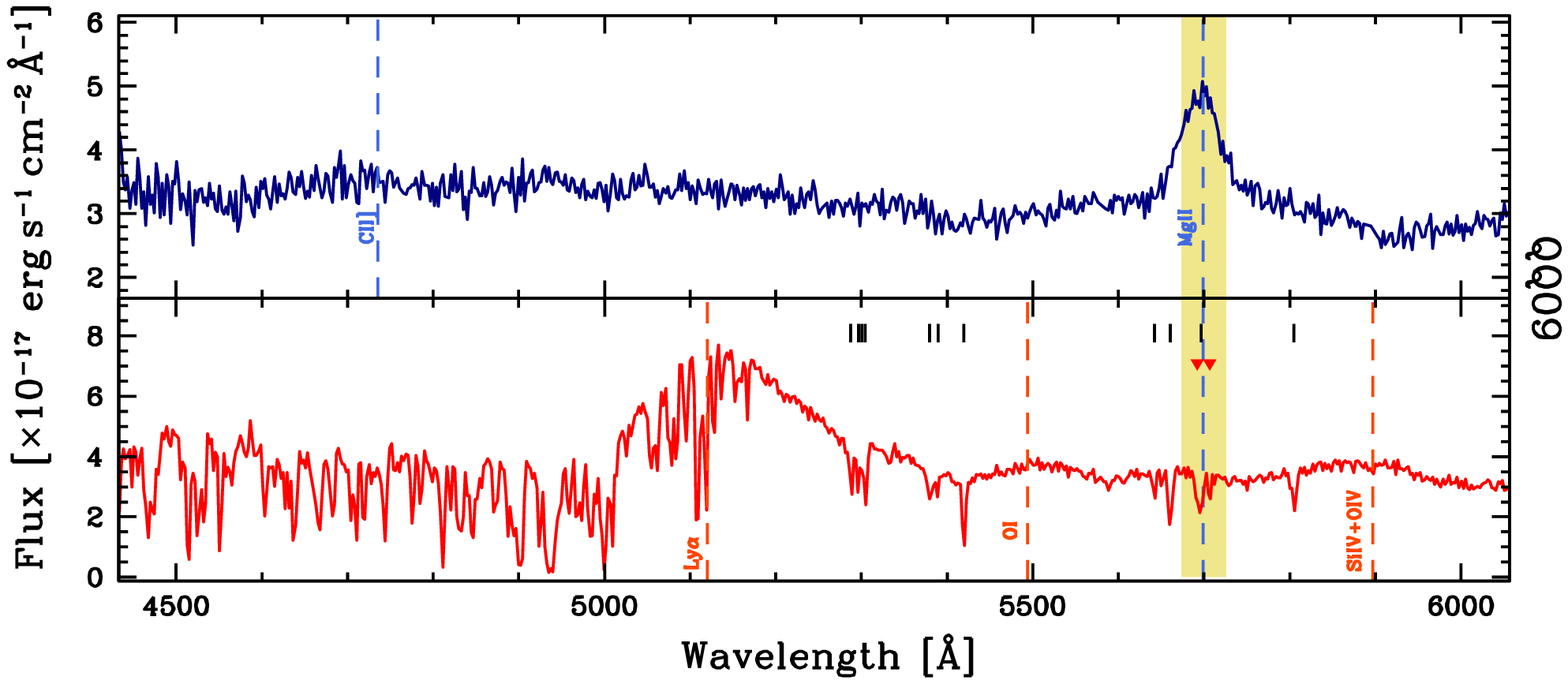}
\includegraphics[width=1.99\columnwidth]{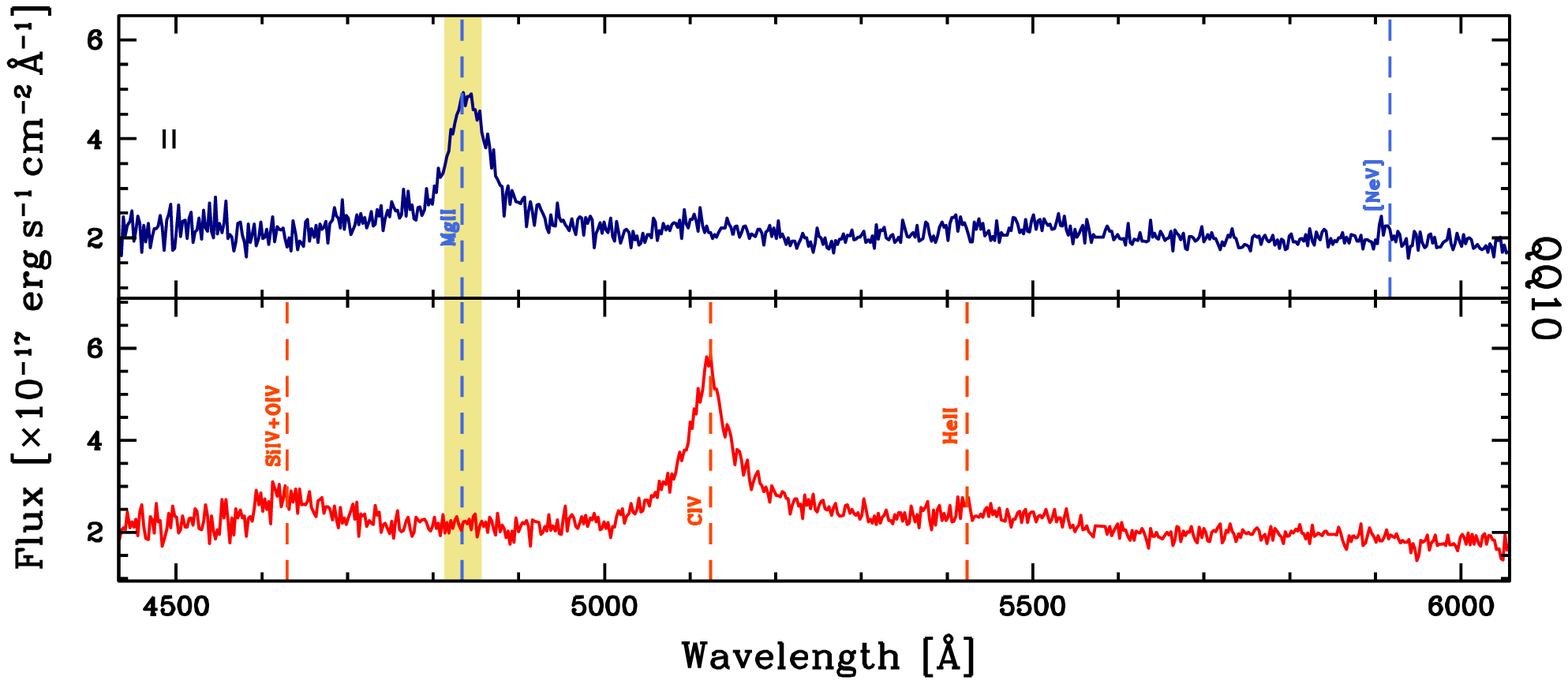}
\caption{ continued.}
\end{figure*}

\addtocounter{figure}{-1}
\begin{figure*}
\centering
\includegraphics[width=1.99\columnwidth]{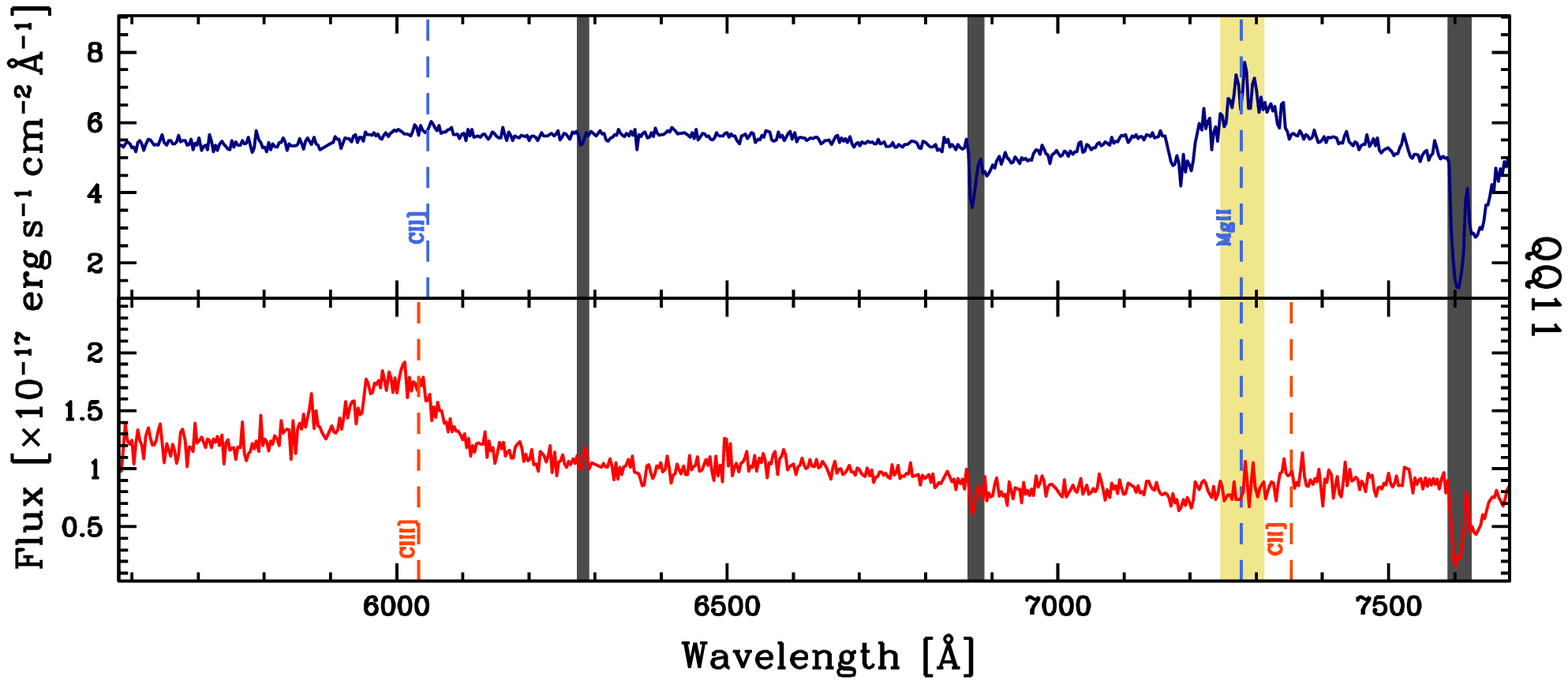}
\includegraphics[width=1.99\columnwidth]{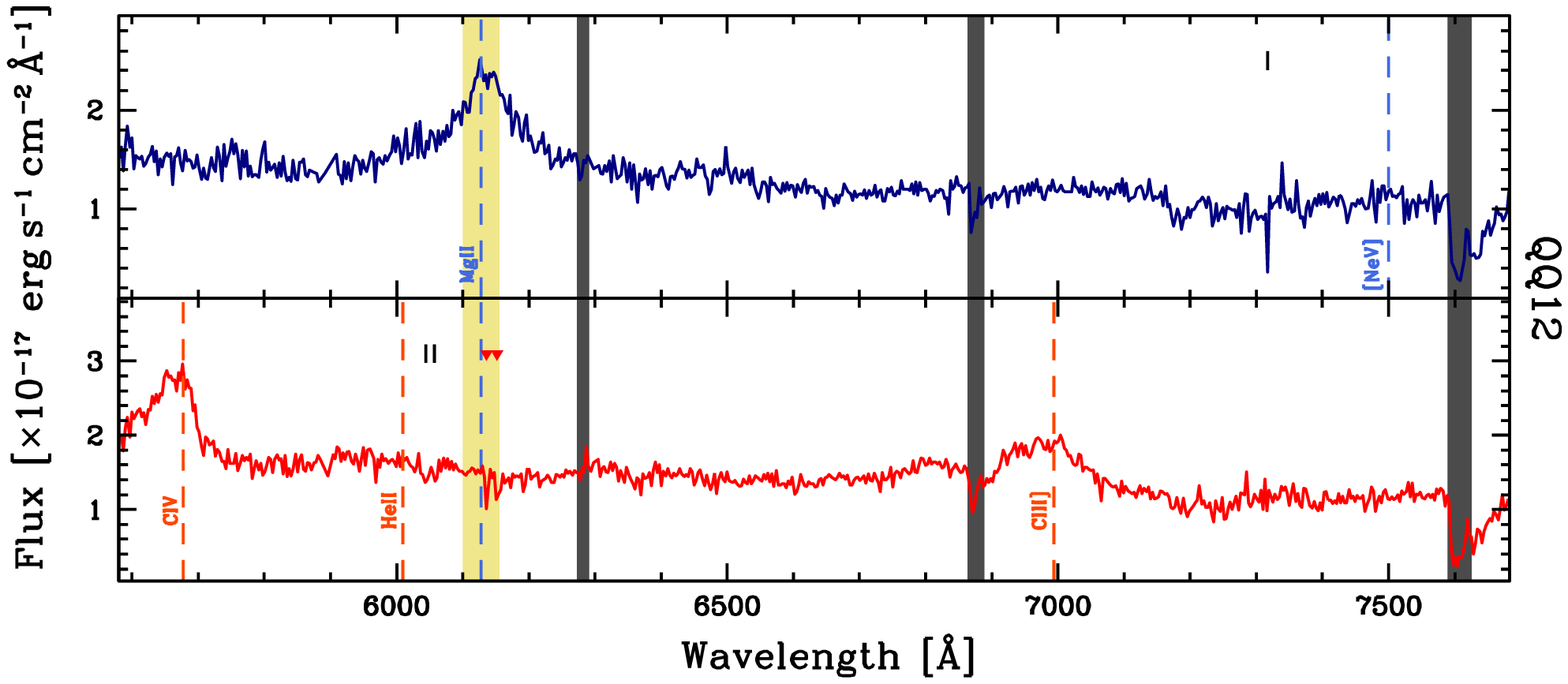}
\includegraphics[width=1.99\columnwidth]{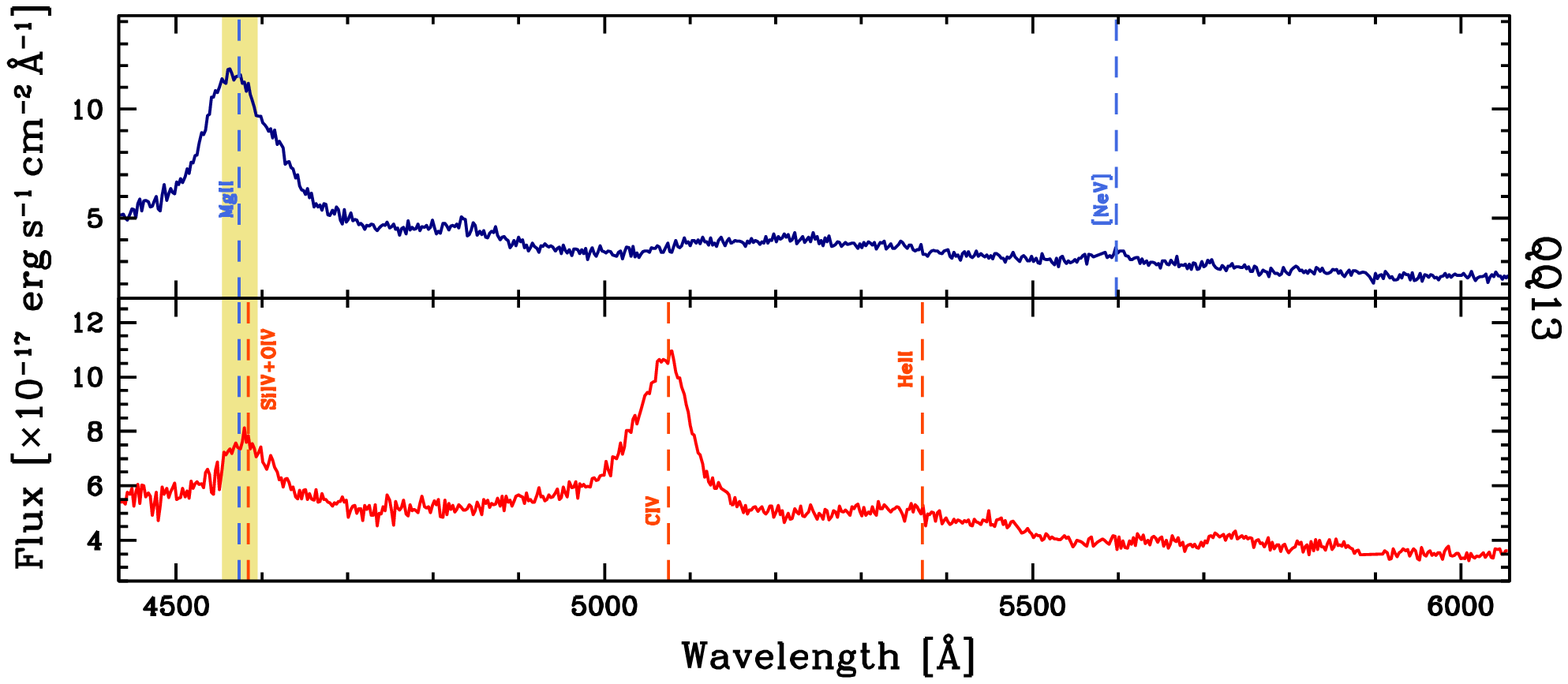}
\caption{ continued.}
\end{figure*}

\addtocounter{figure}{-1}
\begin{figure*}
\centering
\includegraphics[width=1.99\columnwidth]{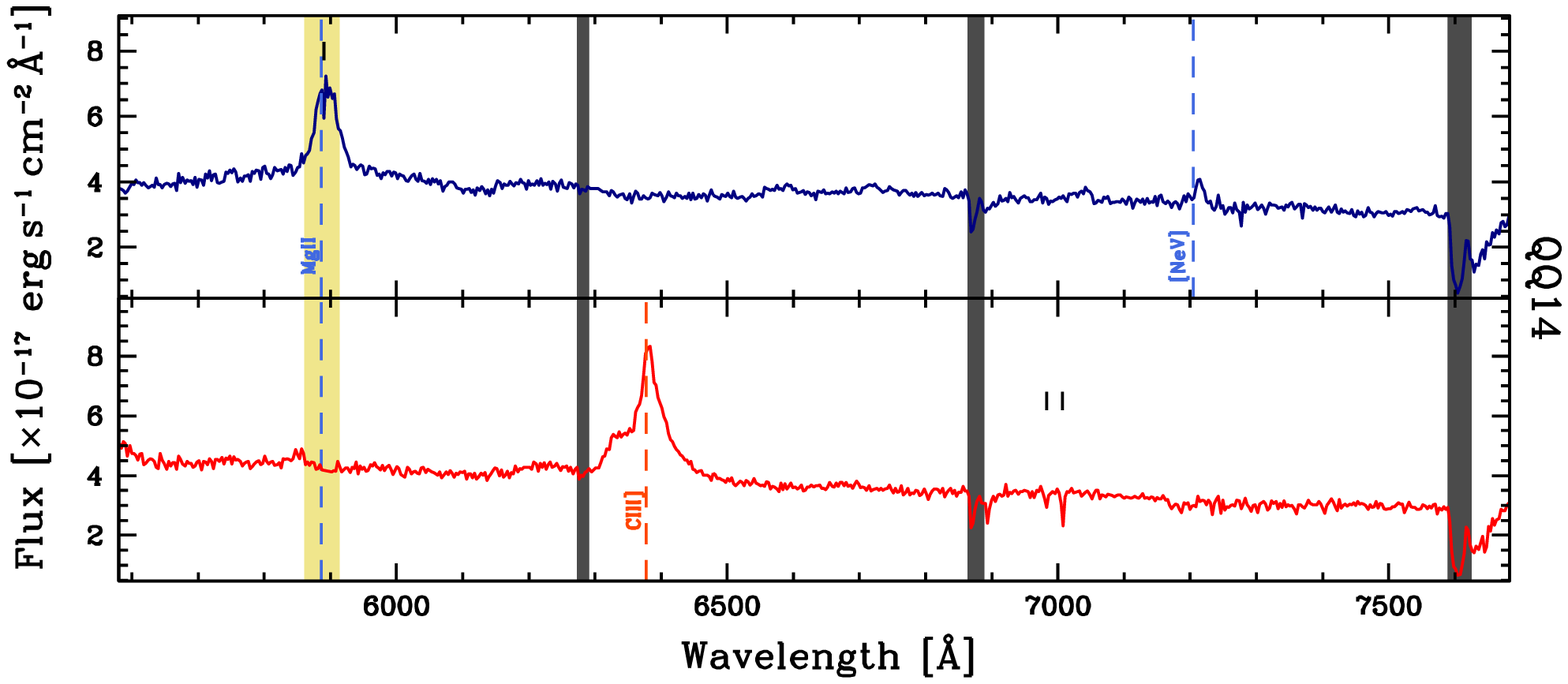}
\includegraphics[width=1.99\columnwidth]{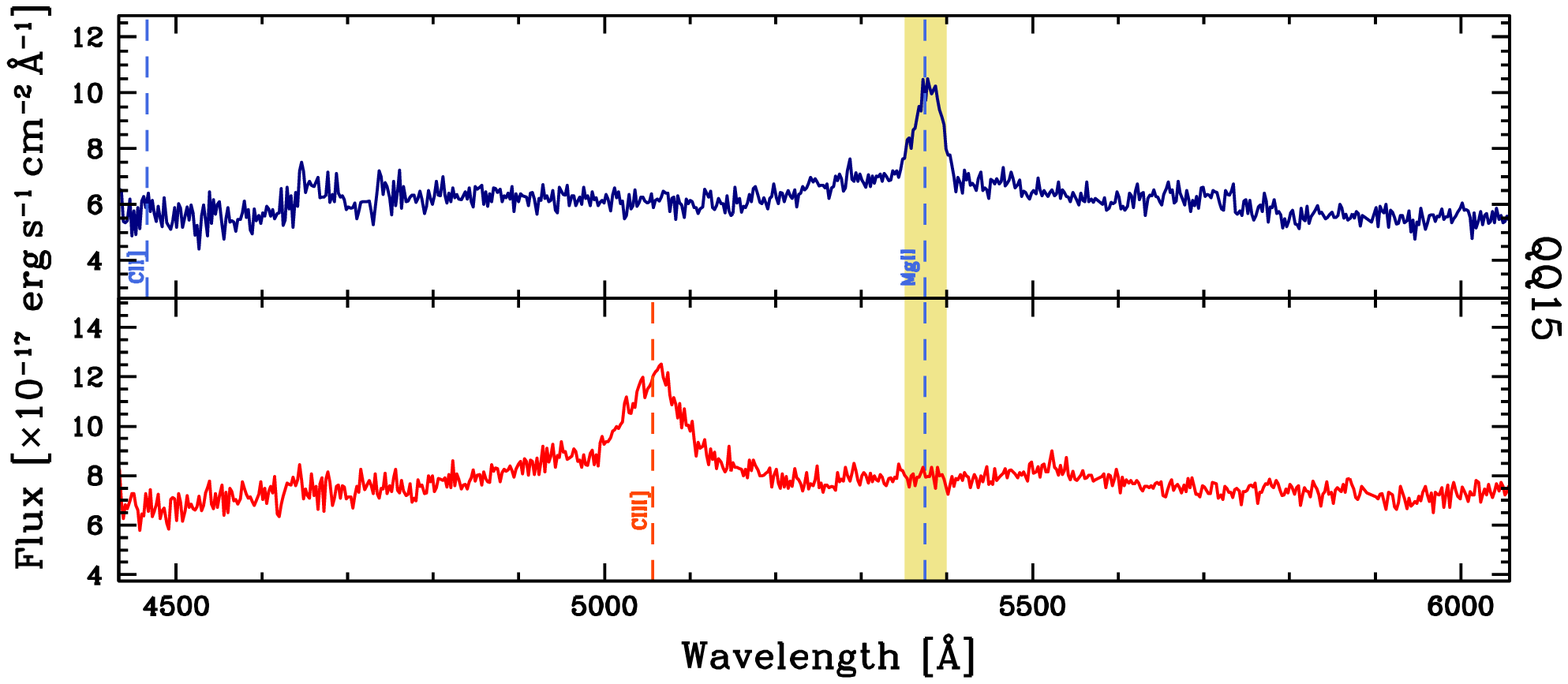}
\includegraphics[width=1.99\columnwidth]{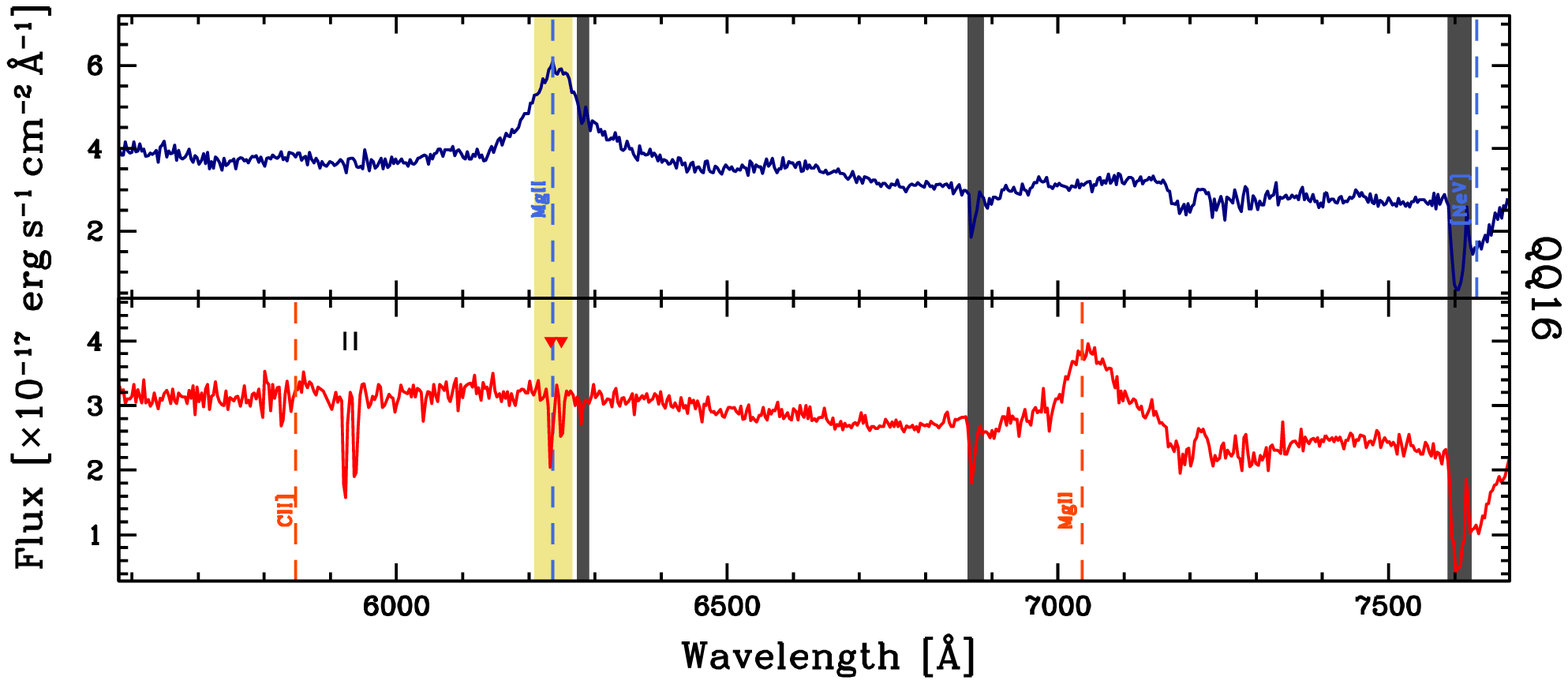}
\caption{ continued.}
\end{figure*}

\addtocounter{figure}{-1}
\begin{figure*}
\centering
\includegraphics[width=1.99\columnwidth]{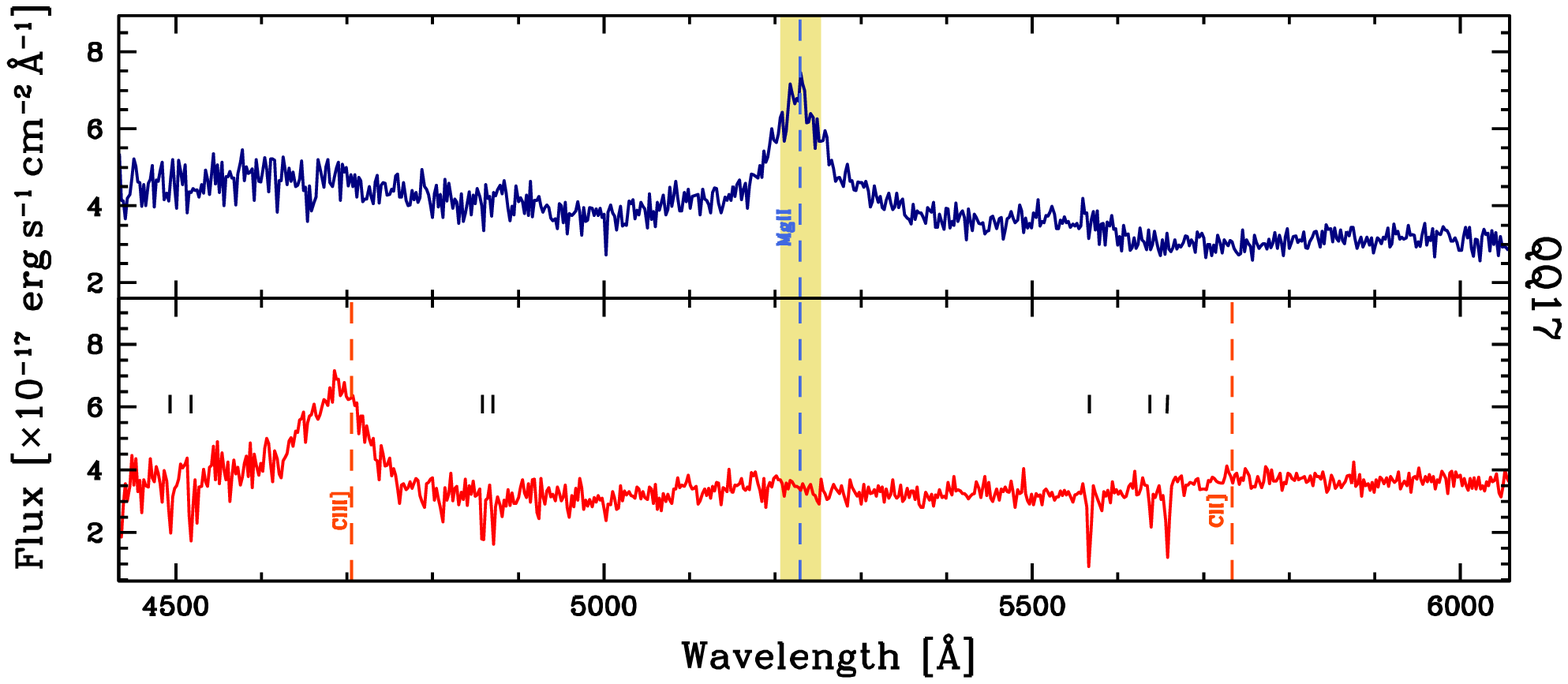}
\includegraphics[width=1.99\columnwidth]{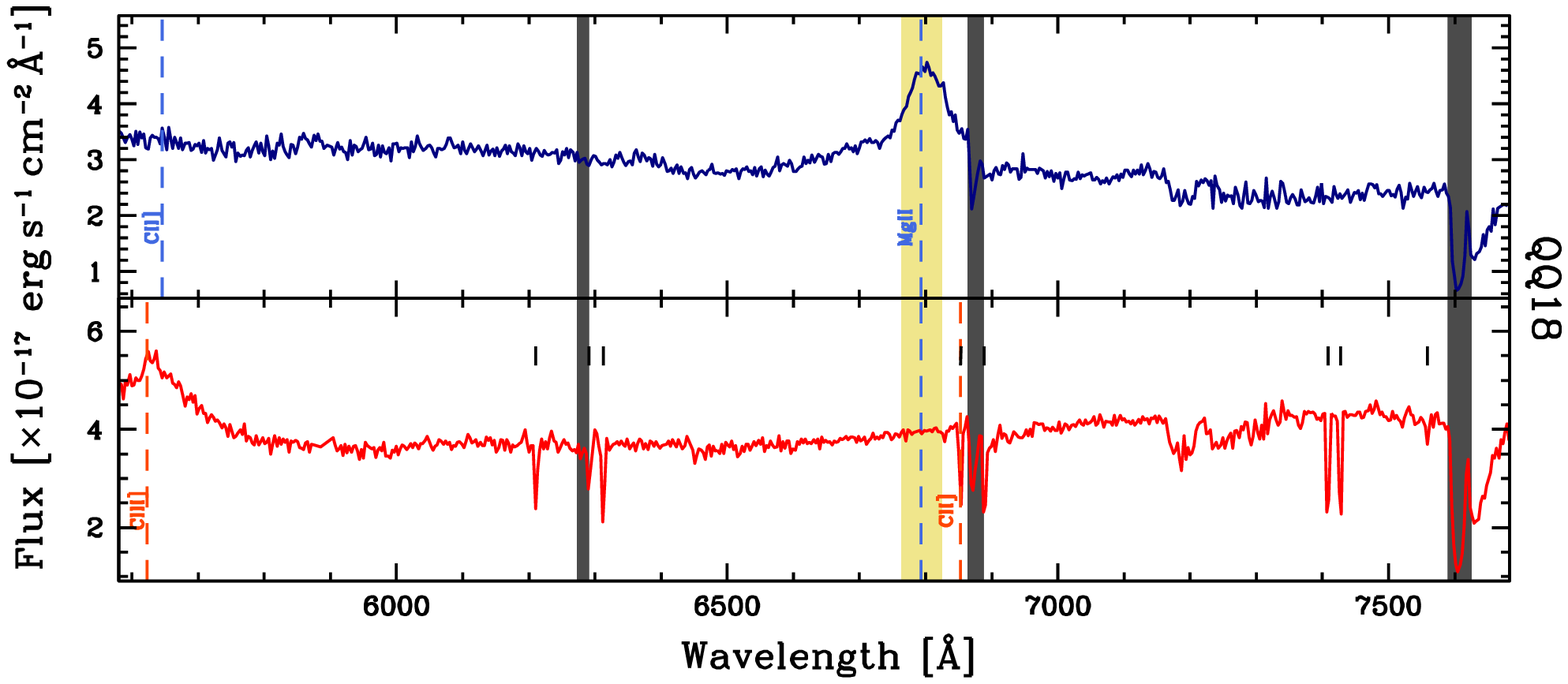}
\caption{ continued.}
\end{figure*}


\section[]{Absorption systems associated to quasars}\label{sec:4}

To investigate the presence of \mgii\ absorption lines in our spectra we 
adopted the procedure described in \citet{Farina2013}.
In summary, we first model the quasar emission by interpolating with a cubic 
spline the median values of the flux estimated in bins of $20$\,\AA\ 
each \citep[e.g.][]{Sbarufatti2005}.
Then, the $1\sigma$ detection limit of a spectral line was calculated from 
the equivalent width and the relative uncertainty of an unresolved absorption 
feature, assuming that it has the shape of a Gaussian with FWHM equal to the 
spectral resolution \citep[see][]{Schneider1993}.
Finally, the properties of the absorption features detected above a $3\sigma$ 
threshold were measured by fitting the lines with a single Gaussian function 
\citep[e.g.][]{Churchill2000}. 
Uncertainties on the derived quantities were estimated through standard error 
propagation, and are dominated by the noise of the continuum close to
the absorption lines.

Hereafter we will refer to an absorbers as {\it transverse} or as 
{\it line--of--sight} (LOS) depending on whether it was detected 
in the \qsob\ or in the \qsof\ spectrum.
A detected absorption system will be considered as {\it associated} 
to the \qsof\ if it lies within $\pm1000$\,\kms\ from the redshift 
derived from \mgii\ broad emission line (see Table~\ref{tab:sample}).
This operational definition was motivated by the large uncertainties
associated to the fit of the broad lines and by the possibility 
that redshifts derived from various emission lines may differ from the 
systemic redshift by even hundreds of \kms\ \citep[e.g.][]{Tytler1992, 
Richards2002, Bonning2007}, and that LOS absorbers within up to 
thousands of~\kms\ from a quasar may be still connected with the 
quasar itself \citep[e.g.][]{Wild2008, Shen2012}. A choice of
a wider velocity range to associate the LOS absorption systems 
to \qsof\ has only marginal effects on our results (see 
Section~\ref{sec:5}).

\begin{figure*}
\centering
\includegraphics[width=1.99\columnwidth]{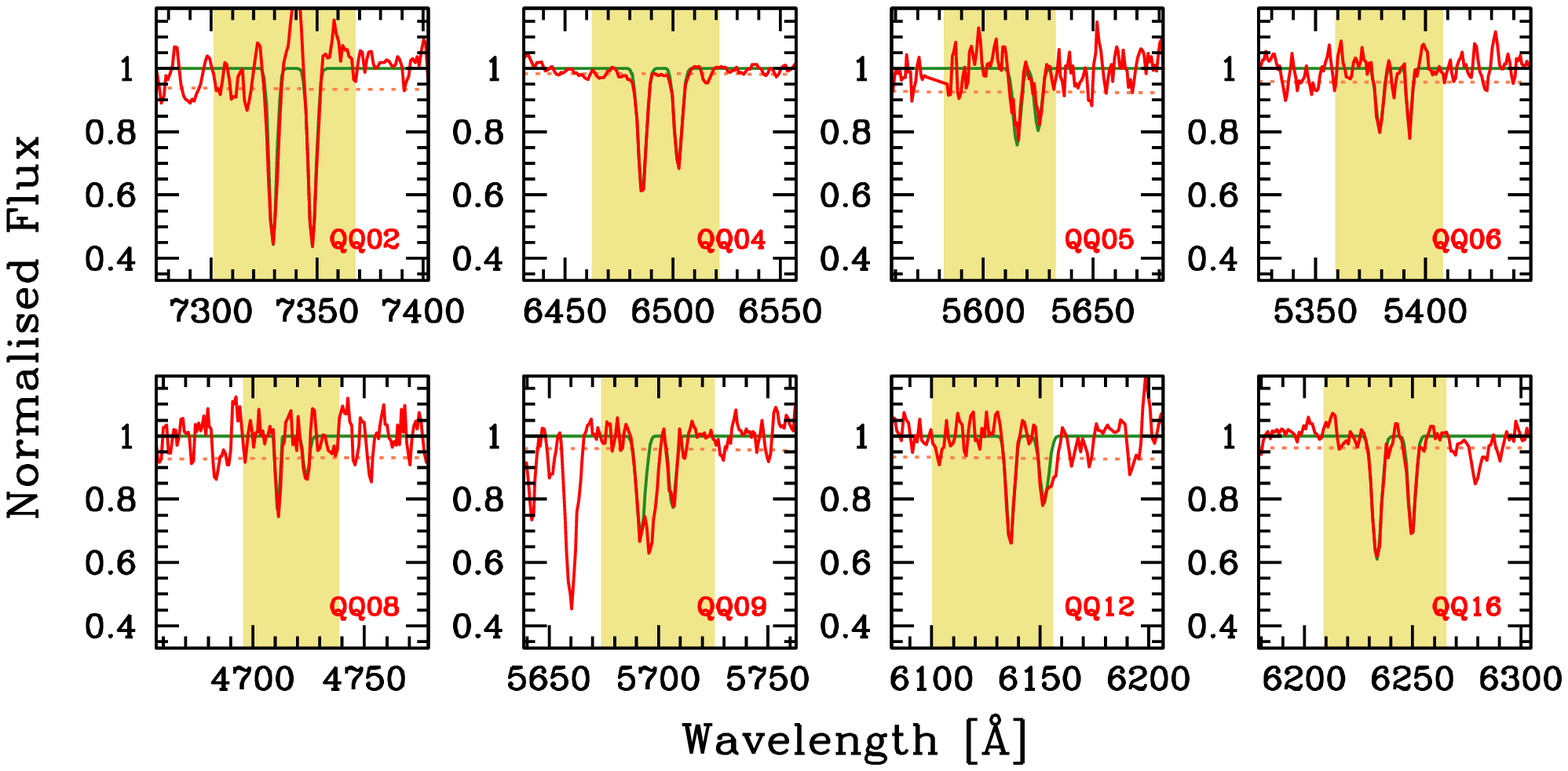}
\caption{
Close up of the normalised \qsob\ spectra where the transverse absorption 
systems are detected (red line). 
The Gaussian fit performed as described in \S\ref{sec:4} and in 
\citet{Farina2013} is marked with a green solid line, and the $1-1\sigma$ 
spectrum with a pale--red dotted line. 
The shaded yellow region shows the windows in which we consider the absorption 
system as associated to \qsof.
The system QQ08 exhibit several strong absorption systems close to the \mgii\ 
doublet that are not associated to \qsof\ (see Appendix~\ref{app:2}).
}\label{fig:abs}
\end{figure*}

In our new sample we detected 8~\mgii\ transverse absorption features 
associated to the \qsof\ (almost doubling the number of known such 
system, see \citealt{Bowen2006} and \citealt{Farina2013}), while no 
associated absorbers are present along the LOS (see Table~\ref{tab:abs} 
and Figure~\ref{fig:abs}).
The average shift between transverse absorbers and \mgii\ broad line 
is $\langle \Delta{\rm V}_{\rm QSO-Abs.}\rangle\sim-250$\,\kms\ with
an RMS of~$\sim380$\,\kms\ confirming the strict association with the
quasars.
This is further supported by the paucity of random absorbers present in 
the proximity of a quasars.
Indeed, integrating over $2000$\,\kms the redshift number density of systems 
with $\ewr(\lambda2796)\geq0.30$\,\AA, only $\sim0.01$ absorption systems are
expected \citep[][]{Nestor2005}. 
These results exclude that the observed \mgii\ lines are due to chance effect.
The rest frame equivalent width of the observed transverse absorption systems
range from $\ewr(\lambda2796)=0.33$\,\AA\ to $\ewr(\lambda2796)=0.92$\,\AA\ 
with an average $\ewr(\lambda2796)/\ewr(\lambda2803)$ doublet ratio of 
$\langle{\rm DR}\rangle=1.35\pm0.03$. 
This suggests that most of our systems are partially saturated as commonly 
observed in intervening \mgii\ doublets detected in quasar spectra 
\citep[e.g.][]{Nestor2005}.
The absence of LOS \mgii\ absorbers agrees with the studies performed by 
\citet{Vandenberk2008} and \citet{Shen2012} on SDSS spectra, which have 
shown that LOS \mgii\ absorbers occur only in a few percent of the 
examined systems, and that are possibly related to an enhanced star 
formation rate of the quasar host galaxy.

\begin{table*}
\centering
\caption{
Properties of \mgii\ absorption features associated to \qsof: 
our identification label of the quasar (ID), 
observed wavelength ($\lambda_{\rm abs}$),
rest frame equivalent width ($\ewr$),
doublet ratio (DR), and
redshift (z$_{\rm abs}$).
If no absorption system is present, the 2$\sigma$ upper limit for the $\ewr$ 
is quoted.
The labels ${\rm F}$ and ${\rm B}$ indicate the foreground and the background 
quasar, respectively.
}\label{tab:abs}
\begin{tabular}{lcccccc}
\hline
ID     &  $\lambda_{\rm abs}(\lambda2796)$ & W$_{\rm r}(\lambda2796)$ &  $\lambda_{\rm abs}(\lambda2803)$ & W$_{\rm r}(\lambda2803)$ & DR & z$_{\rm abs}$ \\ 
       &  (\AA)	                           & (\AA)        	      &  (\AA) 			          & (\AA)                     &    &              \\      	  
\hline  		 
\hline  		 
 QQ01F &  \dots  & $<$0.22	  &   \dots  &   $<$0.22	&  \dots	 &  \dots		\\  
 QQ01B &  \dots  & $<$0.13	  &   \dots  &   $<$0.13	&  \dots	 &  \dots		\\
 QQ02F &  \dots  & $<$0.17	  &   \dots  &   $<$0.17	&  \dots	 &  \dots		\\   
 QQ02B & 7329.1  & 0.92$\pm$0.10  &  7347.7  &   0.98$\pm$0.09  &  0.94$\pm$0.12 &  1.6213$\pm$0.0001	\\
 QQ03F &  \dots  & $<$0.12	  &   \dots  &   $<$0.12	&  \dots	 &  \dots		\\  
 QQ03B &  \dots  & $<$0.15	  &   \dots  &   $<$0.15	&  \dots	 &  \dots		\\
 QQ04F &  \dots  & $<$0.12	  &   \dots  &   $<$0.12	&  \dots	 &  \dots		\\  
 QQ04B & 6485.8  & 0.78$\pm$0.03  &  6502.5  &   0.58$\pm$0.02  &  1.34$\pm$0.04 &  1.3197$\pm$0.0001	\\
 QQ05F &  \dots  & $<$0.21	  &   \dots  &   $<$0.21	&  \dots	 &  \dots		\\  
 QQ05B & 5615.2  & 0.50$\pm$0.06  &  5624.8  &   0.39$\pm$0.05  &  1.29$\pm$0.19 &  1.0075$\pm$0.0008	\\
 QQ06F &  \dots  & $<$0.22	  &   \dots  &   $<$0.22	&  \dots	 &  \dots		\\  
 QQ06B & 5379.1  & 0.39$\pm$0.07  &  5392.7  &   0.28$\pm$0.08  &  1.41$\pm$0.12 &  0.9239$\pm$0.0001	\\
 QQ07F &  \dots  & $<$0.17	  &   \dots  &   $<$0.17	&  \dots	 &  \dots		\\  
 QQ07B &  \dots  & $<$0.14	  &   \dots  &   $<$0.14	&  \dots	 &  \dots		\\
 QQ08F &  \dots  & $<$0.19	  &   \dots  &   $<$0.19	&  \dots	 &  \dots		\\  
 QQ08B & 4711.2  & 0.33$\pm$0.04  &  4724.0  &   0.24$\pm$0.03  &  1.37$\pm$0.19 &  0.6852$\pm$0.0002	\\
 QQ09F &  \dots  & $<$0.17	  &   \dots  &   $<$0.17	&  \dots	 &  \dots		\\  
 QQ09B & 5691.9  & 0.72$\pm$0.05  &  5706.5  &   0.42$\pm$0.03  &  1.71$\pm$0.14 &  1.0358$\pm$0.0001	\\
 QQ10F &  \dots  & $<$0.22	  &   \dots  &   $<$0.22	&  \dots	 &  \dots		\\  
 QQ10B &  \dots  & $<$0.17	  &   \dots  &   $<$0.17	&  \dots	 &  \dots		\\
 QQ11F &  \dots  & $<$0.16	  &   \dots  &   $<$0.16	&  \dots	 &  \dots		\\  
 QQ11B &  \dots  & $<$0.24	  &   \dots  &   $<$0.24	&  \dots	 &  \dots		\\
 QQ12F &  \dots  & $<$0.21	  &   \dots  &   $<$0.21	&  \dots	 &  \dots		\\  
 QQ12B & 6136.2  & 0.64$\pm$0.07  &  6152.0  &   0.47$\pm$0.06  &  1.36$\pm$0.19 &  1.1947$\pm$0.0001	\\
 QQ13F &  \dots  & $<$0.17	  &   \dots  &   $<$0.17  	&  \dots         &  \dots		\\  
 QQ13B &  \dots  & $<$0.11	  &   \dots  &   $<$0.11  	&  \dots         &  \dots		\\
 QQ14F &  \dots  & $<$0.17	  &   \dots  &   $<$0.17  	&  \dots         &  \dots		\\  
 QQ14B &  \dots  & $<$0.11	  &   \dots  &   $<$0.11  	&  \dots         &  \dots		\\
 QQ15F &  \dots  & $<$0.18	  &   \dots  &   $<$0.18  	&  \dots         &  \dots		\\  
 QQ15B &  \dots  & $<$0.14	  &   \dots  &   $<$0.14  	&  \dots         &  \dots		\\
 QQ16F &  \dots  & $<$0.13	  &   \dots  &   $<$0.13  	&  \dots         &  \dots		\\  
 QQ16B & 6233.6  & 0.91$\pm$0.04  &  6249.7  &   0.61$\pm$0.04  &  1.51$\pm$0.11 &  1.2296$\pm$0.0001	\\
 QQ17F &  \dots  & $<$0.21	  &   \dots  &   $<$0.21  	&  \dots         &  \dots		\\  
 QQ17B &  \dots  & $<$0.22	  &   \dots  &   $<$0.22  	&  \dots         &  \dots		\\
 QQ18F &  \dots  & $<$0.13	  &   \dots  &   $<$0.13  	&  \dots         &  \dots		\\  
 QQ18B &  \dots  & $<$0.11	  &   \dots  &   $<$0.11  	&  \dots         &  \dots		\\
\hline
\end{tabular}
\end{table*}


\section[]{Discussion}\label{sec:5}

In this Section, we report on the detected \mgii\ absorption systems and  
relate their properties to the impact parameters, to the mass of the host
galaxies, and to the direction of view (i.e., transverse or LOS).
We also compare our results with the properties of the CGM of normal galaxies, 
for which \mgii\ absorption systems were detected up to $\pd\sim200$\,kpc 
\citep[e.g.][and references therein]{Bergeron1991, Steidel1997, Kacprzak2008, 
Chen2010a, Nielsen2013b}.
In particular we investigate whether galaxies and quasars show different trend
in the well known anti--correlation between $\ewr(\lambda2796)$ and the impact 
parameter \citep[e.g.][]{Lanzetta1990, Steidel1992, Steidel1995}.

\subsection{Covering Fraction}

To study the covering fraction ($\fc$) of \mgii\ absorption systems at 
different impact parameters from the quasar, we define $\fc\equiv\fc
\left(\ew_{\rm lim}\right)$ as the fraction of absorbers with $\ewr$
greater than a given equivalent width ($\ew_{\rm lim}$) detected in 
each bin of projected distance.
If the upper limit on the equivalent width of an absorber (see 
Table~\ref{tab:abs}) is larger than $\ew_{\rm lim}$, this system 
is not considered in the estimate.
The $1\sigma$ uncertainties in $\fc$ are calculated upon the binomial
statistics \citep[e.g.][]{Gehrels1986, Cameron2011}.

\begin{figure*}
\centering
\includegraphics[width=0.99\columnwidth]{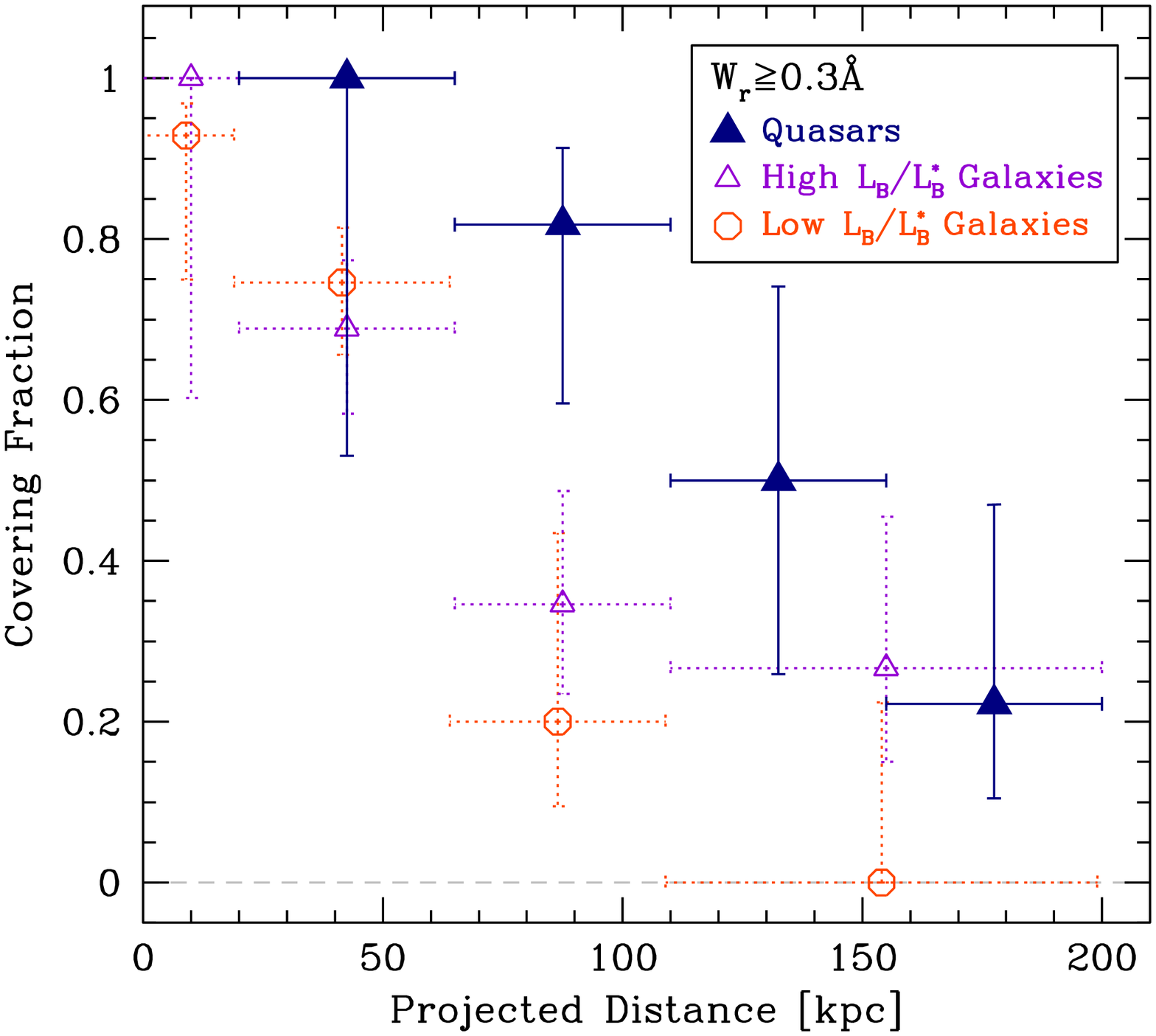}
\qquad
\includegraphics[width=0.99\columnwidth]{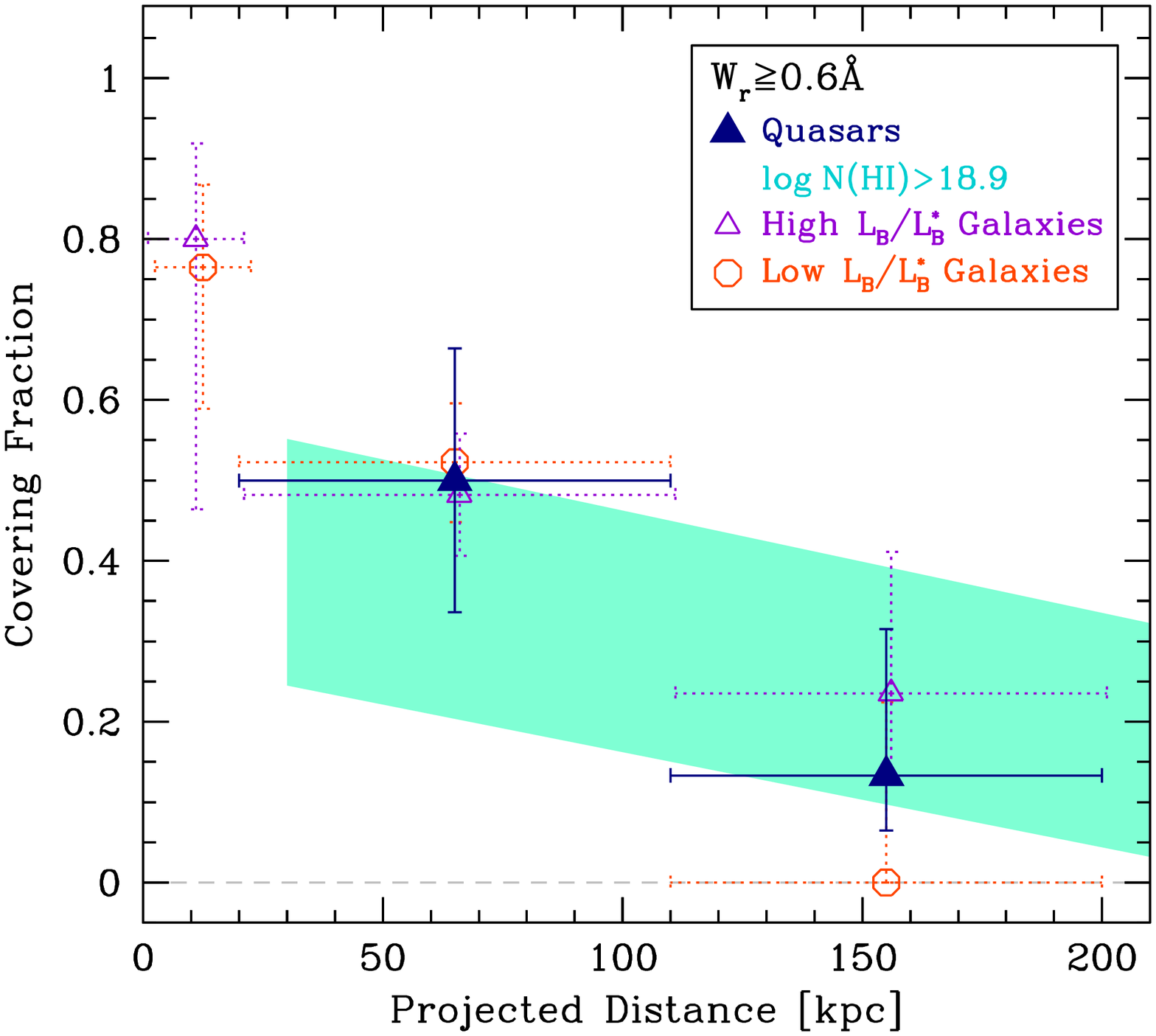}
\caption{
{\it Left Panel:}
Covering fraction profile for transverse absorption systems with $\ewr(\lambda
2796)>0.30$\,\AA\ associated to \qsof\ (blue triangles). 
The horizontal bars indicate the impact parameter bin width and vertical bars 
are the $1\sigma$ binomial uncertainties \citep[e.g.][]{Gehrels1986, 
Cameron2011}.
For comparison we plot also the \mgii\ covering fraction of  182 isolated 
(i.e., without a companion closer than 100\,kpc and 500\,\kms) galaxies 
investigated by \citet[][]{Nielsen2013a} with with B--band luminosities 
larger (violet empty triangles) and smaller (orange empty circles) than 
$L_{\rm B}\sim0.6\,L_{\rm B}^\star$, where $L_{\rm B}^\star$ is the 
characteristic luminosity of galaxies as derived from~\citet{Faber2007}.
We converted the upper limits listed in \citet[][]{Nielsen2013a} to
the 2$\sigma$ limits considered here.
{\it Right Panel:}
Same of Left Panel but for absorption systems with $\ewr(\lambda2796)>0.6$\,\AA.
We also show the $\pm1\sigma$ region of the covering fraction of the \hi\ 
absorbers associated to high redshift quasars ($1.6\lsim\z\lsim3.2$) presented 
by \citet[][pale blue filled area]{Prochaska2013a}.
For the sake of comparison we limit the \hi\ column density to $N_{\rm H\,I}
\gsim10^{18.9}$\,cm$^{-2}$ that roughly corresponds to the considered \mgii\ 
equivalent width limit (see text for details).
}\label{fig:cov}
\end{figure*}

In Figure~\ref{fig:cov} we plot $\fc$ of \mgii\ transverse absorbers 
associated to quasars against the impact parameter, including data 
from \citet{Bowen2006} and \citet{Farina2013}. The covering fraction 
of absorbers with $\ewr(\lambda2796)\geq0.30$\,\AA\ is $\fcmgii(0.30
\,\textrm{\AA})=1.00^{+0.00}_{-0.47}$ in the first bin ($20\,{\rm kpc}
<\pd\leq65\,{\rm kpc}$) and smoothly decrease with the impact parameter
down to $\fcmgii(0.30\,\textrm{\AA})=0.22^{+0.24}_{-0.12}$ at $155\,{\rm kpc}
<\pd\leq 200\,{\rm kpc}$.
In our sample 10 \mgii\ absorbers have $\ewr(\lambda2796)\geq0.60$\,\AA,
the covering fraction of these systems is $\fc(0.60\,\textrm{\AA})=
0.50^{+0.16}_{-0.16}$ between $20$\,kpc and $110$\,kpc and 
$\fc(0.60\,\textrm{\AA})=0.13^{+0.18}_{-0.07}$ in the $110$--$200$\,kpc 
bin. 

Various studies on the incidence of \mgii\ absorbers agree that the
number of such systems increase in the proximity of a galaxy.
However, the derived covering fractions span a broad range of 
values (i.e., from $\sim0.2$ to $\sim1$) depending on the diverse 
sets of explored galaxy properties, impact parameter, and equivalent 
width limit \citep[e.g.][]{Bechtold1992, Tripp2005, Barton2009, 
Gauthier2010, Lovegrove2011, Lundgren2012}.
In order to minimise the possible bias towards a specific galaxy 
population, we will assume as reference the recent estimates of 
\citet{Nielsen2013a}, derived from a large and heterogeneous 
compilation of 182 isolated galaxies at redshift $0.1\lsim\z\lsim1.1$ 
with B--band magnitudes varying from ${\rm M}_{\rm B}=-16.1$ to 
${\rm M}_{\rm B}=-23.1$ \citep[][and references therein]{Nielsen2013b}.
The quoted 3$\sigma$ upper limits in the detection of $\ewr(\lambda2796)$ 
were converted to the 2$\sigma$ limit considered in this work.

We note that, quasars in each bin show, on average, a higher $\fcmgii
(0.30\,\textrm{\AA})$ with respect to galaxies. 
In particular, while {\it low luminosity} galaxies (defined by Nielsen 
et al. as galaxies with B--band luminosity $L_{\rm B}/L_{\rm B}^\star
\lsim0.6$) do not reveal any absorption systems at impact parameter
larger than 110\,kpc, {\it high luminosity} galaxies ($L_{\rm B}/
L_{\rm B}^\star\gsim0.6$) show a behaviour more similar to quasars,
with a CGM extending also at large separations (see Left Panel of 
Figure~\ref{fig:cov}).
The difference between the covering fraction at different impact 
parameters of quasars, and of high-- and low--luminosity galaxies 
almost disappears for systems with $\ewr(\lambda2796)\geq0.6$\,\AA\ 
(see Right Panel of Figure~\ref{fig:cov}).
Since quasars are hosted by luminous galaxies that in some cases
show an excess of blue light \citep[e.g.][]{Bahcall1997, Floyd2013, 
Kotilainen2013, Falomo2014}, our results are qualitatively in 
agreement with a luminosity dependence for the covering fraction of
\mgii\ absorbing gas with $\ewr\geq0.3$\,\AA\ \citep[e.g.][and references 
therein]{Nielsen2013a}.

It is of interest to compare our results with the \hi\ covering fraction
observed in high redshift quasars by \citet{Prochaska2013a}, who investigate 
a sample of 74 close projected quasar pairs with projected separations 
$\pd<300$\,kpc and average redshift $\langle\z\rangle\sim2.2$. 
We convert the \mgii\ equivalent width limit into a \hi\ column density 
($N_\textrm{HI}$) with the empirical relation provided by 
\citet{Menard2009}\footnote{We considered the relation between median 
$N_\textrm{HI}$ and $\ewr(\lambda2796)$.} and derived from the sample of 
low--redshift Lyman absorbers of \citet{Rao2006}. 
Within $\sim100$\,kpc the \hi\ covering fraction is $\fchi\sim
0.33$ for absorbers with $N_\textrm{HI}>10^{18.9}$\,cm$^{-2}$ (roughly 
corresponding to $\ewr(\lambda2796)>0.60$\,\AA) that is lower, but still 
consistent within uncertainties, than the $\fc(0.60\,\textrm{\AA})\sim0.50$ 
we found for lower redshift quasars.
This suggests that the CGM of quasars does not evolve significatively
from redshift $\z\sim2$ to $\z\sim1$ \citep[see also][]{Chen2012, 
Fumagalli2013}. 
This result could be influenced by the large scatter present 
in the $\ewr(\lambda2796)$ versus $N_\textrm{HI}$ plane. Column 
densities of $N_\textrm{HI}\sim10^{18.5}$\,cm$^{-2}$ could still 
be associated to $\ewr(\lambda2796)\sim0.6$\,\AA\ absorption 
systems, suggesting that we are most probably underestimating the 
\mgii\ covering fraction associated to $\z\sim2$ quasars. 
Future direct observations of the \mgii\ absorbing gas associated 
to high redshift quasars are thus needed to give further support to 
this result.

\subsection{W$_{\rm r}$($\lambda2797$) and Impact Parameter}

\begin{figure*}
\centering
\includegraphics[width=0.99\columnwidth]{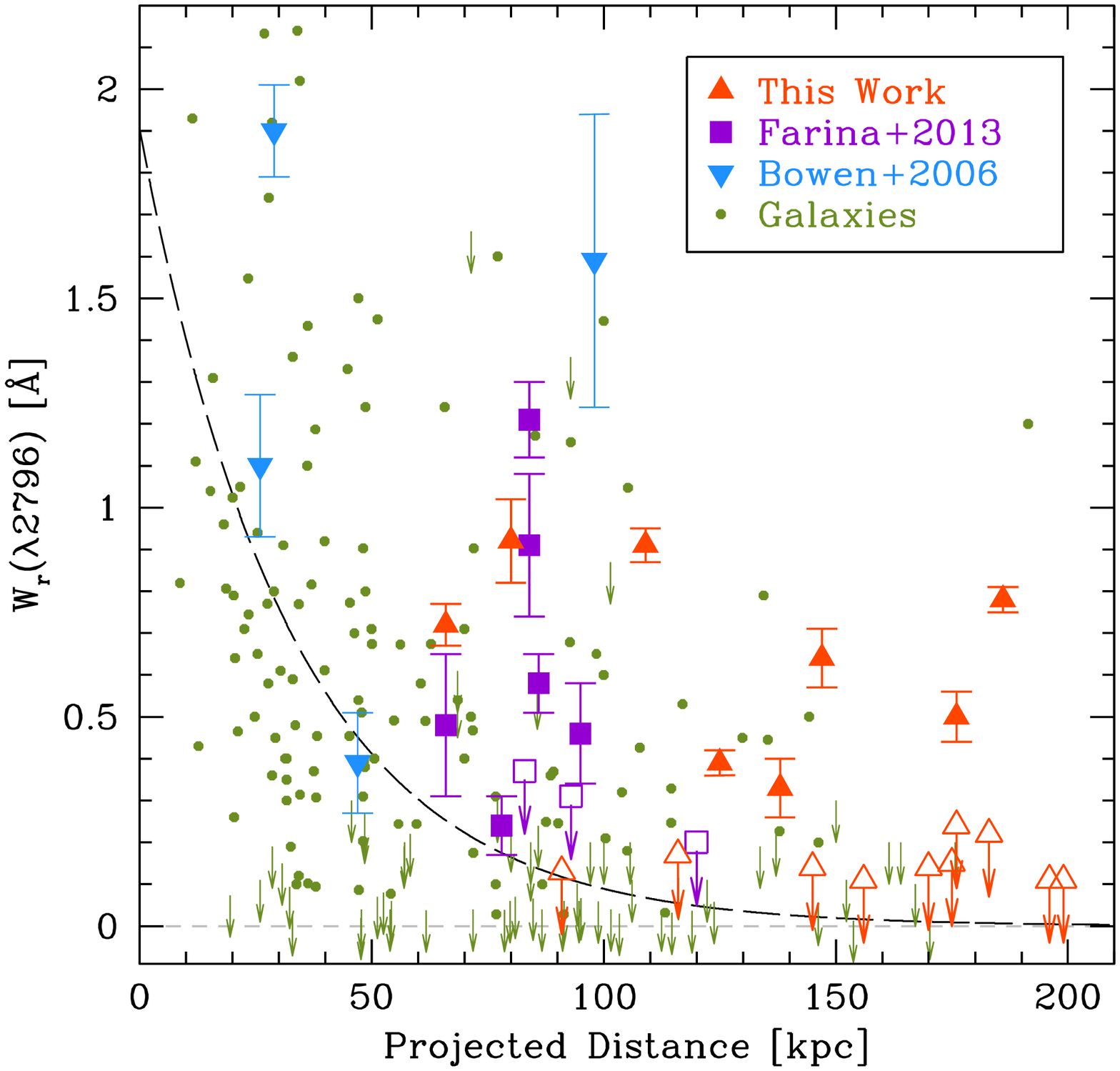}
\qquad
\includegraphics[width=0.99\columnwidth]{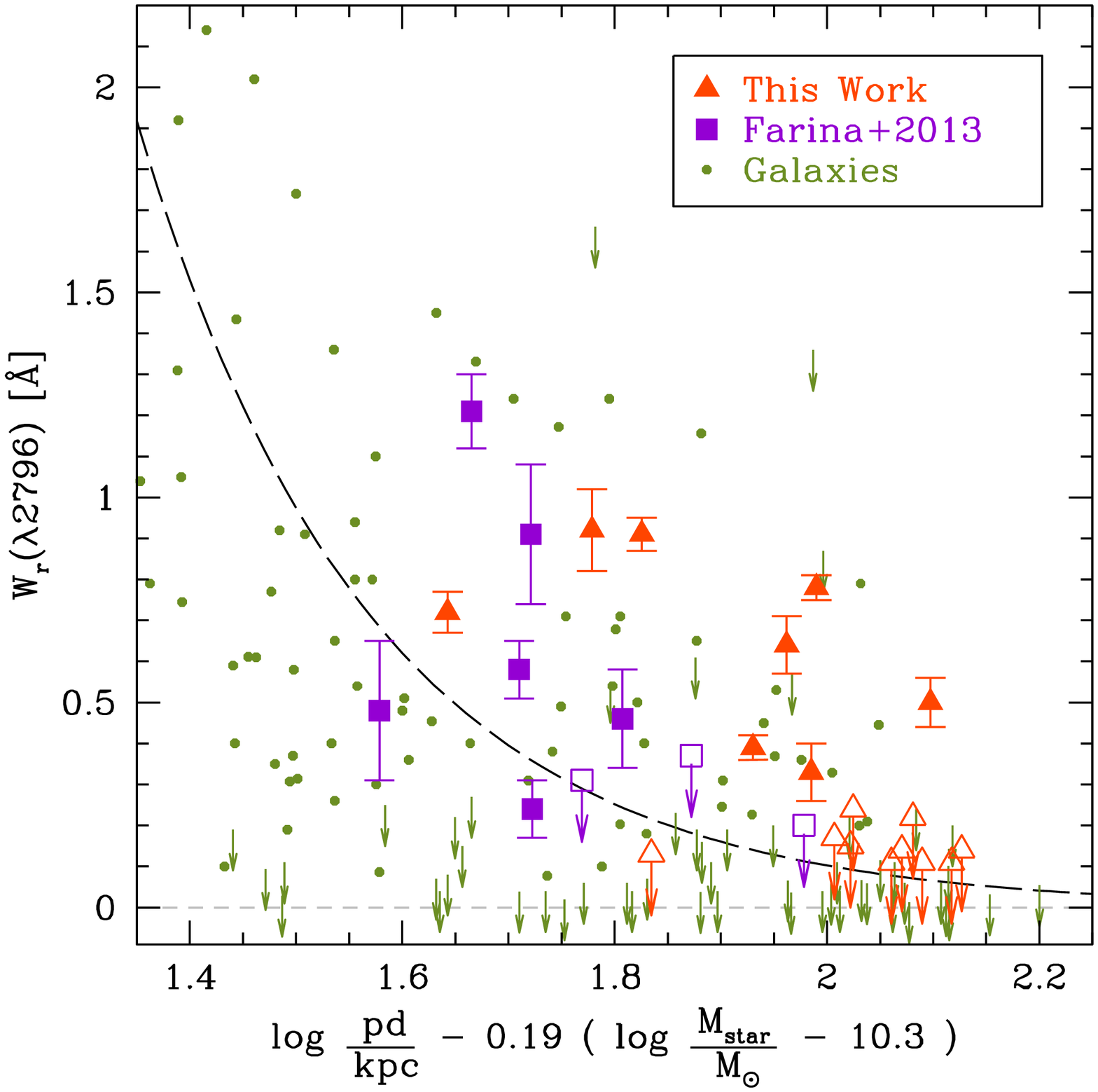}
\caption{
{\it Left Panel:}
Rest frame equivalent width of \mgii($\lambda2796$) absorption line as a 
function of projected distance.
Orange and cyan filled triangles, and violet filled squares represent quasars 
in which an associated transverse absorption system is detected, while the 
empty ones are 2$\sigma$ upper limits (data are from this work, \citealt{
Bowen2006}, and \citealt{Farina2013}, respectively). 
Green points and arrows are absorption features associated to galaxies and 
upper limits from \citet{Werk2013}, \citet{Gauthier2011}, \citet{Kacprzak2011}, 
\citet[][also including the absorption detected in {\it group 
galaxies}]{Chen2010a}, and \citet{Barton2009}. 
Black dashed line shows the best fit of the anti--correlation proposed by 
\citet{Nielsen2013a}.
For the sake of comparison, the upper limits listed by \citet[][]{Barton2009} 
were converted to the considered 2$\sigma$ limits, and all the data were 
rescaled to the considered cosmological model.
{\it Right Panel:}
Rest frame equivalent width of \mgii($\lambda2796$) absorption line as a 
function of projected distance and stellar mass for quasars and galaxies. 
Orange triangles and violet squares are data for quasars from this work and 
\citet{Farina2013}, respectively, while green points and arrows are absorption 
features associated to galaxies and upper limits from \citet{Werk2013},
\citet[][]{Chen2010b}, and \citet{Barton2009}.
Spectra of the four \qsof\ in \citet{Bowen2006} are not publicly available 
and thus we can not give an estimate of the $\mhost$ for these systems.
Black dashed line is the $\ewr$ vs. $\pd$ anti--correlation for galaxies 
including the scaling relation with stellar mass proposed by \citet{Chen2010b}.
For the x--axis we have adopted the same projection of Figure~3 in 
\citet{Chen2010b}.
}\label{fig:pdew}
\end{figure*}

It is well assessed that the equivalent width of \mgii\ absorption features 
anti--correlates with the impact parameter \citep[e.g.][]{Lanzetta1990, 
Bergeron1991, Steidel1995, Chen2010a, Nielsen2013a}. 
In Figure~\ref{fig:pdew} we present the distribution of $\ewr(\lambda2796)$ as 
a function of \pd\ for quasars and for galaxies derived from \citet{Barton2009}, 
\citet{Chen2010a}, \citet{Kacprzak2011}, \citet{Gauthier2011}, and 
\citet{Werk2013}.
While some of the absorbers associated to quasars lie almost in the same
regions of the case of galaxies, at impact parameters larger than $\sim50$\,kpc 
a number of systems with $\ewr(\lambda2796)\gsim0.5$\,\AA\ are present. 
These are rarely found in correspondence of galaxies \citep[see also][]{Farina2013}.
In order to test this qualitative finding we performed a non--parametric 
Kendall's test that include upper limits \citep[e.g.][]{Isobe1986}.
While for galaxies $\ewr(\lambda2796)$ and \pd\ are anti--correlated at 
the 7.9$\sigma$ level \citep[][see also e.g., \citealt{Chen2010a}]{Nielsen2013a}, 
for quasar the significance level is much weaker (2.2$\sigma$). 
In addition, a two dimensions Kolmogorov--Smirnov test \citep{Fasano1987}
performed on the sources with detected \mgii\ absorption, ruled out 
with a probability of $P_{\rm 2D-KS}=99.8\%$ the null hypothesis that, in the 
$20\,{\rm kpc}\leq\pd\leq200\,{\rm kpc}$ region, the absorption systems
associated to galaxies and to quasars are drawn from the same parent 
population.

\subsection[]{The Role of the Mass of Galaxies}

Quasars are generally harboured by massive galaxies with a typical mass 
of few times $10^{11}\,\msun$ \citep{McLure1999, Kukula2001, Falomo2001, 
Falomo2004, Falomo2008, Jahnke2003, Floyd2004, Floyd2013, Hyvonen2007, 
Kotilainen2007, Kotilainen2009, Kotilainen2013}, hence the stellar mass 
\citep[considered as an optimal proxy for the dark halo masses, e.g.][]{
More2011, Moster2010} should have a substantial role in shaping the 
$\ewr$--$\pd$ anti--correlation \citep[e.g.][]{Chelouche2008, Chen2008}.

No deep images of the considered systems are available to directly detect 
the quasar host galaxies. Therefore we derive an estimate of the mass 
($\mhost$) from the $\mbh$--$\mhost$ relation \citep[see e.g.][]{
Marconi2003, Haring2004, Peng2006a, Peng2006b, Decarli2010b, Decarli2012, 
Bennert2011}.
In particular we consider the relation presented by \citet{Decarli2010b}, 
which is based upon the study 96 quasars in the redshift range $0.07<
\z<2.74$:
\begin{equation}
\log{\frac{\mbh}{\mhost}} = \left(0.28\pm0.06\right)\z - \left(2.91\pm0.06\right)
\label{eq:mbhmhost}
\end{equation}
where $\mbh$ is the black hole mass deduced with the virial method as
described in Appendix~\ref{app:1}, and $\z$ is the redshift of the 
foreground quasar. 
Uncertainties associated to the $\mhost$ obtained in this way could 
be as large as $\sim0.6$\,dex \citep[e.g.][]{Decarli2010b}. 
We note that \citet{Decarli2010b} estimated the stellar masses 
assuming bulge--dominated host galaxies and a passive evolution of 
the stellar population from $\z=5$ to $\z=0$. 
Various authors showed that quasars often suffer of intense star 
formation episodes during their lifetime \citep[e.g.][]{Canalizo2013}
and thus masses calculated from equation~\ref{eq:mbhmhost} 
(see Table~\ref{tab:sample}) could be underestimated. 

In the right panel of Figure~\ref{fig:pdew}, we show the distribution of the 
$\ewr(\lambda2796)$ as a function of the impact parameter, rescaled for the 
stellar mass for galaxies from \citet{Barton2009}, \citet{Chen2010b}, and
\citet{Werk2013} (on average, ${\rm M}_{\rm gal}=0.4\times10^{11}\,\msun$) and 
for the quasar hosts (on average, $\mhost\sim2.0\times10^{11}\,\msun$) 
assuming, for the sake of comparison, the same scale on the x--axis presented 
in \citet{Chen2010b}:
\begin{equation}
\pd_{\rm M} = \log{\frac{\pd}{\rm kpc}} - 0.19\,\left(\log \frac{{\rm M}_{\rm star}}{\msun} - 10.3\right).
\end{equation}
Taking into account the mass of the galaxies, the anti--correlation between 
$\ewr$ and \pd\ for quasars is enhanced to the 3.1$\sigma$ level\footnote{A 
Monte Carlo analysis of the anti--correlation shows that, even allowing the 
host galaxy masses to vary of $0.6$\,dex around the calculated values, the 
significance is increased (i.e. it is better than 2.2$\sigma$) in more than 
70\% of the realisations. The large uncertainties associated to the stellar 
mass of the quasar hosts marginally affect this result.} 
and the null hypothesis of a same parent population for absorption systems 
associated to quasars and to galaxies is ruled out with a probability of 
$P_{\rm 2D-KS}=77.4\%$.
The $\widetilde{\chi}^2$ values for our data calculated against the relations 
presented by \citet{Nielsen2013a} and \citet[][see Figure~\ref{fig:pdew}]{Chen2010b} 
prior and after considering the host galaxy masses decrease of $\sim$30\%.

\begin{figure}
\centering
\includegraphics[width=1.0\columnwidth]{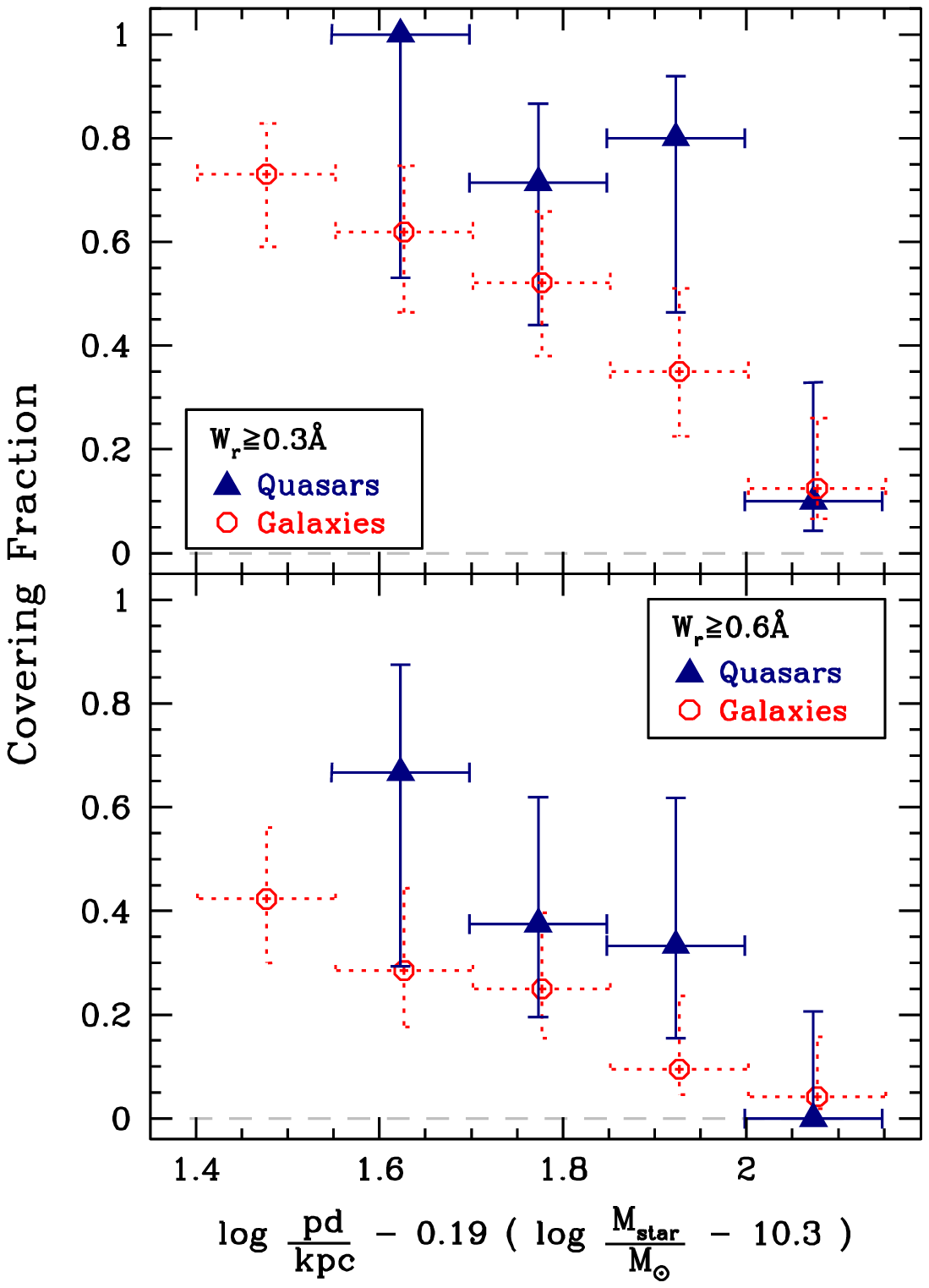}
\caption{
{\it Top Panel:}
Covering fraction profile for transverse absorption systems with 
$\ewr(\lambda2796)>0.30$\,\AA\ associated to \qsof\ (blue filled
triangles) with impact parameter normalised by the host galaxy 
masses as in Figure~\ref{fig:pdew}. We also show the \mgii\ covering 
fraction of galaxies (red empty circles) with public available 
measure of the stellar masses from \citet{Chen2010b}, \citet{Barton2009}, 
and \citet{Werk2013} that constitute a representative subsample of 
roughly half the systems studied by \citet{Nielsen2013a}. Error
bars are calculated as in Figure~\ref{fig:cov}.
{\it Bottom Panel:}
Same of Top Panel but for absorption systems with $\ewr(\lambda2796)>
0.6$\,\AA.
}\label{fig:covmass}
\end{figure}

In Figure~\ref{fig:covmass} we show the \mgii\ covering fraction of 
quasars and galaxies estimated in bins of $\pd_{\rm M}$. 
In spite of the large uncertainties, quasars show a systematically
higher covering fraction than galaxies. This difference is 
more marked for systems with $\ewr(\lambda2796)>0.3$\,\AA, but holds
also for those with $\ewr(\lambda2796)>0.6$\,\AA.

These findings suggest that the stellar mass plays an important role, but 
its effect is not strong enough to reconcile the different properties of 
the CGM of galaxies and of quasars.
As suggested by different studies, other parameters related to the host 
galaxies could be involved, such as: star formation, morphology, or close 
galactic environment \citep[e.g.][and next Section]{Chen2010a, Chen2010b, 
Kacprzak2007, Kacprzak2012, Menard2011, Bordoloi2011, Bordoloi2012}.
Moreover, the {\it patchiness} of the cool gas in the CGM could have
an important effect in the large scatter present in the anti--correlation 
\citep[e.g.,][]{Kacprzak2008}.

Our results are qualitatively in agreement with the systematic segregation 
of the galaxy virial masses on the $\ewr(\lambda2796)$--$\pd$ recently
reported by \citet{Churchill2013a, Churchill2013b}: galaxies with higher 
mass haloes show stronger \mgii\ absorption systems at a given $\pd$ with 
respect to lower mass haloes.
In this context it is of interest to compare our findings with the sample 
of \mgii\ absorbers observed by \citet{Gauthier2011} around Luminous Red
Galaxies (LRGs), which are expected to inhabit haloes with masses comparable 
or larger than those of quasars \citep[e.g.][]{Zheng2009}.
Between 45 and 200\,kpc \citep[where the separation limit of 45\,kpc is 
set by the smallest impact parameter investigated by][]{Gauthier2011}
we calculate a covering fraction of $\fc(0.3\,\textrm{\AA})=0.42^{+0.20}_{-0.16}$
for LRGs that is slightly lower but consistent within the uncertainties with 
the $\fc(0.3\,\textrm{\AA})=0.59^{+0.12}_{-0.13}$ observed for quasars.

\subsection[]{The Role of the Immediate Galactic Environment}

Since quasars are often associated with group of galaxies 
\citep[e.g.][]{Wold2001, Serber2006, Hutchings2009}, it is of 
interest to estimate the contribution of the immediate galactic 
environment to the strength of the observed absorption systems.
Indeed, the presence of a rich environment could induce an over--abundance 
of strong equivalent width systems ($\ewr(\lambda2796)>1$\,\AA) 
compared to field galaxies \citep[e.g.][]{Nestor2007, Lopez2008, 
Kacprzak2010, Andrews2013, Gauthier2013}. 

In order to evaluate this effect, we have first simulated the 
typical galactic environment of a quasar following the quasar--galaxy 
cross--correlation function presented by \citet{Zhang2013}.
Then, to each mock galaxy we have assigned a CGM that reproduce 
the observed properties of the \mgii\ absorption systems reported 
by \citet{Chen2010a} and \citet{Nielsen2013b}.
Finally, the average effect of the environment in terms of strength 
of the absorption and covering fraction was calculated summing up the
contribution of the single galaxies at different impact parameter from
the quasar. 
A detailed description of the simulation and a discussion of the 
possible caveats in our estimate are given in Appendix~\ref{app:3}.
We here summarise our results: 
{\it (i)} less than $25\%$ of the quasar sight lines are covered by the 
absorbing gas associated to companion galaxies;
{\it (ii)} the covering fractions of the absorbing gas associated with 
the quasar's environment show an almost flat profile between 20 and
200\,kpc with $f_{\rm C,\,Env}(0.30\,\textrm{\AA})\sim0.10$ and 
$f_{\rm C,\,Env}(0.60\,\textrm{\AA})\sim0.05$; and 
{\it (iii)} the contribution of galaxies in proximity of quasars to the 
strength of the observed absorption systems is ${\rm W}_{\rm r,\,Env}
(\lambda2796)\lsim0.1$\,\AA, with almost no dependence on the impact 
parameter.
This latter effect is of the same order of magnitude of the uncertainties
in the $\ewr(\lambda2796)$ measurements, thus has a marginal impact on 
our results, especially for the systems with large $\ewr$. 
The contribution of the environment to the covering fraction is also 
negligible, however we notice that at large impact parameter (and small
$\fc$) the presence of satellite galaxies could have enhanced the measured 
$\fc$ up to a factor of~2 (see Figure~\ref{fig:cov}).

\subsection{Fe\,II Transverse Absorption Systems}

\begin{figure}
\centering
\includegraphics[width=0.99\columnwidth]{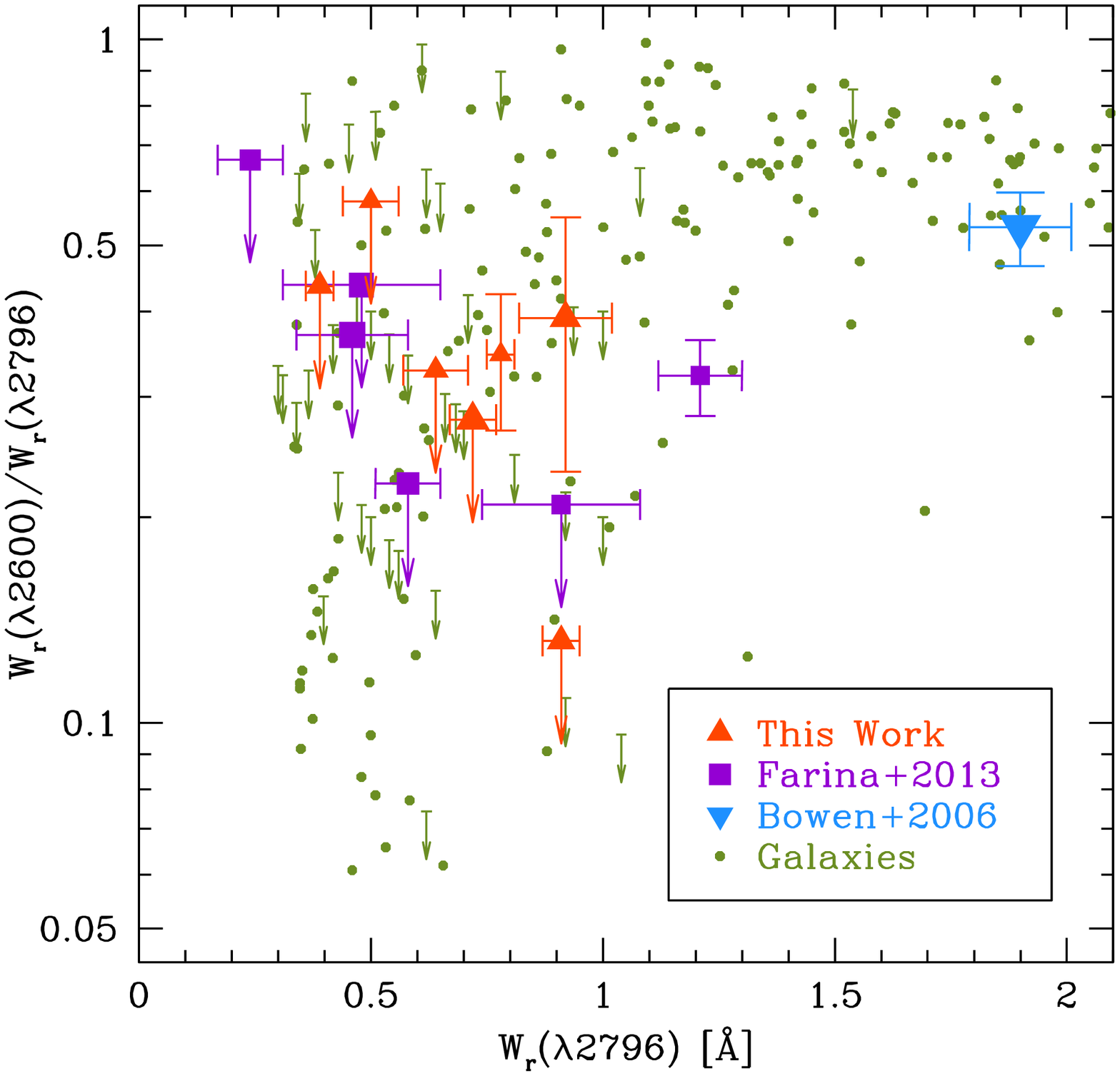}
\caption{
Ratio of the rest frame equivalent width of \feii($\lambda2600$) to 
\mgii($\lambda2796$) absorption lines as a function of $\ewr(\lambda2796)$.
Symbols are equivalent to those in Figure~\ref{fig:pdew}.
Only quasars that shows the presence of an associated \mgii\ transverse 
absorption systems are shown and the size of these points is inversely 
proportional to the logarithm of the impact parameter.
Note that \citet{Bowen2006} report information about the presence of \feii\
associated absorption line only for the quasar SDSS~J083649.45+484150.0, 
which is one of the closer pair considered here ($\pd=29$\,kpc).
Data for galaxies are from the sample of 87 \mgii\ absorption systems with
$\ewr(\lambda2796)>0.3$\,\AA\ investigated by \citet{Rodriguez2012} in the
redshift range $0.2<\z<2.5$ and from the study of 197 \mgii\ absorption
features associated to damped \lya\ systems at $\z<1.65$ performed by 
\citet{Rao2006}.
}\label{fig:mgfe}
\end{figure}

In 21 of the investigated quasar pairs, the spectral coverage of our 
data allows to investigate also for the presence of the \feii\ 
$\lambda\lambda2586,2600$ absorption systems.
We detect 4 \feii($\lambda2600$) absorption lines (see Figure~\ref{fig:mgfe} 
and Table~\ref{tab:allabs}), formally yielding a covering fraction of 
$\fcfeii=0.19^{+0.15}_{-0.08}$ for systems with $\ewr(\lambda2600)\geq0.30$\AA\
and $\pd\leq200$\,kpc.

It is worth noting that absorption systems with $\ewr(\lambda2796)\geq
0.50$\,\AA\ and $\ewr(\lambda2600)/\ewr(\lambda2796)>0.5$ have a $\sim$40\% 
probability to be a damped \lya\ systems (DLAs) with $N_\textrm{HI}>
10^{20}$\,cm$^{-2}$ \citep{Rao2006}. Only two of the detected \mgii\ 
absorptions match these constraints (see Figure~\ref{fig:mgfe}), suggesting
that also the stronger absorptions might not arise in galactic disc, where
high column densities are expected \citep[e.g.][]{Zwaan2005}.
However, we can not exclude the possibility that the \mgii\ absorption
originate from extraplanar neutral gas associated with spiral galaxies 
\citep[see][and references therein]{Sancisi2008}.

The estimated fraction of DLAs present within 200\,kpc from $\z\sim1.1$ 
quasars (i.e., $4^{+5}_{-2}\%$) is consistent with the $10^{+8}_{-4}\%$ 
observed at higher redshift by \citet{Prochaska2013a}. This further 
supports the hypothesis of little, if any, evolution of the quasars' CGM.

\subsection{Anisotropic Distribution of Mg\,II Absorbers}

In the whole sample of quasars we do not detect any LOS \mgii\ absorption lines
of the same strength of the transverse one.
We emphasise that even considering a larger velocity windows to associate the 
LOS absorbers to the quasars \citep[e.g $<5000$\,\kms\ as suggested by][]{
Sharma2013}, only one more systems would be added to our sample (see 
Appendix~\ref{app:2}).
A similar behaviour was also observed for \hi\ absorptions systems \citep[e.g
][]{Hennawi2007}, while \civ\ LOS absorbers were detected in 2 out of the 3 
quasars investigated by \citet{Farina2013}.
This higher incidence of \civ\ LOS absorbers with respect to \mgii\ is in
agreement with the study of \citet{Wild2008} in their SDSS based study of LOS 
absorption systems directly associated to quasars.

The absence of LOS absorbers is possibly a consequence of the SMBH emission, 
that could photoionise the surrounding \mgii\ absorbing clouds.
\citet{Chelouche2008} consider that the CGM of quasars is clumpy and filled by 
clouds having a size of $\sim1$\,pc and a density of $\sim10^{-2}$\,cm$^{-3}$
\citep[see][]{Petitjean1990, Churchill2001, Rauch2002, Churchill2003, 
Ellison2004, Prochaska2009}. 
Under these conditions a quasar with luminosity $\sim10^{46}$\,erg\,s$^{-1}$ 
\citep[the average of our sample, see Table~\ref{tab:sample} and][]{Farina2013}
is able to photoionise the gas of the CGM and to heat it (through photoabsorption)
up to a temperature of ${\rm T}\sim10^5$\,K, allowing the persistence of only few
\mgii\ absorption systems.
This is in agreement with \citet{Wild2008}, who found that the SMBH 
emission photoionise \mgii\ absorbers with $\ewr(\lambda2796)\geq0.30$\,\AA\ 
out to at least 800\,kpc, while, thanks to its higher ionising potential, \civ\ 
absorbing clouds could survive.

In this scenario, the presence of the transverse \mgii\ absorption features is 
explained considering that the quasar radiation is emitted into cones 
\citep[e.g.][]{Antonucci1993, Elvis2000} and thus only marginally affects the gas in 
the transverse direction \citep[e.g.][]{Hennawi2007, Bowen2006, Chelouche2008, 
Prochaska2013a, Prochaska2013b}.
Similarly a non isotropic emission of quasars is invoked to explain the
non detection of the transverse proximity effect \citep[i.e., the expected 
decrease on absorption systems in the \lya\ forest of close projected quasar 
pairs due to the ionising emission of the foreground SMBH, see e.g.][]{
Crotts1989, Dobrzycki1991, Liske2001, Schirber2004}.


\section[]{Summary and Conclusions}\label{sec:6}

We have investigated the properties of the \mgii\ absorbing CGM 
of quasars using a sample of 31~projected quasars pairs with 
impact parameter ranging from 20 to 200\,kpc at $0.5\lsim\z\lsim1.6$.

The main results of our study are:

1.\mbox{ } 
Quasars are surrounded by a large amount of \mgii\ absorbing 
gas with a covering fraction that ranges from $\fc\sim1.0$ at
$\pd\lsim60$\,kpc to $\fc\sim0.2$ at $\pd\gsim150$\,kpc for 
systems with $\ewr(\lambda2796)>0.3$\,\AA.

2.\mbox{ }  
We find a weak anti--correlation between the \mgii\ rest frame 
equivalent width and the impact parameter that is enanched once
the stellar mass of the quasar host galaxy is taken into account.

3.\mbox{ }
While \mgii\ absorbers are frequently observed in the transverse 
direction, such systems are rarely found along the line--of--sight.
This supports a scenario where the ionising emission of the SMBH 
occurs in cones \citep[e.g.][]{Antonucci1993}, thus the CGM is not
illuminated by the central engine in the transverse direction, 
resulting in a non isotropic distribution of the absorption systems.
 
Since quasars are harboured by luminous galaxies, our result supports
a scenario in which galaxies with high luminosity/mass typically possess
a more extended CGM with respect to the fainter ones \citep[e.g.][]{Chelouche2008, 
Nielsen2013a}.
Nevertheless, we observe that quasars are surrounded by a larger amount
of \mgii\ absorbing gas even considering the difference in size.
The presence of this large reservoir of cool gas may be a challenge for 
the cold--mode accretion paradigm that predicts a little amount of 
cool gas around more massive haloes \citep[e.g.][]{Birnboim2003, 
Keres2005, Keres2009, vandeVoort2012, Nelson2013}.
Possibly wind outflows and/or inflows of metal enriched gas associated
to the galaxy interactions responsible for the quasar activity could be
able to supply of cool \mgii\ absorbing gas the CGM. 
Future deep imaging observation of the foreground quasars aimed to 
characterise the quasar hosts and its close environment will help 
to clarify the origin of the \mgii\ absorption systems and to 
investigate which are the most important parameters that regulate
the properties of the CGM of quasars.  

\section*{Acknowledgements}

We are grateful to E.~Lusso and C.~Montuori for useful discussion 
and helpful comments on the manuscript.
We acknowledge financial contribution from the grant PRIN--MIUR
2010NHBSBE by the Italian Ministry of University and Research.
EPF acknowledges funding through the ERC grant ``Cosmic Dawn''.
Support for RD was provided by the DFG priority program 1573 ``The 
physics of the interstellar medium''. 
This work was based on observations made with the GTC Telescope
in Roque de los Muchachos and the ESO/VLT Telescope in Paranal.
Funding for the SDSS and SDSS-II has been provided by the Alfred 
P. Sloan Foundation, the Participating Institutions, the National 
Science Foundation, the U.S. Department of Energy, the National 
Aeronautics and Space Administration, the Japanese Monbukagakusho, 
the Max Planck Society, and the Higher Education Funding Council 
for England. 
The SDSS Web Site is \texttt{http://www.sdss.org/}.

\appendix

\section[]{Black Hole Masses and Bolometric Luminosities of Quasars}\label{app:1}

To determine the black hole masses ($\mbh$) and the bolometric luminosity 
($\lbol$) of the observed quasars, we fitted their spectra following the 
procedure presented in \citet{Decarli2010a} and in \citet{Derosa2011}.
Namely, the quasar emission was composed with a superposition of:
(i) a non--stellar continuum, modelled as a power law;
(ii) the contribution from \feii\ blended multiplets \citep[assuming the 
template of][]{Vestergaard2001};
(iii) the stellar--light from the host galaxy \citep[assuming the elliptical 
template of][]{Mannucci2001};
(iv) the broad lines emission \citep[fitted with two Gaussian curves with the 
same peak wavelength, see][]{Decarli2008}.
The results of the fitting procedure allowed us to estimate the $\lbol$s from 
the monochromatic luminosity at $3000$\,\AA\ \citep[assuming the bolometric 
correction presented in][see Table~\ref{tab:sample}]{Runnoe2012} and the 
$\mbh$s  applying the virial theorem to the gas of the broad--line region 
\citep[e.g.][and references therein, see Table~\ref{tab:sample}]{Kaspi2000, 
McLure2002, Vestergaard2006, Peterson2010, Shen2013}.

\section[]{Notes on Individual Objects}\label{app:2}

In this appendix we present all the absorption lines 
detected in the quasar spectra over a $3\sigma$ threshold. 
The properties of the observed systems are listed 
Table~\ref{tab:allabs}.

\noindent {\bf QQ01} --- 
Several intervening metal absorption lines are present in 
the \qsob\ spectra. 
The most prominent are a \mgii\ and two \civ\ doublets at 
$\z\sim0.734$, $2.173$, and $2.383$, and a \feii\ multiplet 
at $\z\sim1.174$.

\noindent {\bf QQ02} --- 
In the spectra of QQ02F we detected a \mgii\ absorption 
doublet at $\z\sim1.264$.
A further one is present at $\z\sim1.602$, $\sim2000$\,\kms\
blueward of the \mgii\ broad emission line.
Even if this system does not match the considered velocity 
constraint (see~\S\ref{sec:4}), it could be associated 
to an outflow of gas originated from the quasar itself or from 
its host galaxy \citep[e.g.][]{Crenshaw2003, Tremonti2007, Wild2008, 
Shen2012, Sharma2013}. 

In the spectra of QQ02B, we observe the presence a  \civ\ 
absorption system ($\ewr(\lambda1548)=(0.39\pm0.09)$\,\AA)
blueshifted of $\sim$700\,\kms\ with respect to the quasar 
emission frame. 
This kind of features are often detected close the \civ\ 
broad emission lines and are thought to arise in quasar 
outflows \citep[e.g.][]{Vestergaard2003, Nestor2008}.
The identification of the \mgii\ transverse absorption 
features associated to \qsof\ is sustained by the presence 
of the \feii ($\lambda$2382) and \feii ($\lambda$2600) lines 
at the same redshift.

\noindent {\bf QQ04} ---  
In the spectra of QQ04F we detect \feii\ and \mgii\ lines 
produced by an intervening absorption system at $\z\sim1.211$. 
The identification of the \mgii\ transverse absorption system 
is supported by the detection of the associated \feii 
($\lambda$2586) and \feii ($\lambda$2600) lines.

\noindent {\bf QQ05} --- 
An \mgii\ and \feii\ absorption system is present in the 
spectra of QQ04B at $\z\sim1.081$. 

\noindent {\bf QQ07} --- 
Two close absorption lines ($\lambda\lambda$=4641.8\AA,
4650.7\AA) are present in the spectra if QQ07B. We tentatively 
identify these features with a \mgii\ doublet at $\z\sim0.660$.
We also observe a \civ\ doublet at almost the same redshift of 
the \civ\ broad emission line ($\z\sim2.213$). The velocity 
difference between the quasar frame and the absorber ($\lesssim
200$\,\kms) suggests that we are probing cool gas clouds strictly 
connected to the quasar itself. 

\noindent {\bf QQ08} --- 
The detection of the [\nev] forbidden emission line allows us to 
refine the redshift of the \qsof: $\z=0.6853\pm0.0008$.
The redshift of the detected transverse absorption system
is consistent, within the uncertainties, with this value.

\noindent {\bf QQ09} --- 
QQ09B is the highest redshift quasar that we have observed. 
Its spectra is polluted by the \lya\ forest hence recognise the 
detected absorption lines is challenging. We tentatively identify 
\mgii, \civ, \feii\, and \siiv\ doublets at $\z\sim0.891$, $2.421$, 
$1.084$, and $3.062$, respectively. In particular the \siiv 
($\lambda 1402$) line is superimposed to the \mgii ($\lambda 2796$) 
transverse absorption associated to the \qsof\ (see Figure~\ref{fig:abs}). 
We decouple the two contribution by fitting the blended lines with 
two Gaussian at the same time, and matching the width of the \mgii 
($\lambda 2796$) line to that of the \mgii ($\lambda 2803$) one.

\noindent {\bf QQ10} --- 
In the spectra of \qsof\ we detect an intervining \mgii\ absorption 
system at $\z=0.6045\pm0.0002$.

\noindent {\bf QQ12} --- 
A \mgii\ absorption feature is present at $\z=1.1618\pm0.0004$, 
$\sim4500$\kms\ from the transverse system associated to the \qsof. 
As for QQ02F, this doublet does not match our constraint on the 
velocity difference with the quasar frame, but we can not exclude 
that it arise in a strong ouflows of gas originated from region 
close to the SMBH or from its host galaxy.

\noindent {\bf QQ14} --- 
An absorption line superimposed to the \mgii\ emission is present
in the spectra of QQ14F. 
The absence of another close absorption line over a $2\sigma$ threshold 
suggests that this features is not a \mgii\ LOS doublet associated to the 
\qsof.

The \qsof\ redshift derived from the \mgii\ broad emission
line is consistent with the redshift inferred from 
the [\nev] narrow line ($\z=1.1057\pm0.0008$).

A \feii\ doublet is present in the spectra of QQ14B at
$\z\sim1.942$.

\noindent {\bf QQ16} ---
Data for this pair was already gathered with FORS2 at ESO--VLT \citep[see][]{
Farina2013}.
The values of the rest frame equivalent widths quoted in Table~\ref{tab:abs}
are the weighted mean of the two observations.
A \mgii\ doublet is detected at $\z\sim1.118$ in the QQ16B spectra.

\noindent {\bf QQ17} ---
Two absorbing systems at redshift $\z\sim0.737$ and $\z\sim1.375$
are identified in the spectra of QQ17B from the detection of \mgii\
and \feii\ absorption lines.

\noindent {\bf QQ18} ---
A strong interving \mgii\ absorption system ($\ewr(\lambda 2796)=
0.98\pm0.07$) is present at $\z\sim1.650$, futher confirmed by
the detection of the corresponding \feii\ and \mgi\ lines.

\begin{table*}
\centering
\caption{
Properties of absorption features detected in QSOs spectra over a 
3$\sigma$ level: 
our identification label of the quasar (ID), 
observed wavelength ($\lambda_{\rm abs}$),
observed equivalent width ($\ew$),
absorption line (line), 
and redshift (z$_{\rm abs}$).
}\label{tab:allabs}
\scriptsize
\begin{tabular}{lcccc}
\hline
ID     &  $\lambda_{\rm abs}$ & W             & line                                           & z$_{\rm abs}$     \\ 
       &  [\AA]	              & [\AA]         &                                                &                   \\       
\hline  		 
\hline  		 
 QQ01B & 4850.8 	      & 1.01$\pm$0.34 & \mgii\ $\lambda 2796$  & 0.7349$\pm$0.0004 \\
 QQ01B & 4861.1 	      & 1.00$\pm$0.34 & \mgii\ $\lambda 2803$  & 0.7342$\pm$0.0004 \\
 QQ01B & 4910.4	              & 1.22$\pm$0.34 & \civ\  $\lambda 1548$  & 2.1721$\pm$0.0006 \\
 QQ01B & 4918.1	              & 1.07$\pm$0.35 & \civ\  $\lambda 1550$  & 2.1730$\pm$0.0006 \\
 QQ01B & 5096.3               & 0.61$\pm$0.25 & \feii\ $\lambda 2344$  & 1.1742$\pm$0.0004 \\
 QQ01B & 5180.3 	      & 0.72$\pm$0.29 & \feii\ $\lambda 2382$  & 1.1748$\pm$0.0004 \\
 QQ01B & 5235.4 	      & 1.02$\pm$0.35 & \civ\  $\lambda 1548$  & 2.3820$\pm$0.0006 \\
 QQ01B & 5244.2 	      & 1.01$\pm$0.35 & \civ\  $\lambda 1550$  & 2.3834$\pm$0.0006 \\
 QQ01B & 5444.6 	      & 0.70$\pm$0.18 & \dots		       & \dots  	   \\
 QQ01B & 5652.3 	      & 0.82$\pm$0.26 & \feii\ $\lambda 2600$  & 1.1740$\pm$0.0004 \\
 QQ01B & 5869.9 	      & 0.82$\pm$0.23 & \mgii\ $\lambda 2796$  & 1.0994$\pm$0.0005 \\
 QQ01B & 5881.0 	      & 0.52$\pm$0.38 & \mgii\ $\lambda 2803$  & 1.0981$\pm$0.0005 \\

 QQ02F & 6330.8               & 0.84$\pm$0.20 & \mgii\ $\lambda 2796$  & 1.2642$\pm$0.0004 \\
 QQ02F & 6330.8               & 0.80$\pm$0.23 & \mgii\ $\lambda 2803$  & 1.2644$\pm$0.0004 \\
 QQ02F & 7275.6               & 1.34$\pm$0.42 & \mgii\ $\lambda 2796$  & 1.6021$\pm$0.0005 \\
 QQ02F & 7291.8               & 0.85$\pm$0.42 & \mgii\ $\lambda 2803$  & 1.6014$\pm$0.0005 \\
 QQ02B & 5785.9               & 1.45$\pm$0.35 & \civ\  $\lambda 1548$  & 2.7377$\pm$0.0010 \\
 QQ02B & 5795.7               & 1.18$\pm$0.32 & \civ\  $\lambda 1550$  & 2.7392$\pm$0.0009 \\
 QQ02B & 6086.7               & 1.03$\pm$0.34 & \dots                  & \dots		   \\
 QQ02B & 6245.1               & 0.97$\pm$0.31 & \feii\ $\lambda 2382$  & 1.6218$\pm$0.0003 \\
 QQ02B & 6814.9               & 0.95$\pm$0.36 & \feii\ $\lambda 2600$  & 1.6211$\pm$0.0003 \\
 QQ04F & 5718.7 	      & 0.76$\pm$0.28 & \feii\ $\lambda 2586$  & 1.2114$\pm$0.0004 \\
 QQ04F & 5748.2 	      & 1.12$\pm$0.32 & \feii\ $\lambda 2600$  & 1.2108$\pm$0.0003 \\
 QQ04F & 6181.9 	      & 2.93$\pm$0.22 & \mgii\ $\lambda 2796$  & 1.2110$\pm$0.0002 \\
 QQ04F & 6198.0 	      & 2.36$\pm$0.19 & \mgii\ $\lambda 2803$  & 1.2112$\pm$0.0002 \\
 QQ04B & 5999.6 	      & 0.27$\pm$0.13 & \feii\ $\lambda 2586$  & 1.3200$\pm$0.0004 \\
 QQ04B & 6031.0 	      & 0.63$\pm$0.12 & \feii\ $\lambda 2600$  & 1.3196$\pm$0.0004 \\
 QQ05B & 4878.1 	      & 1.61$\pm$0.56 & \feii\ $\lambda 2344$  & 1.0811$\pm$0.0005 \\
 QQ05B & 5383.0 	      & 0.83$\pm$0.32 & \feii\ $\lambda 2586$  & 1.0816$\pm$0.0005 \\
 QQ05B & 5410.7 	      & 1.42$\pm$0.30 & \feii\ $\lambda 2600$  & 1.0810$\pm$0.0003 \\
 QQ05B & 5818.3 	      & 2.60$\pm$0.43 & \mgii\ $\lambda 2796$  & 1.0809$\pm$0.0002 \\
 QQ05B & 5833.4 	      & 2.30$\pm$0.32 & \mgii\ $\lambda 2803$  & 1.0811$\pm$0.0002 \\
 QQ07B & 4641.8 	      & 3.15$\pm$0.43 & \mgii\ $\lambda 2796$? & 0.6602$\pm$0.0006 \\
 QQ07B & 4650.7 	      & 2.15$\pm$0.46 & \mgii\ $\lambda 2803$? & 0.6592$\pm$0.0006 \\
 QQ07B & 4973.9 	      & 1.02$\pm$0.39 & \civ\  $\lambda 1548$  & 2.2131$\pm$0.0006 \\
 QQ07B & 4981.6 	      & 0.70$\pm$0.18 & \civ\  $\lambda 1550$  & 2.2139$\pm$0.0006 \\
 QQ09B & 5287.3 	      & 1.61$\pm$0.38 & \mgii\ $\lambda 2796$? & 0.8910$\pm$0.0005 \\
 QQ09B & 5296.4 	      & 1.35$\pm$0.29 & \civ\  $\lambda 1548$? & 2.4214$\pm$0.0005 \\
 QQ09B & 5300.3 	      & 0.68$\pm$0.22 & \mgii\ $\lambda 2803$? & 0.8909$\pm$0.0006 \\
 QQ09B & 5304.4 	      & 1.70$\pm$0.38 & \civ\  $\lambda 1550$? & 2.4222$\pm$0.0005 \\
 QQ09B & 5379.6 	      & 1.57$\pm$0.20 & \dots		       & \dots  	   \\
 QQ09B & 5389.4 	      & 0.92$\pm$0.30 & \feii\ $\lambda 2586$? & 1.0841$\pm$0.0003 \\
 QQ09B & 5419.5 	      & 3.39$\pm$0.28 & \feii\ $\lambda 2600$? & 1.0844$\pm$0.0003 \\
 QQ09B & 5642.2 	      & 0.99$\pm$0.28 & \dots		       & \dots  	   \\
 QQ09B & 5660.4 	      & 2.97$\pm$0.38 & \siiv\ $\lambda 1393$? & 3.0622$\pm$0.0007 \\
 QQ09B & 5696.3 	      & 1.94$\pm$0.37 & \siiv\ $\lambda 1402$? & 3.0622$\pm$0.0007 \\
 QQ09B & 5805.1 	      & 2.14$\pm$0.37 & \dots	    	       & \dots	           \\
 QQ10F & 4486.2 	      & 0.48$\pm$0.16 & \mgii\ $\lambda 2796$  & 0.6045$\pm$0.0003 \\
 QQ10F & 4497.4 	      & 0.50$\pm$0.17 & \mgii\ $\lambda 2803$  & 0.6045$\pm$0.0003 \\
 QQ12F & 7317.0 	      & 2.15$\pm$0.56 & \dots		       & \dots  	   \\
 QQ12B & 6044.5 	      & 0.92$\pm$0.31 & \mgii\ $\lambda 2796$? & 1.1618$\pm$0.0005 \\
 QQ12B & 6057.4 	      & 0.94$\pm$0.34 & \mgii\ $\lambda 2803$? & 1.1610$\pm$0.0005 \\
 QQ14F & 5890.3 	      & 0.38$\pm$0.12 & \dots		       & \dots  	   \\
 QQ14B & 6983.3 	      & 0.77$\pm$0.24 & \feii\ $\lambda 2374$  & 1.9416$\pm$0.0003 \\
 QQ14B & 7007.2 	      & 1.46$\pm$0.22 & \feii\ $\lambda 2382$  & 1.9417$\pm$0.0003 \\
 QQ16B & 5922.1 	      & 3.37$\pm$0.51 & \mgii\ $\lambda 2796$  & 1.1181$\pm$0.0003 \\
 QQ16B & 5938.0 	      & 2.75$\pm$0.53 & \mgii\ $\lambda 2803$  & 1.1184$\pm$0.0003 \\
 QQ17B & 4493.1 	      & 1.45$\pm$0.49 & \feii\ $\lambda 2586$  & 0.7375$\pm$0.0005 \\
 QQ17B & 4517.9 	      & 2.53$\pm$0.99 & \feii\ $\lambda 2600$  & 0.7377$\pm$0.0007 \\
 QQ17B & 4858.1 	      & 2.22$\pm$0.61 & \mgii\ $\lambda 2796$  & 0.7375$\pm$0.0004 \\
 QQ17B & 4870.6 	      & 1.88$\pm$0.81 & \mgii\ $\lambda 2803$  & 0.7376$\pm$0.0004 \\
 QQ17B & 5566.4 	      & 2.62$\pm$0.44 & \feii\ $\lambda 2344$  & 1.3748$\pm$0.0004 \\
 QQ17B & 5637.6 	      & 1.22$\pm$0.29 & \feii\ $\lambda 2374$  & 1.3747$\pm$0.0004 \\
 QQ17B & 5657.7 	      & 3.19$\pm$0.46 & \feii\ $\lambda 2382$  & 1.3752$\pm$0.0004 \\
 QQ18B & 6210.9               & 1.67$\pm$0.16 & \feii\ $\lambda 2344$  & 1.6497$\pm$0.0004 \\
 QQ18B & 6291.1               & 1.10$\pm$0.26 & \feii\ $\lambda 2374$  & 1.6500$\pm$0.0004 \\
 QQ18B & 6313.1               & 2.09$\pm$0.29 & \feii\ $\lambda 2382$  & 1.6503$\pm$0.0004 \\
 QQ18B & 6853.3               & 1.58$\pm$0.48 & \feii\ $\lambda 2586$  & 1.6502$\pm$0.0004 \\
 QQ18B & 6888.7               & 2.86$\pm$0.46 & \feii\ $\lambda 2600$  & 1.6495$\pm$0.0004 \\
 QQ18B & 7408.4               & 2.60$\pm$0.19 & \mgii\ $\lambda 2796$  & 1.6496$\pm$0.0004 \\
 QQ18B & 7427.6 	      & 2.67$\pm$0.13 & \mgii\ $\lambda 2803$  & 1.6499$\pm$0.0004 \\
 QQ18B & 7558.5               & 0.75$\pm$0.44 & \mgi\  $\lambda 2853$  & 1.6493$\pm$0.0004 \\

\hline
\end{tabular}
\end{table*}

\section[]{Estimate of the Galactic Environment}\label{app:3}

\begin{figure}
\centering
\includegraphics[width=0.99\columnwidth]{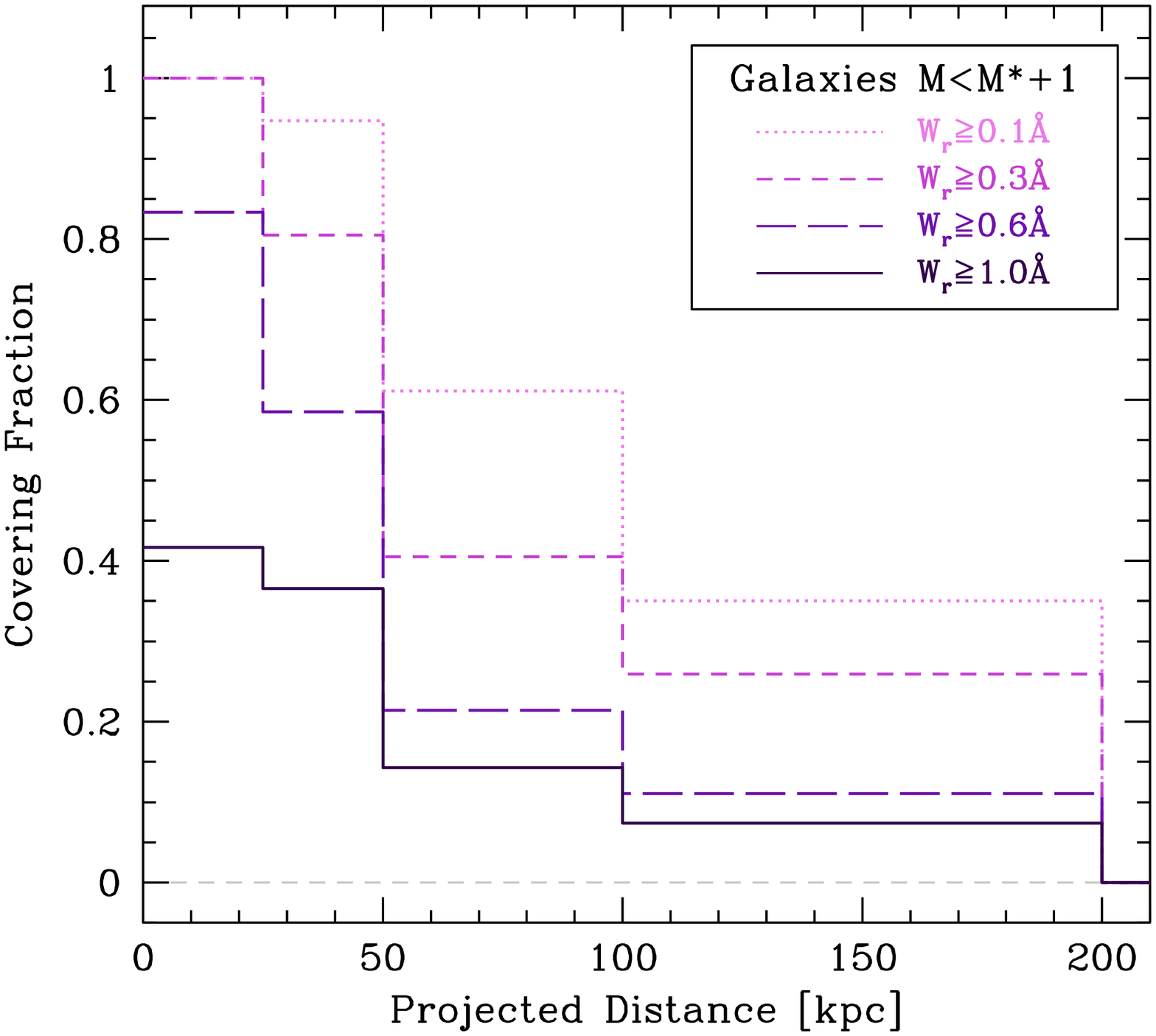}
\caption{
Covering fraction profiles for galaxies brighter than ${\rm M}^\star+1$
estimated from the sample of \citet{Nielsen2013a}. The various shades 
of violet indicate the different equivalent width limit considered.
}\label{fig:cfgal}
\end{figure}

In order to evaluate the contribution of the close galactic environment
to the observed absorption systems we created a set of 10000~mock projected
quasar pairs with impact parameter uniformly distributed between 20~and 
200\,kpc.
For each foreground quasar we randomly generated a distribution of galaxies
that mimic the projected quasar--galaxy cross--correlation function reported
by \citet{Zhang2013}, who have explored the fields of quasars at $\z\sim1.1$
down to $\sim{\rm M}^\star + 1$ (where ${\rm M}^\star$ is the characteristic
absolute magnitude of galaxies). 
This allowed us to calculate that, on average, less then 2~galaxies associated
to a quasar are present within a projected distance of~$400$\,kpc. 
The contribution of these galaxies to the strength of the \mgii\ absorption 
systems depends on their separation from the background quasar, that is indeed 
the pencil--beam used to investigate the CGM.
This distance permitted us to calculate the $\ewr(\lambda2796)$ associated to
each mock galaxy considering the $\ewr$ versus $\pd$ anti--correlation presented
by \citet{Chen2010a}:
\begin{equation}\label{eq:ewpd}
\log{\ewr(\lambda2796)} = -(1.17 \pm 0.10)\,\log{\pd} + (1.28 \pm 0.13).
\end{equation}
In order to reproduce the real distribution of absorbers, to the
$\ewr(\lambda2796)$ derived from Equation~\ref{eq:ewpd} we added 
a random scatter term of $0.35$\,dex that take into account the 
differences between the observed and the estimated \mgii\ absorber 
strength \citep{Chen2010a}.
We also considered that the covering fraction of absorbing gas decrease 
with the increase of the distance from a galaxy, almost vanishing at 
separation larger than 200\,kpc \citep[e.g.][]{Barton2009, Chen2010a, 
Steidel2010, Bordoloi2011, Nielsen2013b}. 
To simulate this effect, the fraction of absorbers present at a certain 
galactocentric radius and with a given equivalent width was tuned to 
reproduce the covering fraction profiles of the galaxies brighter than 
${\rm M}^\star+1$ present in the sample of \citet{Nielsen2013a} (see 
Figure~\ref{fig:cfgal}). 
Finally, if more than one galaxy intercept the same line--of--sight we 
summed on all the $\ewr(\lambda2796)$ of the contribution \citep{Bordoloi2011}.

In this estimate we have not considered any dependence of the absorber 
equivalent widths with the galaxy properties. This represent a second
order correction to the $\ewr$ versus $\pd$ anti--correlation, and
has a negligible influence to our results.
We also point out that, in principle, the gas associated to faint galaxies
could contribute to observed absorption systems. 
Indeed galaxies with absolute magnitude between ${\rm M}^\star+1$ 
and ${\rm M}^\star+2$ are almost 2 times more abundant than luminous 
galaxies\footnote{Estimated considering a Schechter luminosity function
\citep{Schechter1976} with a faint--end slope $\alpha=-1.30$ 
\citep[][]{Faber2007}.} (i.e., brighter than ${\rm M}^\star + 1$).
However, these faint galaxies rarely show \mgii\ absorbers with $\ewr(\lambda2796)
\geq0.1$\,\AA\ at impact parameter larger than 60\,kpc \citep[e.g., 1 out of 13
galaxies in the sample of][]{Nielsen2013a}. 
Given the different size of the CGM, the presence of faint galaxies could affect
our estimates for less than $\sim15\%$.
In addition, it is worth noting that the sample of isolated galaxies used as 
comparison in this work derives from different surveys and the magnitude limit
of the imaged quasar fields are not uniforms.
For instance, $\sim40\%$ of the galaxies studied by \citet{Nielsen2013a} comes
from the sample of \citet{Chen2010a}, who used SDSS images to discriminate 
between isolated and group galaxies, while for $\sim30\%$ of the sample deep
Hubble Space Telescope images are available.
This suggests that even some of the galaxies considered as {\it isolated} 
could be affected by the contamination of fainter sources.

\begin{figure}
\centering
\includegraphics[width=0.99\columnwidth]{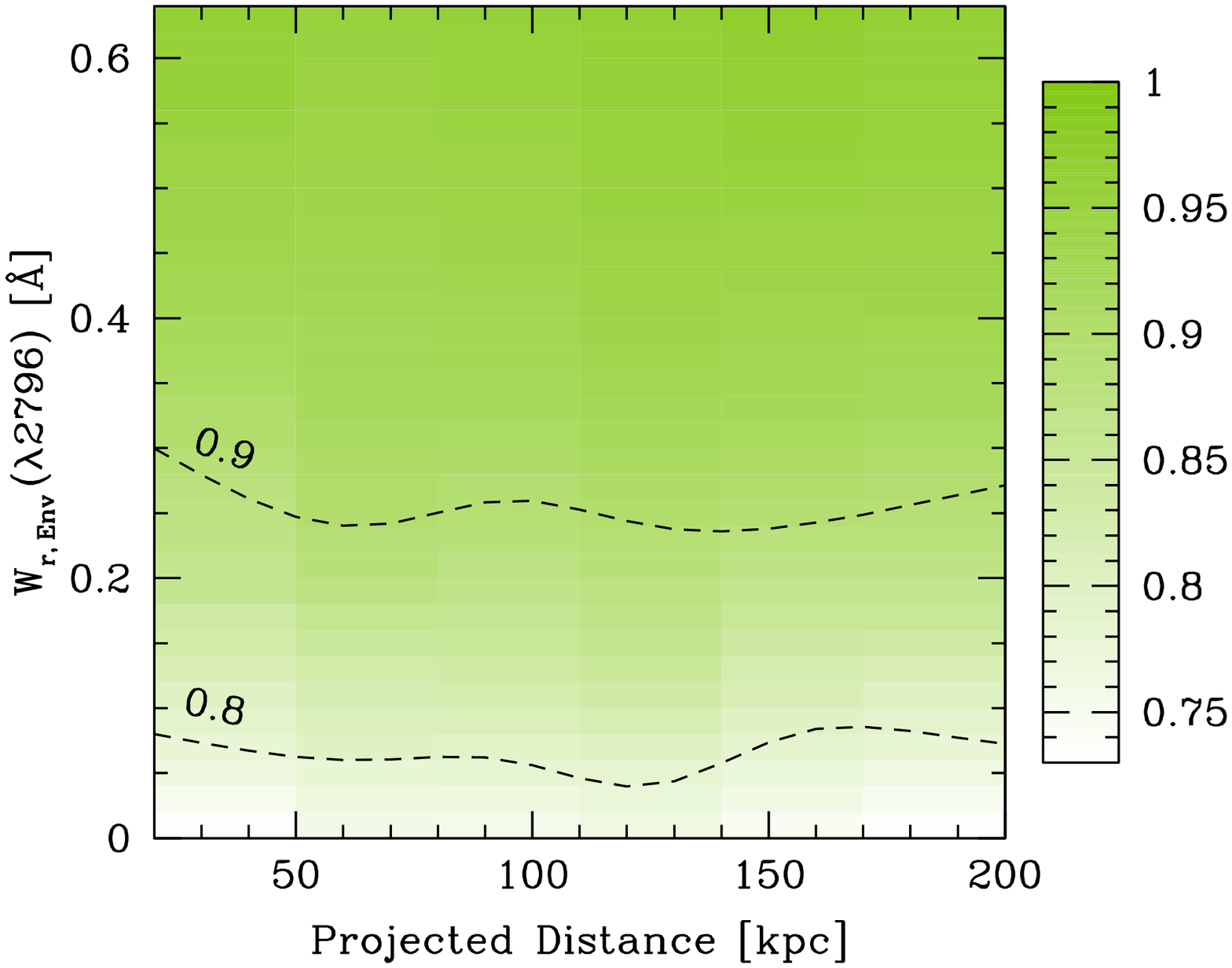}
\caption{
Fraction of simulated galactic environments that contribute to the
strength of the transverse absorption systems with an equivalent width 
${\rm W}_{\rm r}\leq{\rm W}_{\rm r,Env}(\lambda2796)$ (see
Appendix~\ref{app:3} for details).
The cumulative distribution is estimated in bins of $30$\,kpc each,
and the different fractions are colour--coded in the bar at the right
side of the plot.
The two black dashed lines show that the 80\% of the systems influence
the equivalent width of the detected absorption for less than $0.1$\,\AA,
and the 90\% for less than $0.3$\,\AA.
}\label{fig:ewwnv}
\end{figure}

With these caveats in mind, we estimated that the influence of the
environment is of the same order of magnitude of the uncertainties 
in the $\ewr(\lambda2796)$ measurements, and has a negligible influence
on our results.
Indeed the average contribution of the environment to the associated 
absorption systems is ${\rm W}_{\rm r,\,Env}(\lambda2796)\lsim0.1$\,\AA, 
with almost no dependence on the impact parameter.
Only $\sim25\%$ of the sight lines are covered by absorbing gas associated
to satellite galaxies and the covering fractions are $f_{\rm C,\,Env}
(0.30\,\textrm{\AA})\sim0.10$, and $f_{\rm C,\,Env}(0.60\,\textrm{\AA})
\sim0.05$. As shown in Figure~\ref{fig:ewwnv} these values are almost
constant in the range of impact parameter explored.

These results confirm that, in average, the influence of the galactic 
environment is not strong enough to reconcile the differences observed
in the CGM of quasars and galaxies. However, deep images of the \qsof\
fields are needed to assess the effects of galaxies associated to the 
quasars on individual absorption systems. 

\label{lastpage}


\begin{thebibliography}{99}


\bibitem[\protect\citeauthoryear{Antonucci}{1993}]{Antonucci1993} Antonucci R., 1993, ARA\&A, 31, 473 
\bibitem[\protect\citeauthoryear{Appenzeller et al.}{1998}]{Appenzeller1998} Appenzeller I., et al., 1998, Msngr, 94, 1
\bibitem[\protect\citeauthoryear{Abazajian et al.}{2009}]{Abazajian2009} Abazajian K.~N., et al., 2009, ApJS, 182, 543
\bibitem[\protect\citeauthoryear{Andrews et al.}{2013}]{Andrews2013} Andrews H., et al., 2013, ApJ, 774, 40 
%
\bibitem[\protect\citeauthoryear{Bahcall \& Spitzer}{1969}]{Bahcall1969} Bahcall J.~N., Spitzer L., Jr., 1969, ApJ, 156, L63 
\bibitem[\protect\citeauthoryear{Bahcall et al.}{1997}]{Bahcall1997} Bahcall J.~N., Kirhakos S., Saxe D.~H., Schneider D.~P., 1997, ApJ, 479, 642 
\bibitem[\protect\citeauthoryear{Barton \& Cooke}{2009}]{Barton2009} Barton E.~J., Cooke J., 2009, AJ, 138, 1817 
\bibitem[\protect\citeauthoryear{Bechtold \& Ellingson}{1992}]{Bechtold1992} Bechtold J., Ellingson E., 1992, ApJ, 396, 20 
\bibitem[\protect\citeauthoryear{Bennert et al.}{2008}]{Bennert2008} Bennert N., Canalizo G., Jungwiert B., Stockton A., Schweizer F., Peng C.~Y., Lacy M., 2008, ApJ, 677, 846 
\bibitem[\protect\citeauthoryear{Bennert et al.}{2011}]{Bennert2011} Bennert V.~N., Auger M.~W., Treu T., Woo J.-H., Malkan M.~A., 2011, ApJ, 742, 107 
\bibitem[\protect\citeauthoryear{Bergeron \& Stasi{\'n}ska}{1986}]{Bergeron1986} Bergeron J., Stasi{\'n}ska G., 1986, A\&A, 169, 1 
\bibitem[\protect\citeauthoryear{Bergeron \& Boiss{\'e}}{1991}]{Bergeron1991} Bergeron J., Boiss{\'e} P., 1991, A\&A, 243, 344 
\bibitem[\protect\citeauthoryear{Birnboim \& Dekel}{2003}]{Birnboim2003} Birnboim Y., Dekel A., 2003, MNRAS, 345, 349 
\bibitem[\protect\citeauthoryear{Boksenberg \& Sargent}{1978}]{Boksenberg1978} Boksenberg A., Sargent W.~L.~W., 1978, ApJ, 220, 42
\bibitem[\protect\citeauthoryear{Bonning, Shields, \& Salviander}{2007}]{Bonning2007} Bonning E.~W., Shields G.~A., Salviander S., 2007, ApJ, 666, L13 
\bibitem[\protect\citeauthoryear{Bordoloi et al.}{2011}]{Bordoloi2011} Bordoloi R., et al., 2011, ApJ, 743, 10
\bibitem[\protect\citeauthoryear{Bordoloi et al.}{2012}]{Bordoloi2012} Bordoloi R., Lilly S.~J., Kacprzak G.~G., Churchill C.~W., 2012, arXiv, arXiv:1211.3774 
\bibitem[\protect\citeauthoryear{Bouch{\'e} et al.}{2006}]{Bouche2006} Bouch{\'e} N., Murphy M.~T., P{\'e}roux C., Csabai I., Wild V., 2006, MNRAS, 371, 495
\bibitem[\protect\citeauthoryear{Bowen et al.}{2006}]{Bowen2006} Bowen D.~V., et al., 2006, ApJ, 645, L105 
\bibitem[\protect\citeauthoryear{Bowen \& Chelouche}{2011}]{Bowen2011} Bowen D.~V., Chelouche D., 2011, ApJ, 727, 47 
%
\bibitem[\protect\citeauthoryear{Cameron}{2011}]{Cameron2011} Cameron E., 2011, PASA, 28, 128 
\bibitem[\protect\citeauthoryear{Canalizo \& Stockton}{2001}]{Canalizo2001} Canalizo G., Stockton A., 2001, ApJ, 555, 719 
\bibitem[\protect\citeauthoryear{Canalizo \& Stockton}{2013}]{Canalizo2013} Canalizo G., Stockton A., 2013, ApJ, 772, 132 
\bibitem[\protect\citeauthoryear{Cardelli, Clayton, \& Mathis}{1989}]{Cardelli1989} Cardelli J.~A., Clayton G.~C., Mathis J.~S., 1989, ApJ, 345, 245 
\bibitem[\protect\citeauthoryear{Cepa et al.}{2000}]{Cepa2000} Cepa J., et al., 2000, SPIE, 4008, 623 
\bibitem[\protect\citeauthoryear{Cepa et al.}{2003}]{Cepa2003} Cepa J., et al., 2003, SPIE, 4841, 1739 
\bibitem[\protect\citeauthoryear{Charlton \& Churchill}{1998}]{Charlton1998} Charlton J.~C., Churchill C.~W., 1998, ApJ, 499, 181 
\bibitem[\protect\citeauthoryear{Charlton et al.}{2003}]{Charlton2003} Charlton J.~C., Ding J., Zonak S.~G., Churchill C.~W., Bond N.~A., Rigby J.~R., 2003, ApJ, 589, 111 
\bibitem[\protect\citeauthoryear{Chelouche et al.}{2008}]{Chelouche2008} Chelouche D., M{\'e}nard B., Bowen D.~V., Gnat O., 2008, ApJ, 683, 55 
\bibitem[\protect\citeauthoryear{Chen \& Tinker}{2008}]{Chen2008} Chen H.-W., Tinker J.~L., 2008, ApJ, 687, 745
\bibitem[\protect\citeauthoryear{Chen et al.}{2010a}]{Chen2010a} Chen H.-W., Helsby J.~E., Gauthier J.-R., Shectman S.~A., Thompson I.~B., Tinker J.~L., 2010a, ApJ, 714, 1521 
\bibitem[\protect\citeauthoryear{Chen et al.}{2010b}]{Chen2010b} Chen H.-W., Wild V., Tinker J.~L., Gauthier J.-R., Helsby J.~E., Shectman S.~A., Thompson I.~B., 2010b, ApJ, 724, L176 
\bibitem[\protect\citeauthoryear{Chen}{2012}]{Chen2012} Chen H.-W., 2012, MNRAS, 427, 1238
\bibitem[\protect\citeauthoryear{Churchill et al.}{2000}]{Churchill2000} Churchill C.~W., Mellon R.~R., Charlton J.~C., Jannuzi B.~T., Kirhakos S., Steidel C.~C., Schneider D.~P., 2000, ApJS, 130, 91 
\bibitem[\protect\citeauthoryear{Churchill \& Vogt}{2001}]{Churchill2001} Churchill C.~W., Vogt S.~S., 2001, AJ, 122, 679 
\bibitem[\protect\citeauthoryear{Churchill, Vogt, \& Charlton}{2003}]{Churchill2003} Churchill C.~W., Vogt S.~S., Charlton J.~C., 2003, AJ, 125, 98
\bibitem[\protect\citeauthoryear{Churchill, Kacprzak, \& Steidel}{2005}]{Churchill2005} Churchill C.~W., Kacprzak G.~G., Steidel C.~C., 2005, pgqa.conf, 24 
\bibitem[\protect\citeauthoryear{Churchill et al.}{2013a}]{Churchill2013a} Churchill C.~W., Nielsen N.~M., Kacprzak G.~G., Trujillo-Gomez S., 2013, ApJ, 763, L42 
\bibitem[\protect\citeauthoryear{Churchill et al.}{2013b}]{Churchill2013b} Churchill C.~W., Trujillo-Gomez S., Nielsen N.~M., Kacprzak G.~G., 2013, ApJ, 779, 87
\bibitem[\protect\citeauthoryear{Cirasuolo et al.}{2007}]{Cirasuolo2007} Cirasuolo M., et al., 2007, MNRAS, 380, 585 
\bibitem[\protect\citeauthoryear{Crenshaw, Kraemer, \& George}{2003}]{Crenshaw2003} Crenshaw D.~M., Kraemer S.~B., George I.~M., 2003, ARA\&A, 41, 117
\bibitem[\protect\citeauthoryear{Crotts}{1989}]{Crotts1989} Crotts A.~P.~S., 1989, ApJ, 336, 550 
%
\bibitem[\protect\citeauthoryear{Decarli et al.}{2008}]{Decarli2008} Decarli R., Labita M., Treves A., Falomo R., 2008, MNRAS, 387, 1237 
\bibitem[\protect\citeauthoryear{Decarli, Treves, \& Falomo}{2009}]{Decarli2009} Decarli R., Treves A., Falomo R., 2009, MNRAS, 396, L31
\bibitem[\protect\citeauthoryear{Decarli et al.}{2010a}]{Decarli2010a} Decarli R., Falomo R., Treves A., Kotilainen J.~K., Labita M., Scarpa R., 2010a, MNRAS, 402, 2441 
\bibitem[\protect\citeauthoryear{Decarli et al.}{2010b}]{Decarli2010b} Decarli R., Falomo R., Treves A., Labita M., Kotilainen J.~K., Scarpa R., 2010b, MNRAS, 402, 2453 
\bibitem[\protect\citeauthoryear{Decarli et al.}{2012}]{Decarli2012} Decarli R., Falomo R., Kotilainen J.~K., Hyv{\"o}nen T., Uslenghi M., Treves A., 2012, AdAst, 2012,
\bibitem[\protect\citeauthoryear{De Rosa et al.}{2011}]{Derosa2011} De Rosa G., Decarli R., Walter F., Fan X., Jiang L., Kurk J., Pasquali A., Rix H.~W., 2011, ApJ, 739, 56 
\bibitem[\protect\citeauthoryear{Di Matteo, Springel, \& Hernquist}{2005}]{Dimatteo2005} Di Matteo T., Springel V., Hernquist L., 2005, Natur, 433, 604 
\bibitem[\protect\citeauthoryear{Dobrzycki \& Bechtold}{1991}]{Dobrzycki1991} Dobrzycki A., Bechtold J., 1991, ApJ, 377, L69 
%
\bibitem[\protect\citeauthoryear{Ellison et al.}{2004}]{Ellison2004} Ellison S.~L., Ibata R., Pettini M., Lewis G.~F., Aracil B., Petitjean P., Srianand R., 2004, A\&A, 414, 79
\bibitem[\protect\citeauthoryear{Elvis}{2000}]{Elvis2000} Elvis M., 2000, ApJ, 545, 63
%
\bibitem[\protect\citeauthoryear{Faber et al.}{2007}]{Faber2007} Faber S.~M., et al., 2007, ApJ, 665, 265 
\bibitem[\protect\citeauthoryear{Falomo, Kotilainen, \& Treves}{2001}]{Falomo2001} Falomo R., Kotilainen J., Treves A., 2001, ApJ, 547, 124 
\bibitem[\protect\citeauthoryear{Falomo et al.}{2004}]{Falomo2004} Falomo R., Kotilainen J.~K., Pagani C., Scarpa R., Treves A., 2004, ApJ, 604, 495 
\bibitem[\protect\citeauthoryear{Falomo et al.}{2008}]{Falomo2008} Falomo R., Treves A., Kotilainen J.~K., Scarpa R., Uslenghi M., 2008, ApJ, 673, 694 
\bibitem[\protect\citeauthoryear{Falomo et al.}{2014}]{Falomo2014} Falomo R., Bettoni D., Karhunen K., Kotilainen J.~K., Uslenghi M., 2014, arXiv, arXiv:1402.4300 
\bibitem[\protect\citeauthoryear{Farina et al.}{2013}]{Farina2013} Farina E.~P., Falomo R., Decarli R., Treves A., Kotilainen J.~K., 2013, MNRAS, 429, 1267 
\bibitem[\protect\citeauthoryear{Fasano \& Franceschini}{1987}]{Fasano1987} Fasano G., Franceschini A., 1987, MNRAS, 225, 155
\bibitem[\protect\citeauthoryear{Fumagalli et al.}{2013}]{Fumagalli2013} Fumagalli M., O'Meara J.~M., Prochaska J.~X., Worseck G., 2013, ApJ, 775, 78
\bibitem[\protect\citeauthoryear{Floyd et al.}{2004}]{Floyd2004} Floyd D.~J.~E., Kukula M.~J., Dunlop J.~S., McLure R.~J., Miller L., Percival W.~J., Baum S.~A., O'Dea C.~P., 2004, MNRAS, 355, 196 
\bibitem[\protect\citeauthoryear{Floyd et al.}{2013}]{Floyd2013} Floyd D.~J.~E., Dunlop J.~S., Kukula M.~J., Brown M.~J.~I., McLure R.~J., Baum S.~A., O'Dea C.~P., 2013, MNRAS, 429, 2 
%
\bibitem[\protect\citeauthoryear{Gauthier, Chen, \& Tinker}{2010}]{Gauthier2010} Gauthier J.-R., Chen H.-W., Tinker J.~L., 2010, ApJ, 716, 1263 
\bibitem[\protect\citeauthoryear{Gauthier \& Chen}{2011}]{Gauthier2011} Gauthier J.-R., Chen H.-W., 2011, MNRAS, 418, 2730
\bibitem[\protect\citeauthoryear{Gauthier}{2013}]{Gauthier2013} Gauthier J.-R., 2013, MNRAS, 432, 1444
\bibitem[\protect\citeauthoryear{Gehrels}{1986}]{Gehrels1986} Gehrels N., 1986, ApJ, 303, 336 
%
\bibitem[\protect\citeauthoryear{H{\"a}ring \& Rix}{2004}]{Haring2004} H{\"a}ring N., Rix H.-W., 2004, ApJ, 604, L89 
\bibitem[\protect\citeauthoryear{Hennawi et al.}{2006a}]{Hennawi2006bin} Hennawi J.~F., et al., 2006a, AJ, 131, 1 
\bibitem[\protect\citeauthoryear{Hennawi et al.}{2006b}]{Hennawi2006} Hennawi J.~F., et al., 2006b, ApJ, 651, 61 
\bibitem[\protect\citeauthoryear{Hennawi \& Prochaska}{2007}]{Hennawi2007} Hennawi J.~F., Prochaska J.~X., 2007, ApJ, 655, 735 
\bibitem[\protect\citeauthoryear{Hennawi \& Prochaska}{2013}]{Hennawi2013} Hennawi J.~F., Prochaska J.~X., 2013, ApJ, 766, 58
\bibitem[\protect\citeauthoryear{Hernquist}{1989}]{Hernquist1989} Hernquist L., 1989, Natur, 340, 687 
\bibitem[\protect\citeauthoryear{Hutchings, Scholz, \& Bianchi}{2009}]{Hutchings2009} Hutchings J.~B., Scholz P., Bianchi L., 2009, AJ, 137, 3533 
\bibitem[\protect\citeauthoryear{Hyv{\"o}nen et al.}{2007}]{Hyvonen2007} Hyv{\"o}nen T., Kotilainen J.~K., {\"O}rndahl E., Falomo R., Uslenghi M., 2007, A\&A, 462, 525
%
\bibitem[\protect\citeauthoryear{Isobe, Feigelson, \& Nelson}{1986}]{Isobe1986} Isobe T., Feigelson E.~D., Nelson P.~I., 1986, ApJ, 306, 490 
%
\bibitem[\protect\citeauthoryear{Jahnke \& Wisotzki}{2003}]{Jahnke2003} Jahnke K., Wisotzki L., 2003, MNRAS, 346, 304
%
\bibitem[\protect\citeauthoryear{Kacprzak et al.}{2007}]{Kacprzak2007} Kacprzak G.~G., Churchill C.~W., Steidel C.~C., Murphy M.~T., Evans J.~L., 2007, ApJ, 662, 909 
\bibitem[\protect\citeauthoryear{Kacprzak et al.}{2008}]{Kacprzak2008} Kacprzak G.~G., Churchill C.~W., Steidel C.~C., Murphy M.~T., 2008, AJ, 135, 922
\bibitem[\protect\citeauthoryear{Kacprzak, Murphy, \& Churchill}{2010}]{Kacprzak2010} Kacprzak G.~G., Murphy M.~T., Churchill C.~W., 2010, MNRAS, 406, 445 
\bibitem[\protect\citeauthoryear{Kacprzak et al.}{2011}]{Kacprzak2011} Kacprzak G.~G., Churchill C.~W., Evans J.~L., Murphy M.~T., Steidel C.~C., 2011, MNRAS, 416, 3118 
\bibitem[\protect\citeauthoryear{Kacprzak, Churchill, \& Nielsen}{2012}]{Kacprzak2012} Kacprzak G.~G., Churchill C.~W., Nielsen N.~M., 2012, arXiv, arXiv:1205.0245 
\bibitem[\protect\citeauthoryear{Kaspi et al.}{2000}]{Kaspi2000} Kaspi S., Smith P.~S., Netzer H., Maoz D., Jannuzi B.~T., Giveon U., 2000, ApJ, 533, 631 
\bibitem[\protect\citeauthoryear{Kauffmann \& Haehnelt}{2000}]{Kauffmann2000} Kauffmann G., Haehnelt M., 2000, MNRAS, 311, 576 
\bibitem[\protect\citeauthoryear{Kere{\v s} et al.}{2005}]{Keres2005} Kere{\v s} D., Katz N., Weinberg D.~H., Dav{\'e} R., 2005, MNRAS, 363, 2 
\bibitem[\protect\citeauthoryear{Kere{\v s} et al.}{2009}]{Keres2009} Kere{\v s} D., Katz N., Fardal M., Dav{\'e} R., Weinberg D.~H., 2009, MNRAS, 395, 160 
\bibitem[\protect\citeauthoryear{Kotilainen et al.}{2007}]{Kotilainen2007} Kotilainen J.~K., Falomo R., Labita M., Treves A., Uslenghi M., 2007, ApJ, 660, 1039 
\bibitem[\protect\citeauthoryear{Kotilainen et al.}{2009}]{Kotilainen2009} Kotilainen J.~K., Falomo R., Decarli R., Treves A., Uslenghi M., Scarpa R., 2009, ApJ, 703, 1663 
\bibitem[\protect\citeauthoryear{Kotilainen et al.}{2013}]{Kotilainen2013} Kotilainen J., Falomo R., Bettoni D., Karhunen K., Uslenghi M., 2013, arXiv, arXiv:1302.1366
\bibitem[\protect\citeauthoryear{Kukula et al.}{2001}]{Kukula2001} Kukula M.~J., Dunlop J.~S., McLure R.~J., Miller L., Percival W.~J., Baum S.~A., O'Dea C.~P., 2001, MNRAS, 326, 1533 
%
\bibitem[\protect\citeauthoryear{Lanzetta \& Bowen}{1990}]{Lanzetta1990} Lanzetta K.~M., Bowen D., 1990, ApJ, 357, 321 
\bibitem[\protect\citeauthoryear{Liske \& Williger}{2001}]{Liske2001} Liske J., Williger G.~M., 2001, MNRAS, 328, 653
\bibitem[\protect\citeauthoryear{Lopez et al.}{2008}]{Lopez2008} Lopez S., et al., 2008, ApJ, 679, 1144 
\bibitem[\protect\citeauthoryear{Lovegrove \& Simcoe}{2011}]{Lovegrove2011} Lovegrove E., Simcoe R.~A., 2011, ApJ, 740, 3
\bibitem[\protect\citeauthoryear{Lundgren et al.}{2012}]{Lundgren2012} Lundgren B.~F., et al., 2012, ApJ, 760, 49
%
\bibitem[\protect\citeauthoryear{Mannucci et al.}{2001}]{Mannucci2001} Mannucci F., Basile F., Poggianti B.~M., Cimatti A., Daddi E., Pozzetti L., Vanzi L., 2001, MNRAS, 326, 745
\bibitem[\protect\citeauthoryear{McLure et al.}{1999}]{McLure1999} McLure R.~J., Kukula M.~J., Dunlop J.~S., Baum S.~A., O'Dea C.~P., Hughes D.~H., 1999, MNRAS, 308, 377
\bibitem[\protect\citeauthoryear{McLure \& Jarvis}{2002}]{McLure2002} McLure R.~J., Jarvis M.~J., 2002, MNRAS, 337, 109 
\bibitem[\protect\citeauthoryear{Marconi \& Hunt}{2003}]{Marconi2003} Marconi A., Hunt L.~K., 2003, ApJ, 589, L21 
\bibitem[\protect\citeauthoryear{M{\'e}nard \& Chelouche}{2009}]{Menard2009} M{\'e}nard B., Chelouche D., 2009, MNRAS, 393, 808 
\bibitem[\protect\citeauthoryear{M{\'e}nard et al.}{2011}]{Menard2011} M{\'e}nard B., Wild V., Nestor D., Quider A., Zibetti S., Rao S., Turnshek D., 2011, MNRAS, 417, 801 
\bibitem[\protect\citeauthoryear{Monet et al.}{2003}]{Monet2003} Monet D.~G., et al., 2003, AJ, 125, 984
\bibitem[\protect\citeauthoryear{More et al.}{2011}]{More2011} More S., van den Bosch F.~C., Cacciato M., Skibba R., Mo H.~J., Yang X., 2011, MNRAS, 410, 210 
\bibitem[\protect\citeauthoryear{Moster et al.}{2010}]{Moster2010} Moster B.~P., Somerville R.~S., Maulbetsch C., van den Bosch F.~C., Macci{\`o} A.~V., Naab T., Oser L., 2010, ApJ, 710, 903 
%
\bibitem[\protect\citeauthoryear{Nelson et al.}{2013}]{Nelson2013} Nelson D., Vogelsberger M., Genel S., Sijacki D., Kere{\v s} D., Springel V., Hernquist L., 2013, MNRAS, 429, 3353 
\bibitem[\protect\citeauthoryear{Nestor, Turnshek, \& Rao}{2005}]{Nestor2005} Nestor D.~B., Turnshek D.~A., Rao S.~M., 2005, ApJ, 628, 637 
\bibitem[\protect\citeauthoryear{Nestor et al.}{2007}]{Nestor2007} Nestor D.~B., Turnshek D.~A., Rao S.~M., Quider A.~M., 2007, ApJ, 658, 185 
\bibitem[\protect\citeauthoryear{Nestor, Hamann, \& Rodriguez Hidalgo}{2008}]{Nestor2008} Nestor D., Hamann F., Rodriguez Hidalgo P., 2008, MNRAS, 386, 2055
\bibitem[\protect\citeauthoryear{Nestor et al.}{2011}]{Nestor2011} Nestor D.~B., Johnson B.~D., Wild V., M{\'e}nard B., Turnshek D.~A., Rao S., Pettini M., 2011, MNRAS, 412, 1559 
\bibitem[\protect\citeauthoryear{Nielsen, Churchill, \& Kacprzak}{2013}]{Nielsen2013a} Nielsen N.~M., Churchill C.~W., Kacprzak G.~G., 2013, ApJ, 776, 115 
\bibitem[\protect\citeauthoryear{Nielsen et al.}{2013}]{Nielsen2013b} Nielsen N.~M., Churchill C.~W., Kacprzak G.~G., Murphy M.~T., 2013, ApJ, 776, 114 
%
\bibitem[\protect\citeauthoryear{Peng et al.}{2006a}]{Peng2006a} Peng C.~Y., Impey C.~D., Ho L.~C., Barton E.~J., Rix H.-W., 2006, ApJ, 640, 114
\bibitem[\protect\citeauthoryear{Peng et al.}{2006b}]{Peng2006b} Peng C.~Y., Impey C.~D., Rix H.-W., Kochanek C.~S., Keeton C.~R., Falco E.~E., Leh{\'a}r J., McLeod B.~A., 2006, ApJ, 649, 616 
\bibitem[\protect\citeauthoryear{Peterson}{2010}]{Peterson2010} Peterson B.~M., 2010, IAUS, 267, 151 
\bibitem[\protect\citeauthoryear{Petitjean \& Bergeron}{1990}]{Petitjean1990} Petitjean P., Bergeron J., 1990, A\&A, 231, 309
\bibitem[\protect\citeauthoryear{Prochaska, Hennawi, \& Simcoe}{2013}]{Prochaska2013a} Prochaska J.~X., Hennawi J.~F., Simcoe R.~A., 2013, ApJ, 762, L19 
\bibitem[\protect\citeauthoryear{Prochaska et al.}{2013}]{Prochaska2013b} Prochaska J.~X., et al., 2013, ApJ, 776, 136 
\bibitem[\protect\citeauthoryear{Prochaska \& Hennawi}{2009}]{Prochaska2009} Prochaska J.~X., Hennawi J.~F., 2009, ApJ, 690, 1558 
\bibitem[\protect\citeauthoryear{Prochter, Prochaska, \& Burles}{2006}]{Prochter2006} Prochter G.~E., Prochaska J.~X., Burles S.~M., 2006, ApJ, 639, 766 
%
\bibitem[\protect\citeauthoryear{Rao \& Turnshek}{2000}]{Rao2000} Rao S.~M., Turnshek D.~A., 2000, ApJS, 130, 1
\bibitem[\protect\citeauthoryear{Rao, Turnshek, \& Nestor}{2006}]{Rao2006} Rao S.~M., Turnshek D.~A., Nestor D.~B., 2006, ApJ, 636, 610 
\bibitem[\protect\citeauthoryear{Rauch et al.}{2002}]{Rauch2002} Rauch M., Sargent W.~L.~W., Barlow T.~A., Simcoe R.~A., 2002, ApJ, 576, 45
\bibitem[\protect\citeauthoryear{Ribaudo et al.}{2011}]{Ribaudo2011} Ribaudo J., Lehner N., Howk J.~C., Werk J.~K., Tripp T.~M., Prochaska J.~X., Meiring J.~D., Tumlinson J., 2011, ApJ, 743, 20
%
\bibitem[\protect\citeauthoryear{Richards et al.}{2002}]{Richards2002} Richards G.~T., Vanden Berk D.~E., Reichard T.~A., Hall P.~B., Schneider D.~P., SubbaRao M., Thakar A.~R., York D.~G., 2002, AJ, 124, 
\bibitem[\protect\citeauthoryear{Rodr{\'{\i}}guez Hidalgo et al.}{2012}]{Rodriguez2012} Rodr{\'{\i}}guez Hidalgo P., Wessels K., Charlton J.~C., Narayanan A., Mshar A., Cucchiara A., Jones T., 2012, MNRAS, 427, 1801
\bibitem[\protect\citeauthoryear{Runnoe, Brotherton, \& Shang}{2012}]{Runnoe2012} Runnoe J.~C., Brotherton M.~S., Shang Z., 2012, MNRAS, 422, 478 
\bibitem[\protect\citeauthoryear{Rubin et al.}{2010}]{Rubin2010} Rubin K.~H.~R., Weiner B.~J., Koo D.~C., Martin C.~L., Prochaska J.~X., Coil A.~L., Newman J.~A., 2010, ApJ, 719, 1503
\bibitem[\protect\citeauthoryear{Rubin et al.}{2012}]{Rubin2012} Rubin K.~H.~R., Prochaska J.~X., Koo D.~C., Phillips A.~C., 2012, ApJ, 747, L26
\bibitem[\protect\citeauthoryear{Sancisi et al.}{2008}]{Sancisi2008} Sancisi R., Fraternali F., Oosterloo T., van der Hulst T., 2008, A\&ARv, 15, 18
\bibitem[\protect\citeauthoryear{Sbarufatti et al.}{2005}]{Sbarufatti2005} Sbarufatti B., Treves A., Falomo R., Heidt J., Kotilainen J., Scarpa R., 2005, AJ, 129, 559
\bibitem[\protect\citeauthoryear{Schechter}{1976}]{Schechter1976} Schechter P., 1976, ApJ, 203, 297 
\bibitem[\protect\citeauthoryear{Schirber, Miralda-Escud{\'e}, \& McDonald}{2004}]{Schirber2004} Schirber M., Miralda-Escud{\'e} J., McDonald P., 2004, ApJ, 610, 105 
\bibitem[\protect\citeauthoryear{Schlegel, Finkbeiner, \& Davis}{1998}]{Schlegel1998} Schlegel D.~J., Finkbeiner D.~P., Davis M., 1998, ApJ, 500, 525 
\bibitem[\protect\citeauthoryear{Schneider et al.}{1993}]{Schneider1993} Schneider D.~P., et al., 1993, ApJS, 87, 45 
\bibitem[\protect\citeauthoryear{Schneider et al.}{2010}]{Schneider2010} Schneider D.~P., et al., 2010, AJ, 139, 2360 
\bibitem[\protect\citeauthoryear{Serber et al.}{2006}]{Serber2006} Serber W., Bahcall N., M{\'e}nard B., Richards G., 2006, ApJ, 643, 68
\bibitem[\protect\citeauthoryear{Sharma, Nath, \& Chand}{2013}]{Sharma2013} Sharma M., Nath B.~B., Chand H., 2013, MNRAS, 431, L93 
\bibitem[\protect\citeauthoryear{Shaver, Boksenberg, \& Robertson}{1982}]{Shaver1982} Shaver P.~A., Boksenberg A., Robertson J.~G., 1982, ApJ, 261, L7 
\bibitem[\protect\citeauthoryear{Shaver \& Robertson}{1983}]{Shaver1983} Shaver P.~A., Robertson J.~G., 1983, ApJ, 268, L57 
\bibitem[\protect\citeauthoryear{Shaver \& Robertson}{1985}]{Shaver1985} Shaver P.~A., Robertson J.~G., 1985, MNRAS, 212, 15P 
\bibitem[\protect\citeauthoryear{Shen et al.}{2012}]{Shen2012a} Shen S., Madau P., Aguirre A., Guedes J., Mayer L., Wadsley J., 2012, ApJ, 760, 50 
\bibitem[\protect\citeauthoryear{Shen \& M{\'e}nard}{2012}]{Shen2012} Shen Y., M{\'e}nard B., 2012, ApJ, 748, 131 
\bibitem[\protect\citeauthoryear{Shen}{2013}]{Shen2013} Shen Y., 2013, arXiv, arXiv:1302.2643
\bibitem[\protect\citeauthoryear{Steidel \& Sargent}{1992}]{Steidel1992} Steidel C.~C., Sargent W.~L.~W., 1992, ApJS, 80, 1
\bibitem[\protect\citeauthoryear{Steidel, Dickinson, \& Persson}{1994}]{Steidel1994} Steidel C.~C., Dickinson M., Persson S.~E., 1994, ApJ, 437, L75 
\bibitem[\protect\citeauthoryear{Steidel}{1995}]{Steidel1995} Steidel C.~C., 1995, qal..conf, 139  
\bibitem[\protect\citeauthoryear{Steidel et al.}{1997}]{Steidel1997} Steidel C.~C., Dickinson M., Meyer D.~M., Adelberger K.~L., Sembach K.~R., 1997, ApJ, 480, 568
\bibitem[\protect\citeauthoryear{Steidel et al.}{2010}]{Steidel2010} Steidel C.~C., Erb D.~K., Shapley A.~E., Pettini M., Reddy N., Bogosavljevi{\'c} M., Rudie G.~C., Rakic O., 2010, ApJ, 717, 289 
%
\bibitem[\protect\citeauthoryear{Tody}{1986}]{Tody1986} Tody D., 1986, SPIE, 627, 733 
\bibitem[\protect\citeauthoryear{Tremonti, Moustakas, \& Diamond-Stanic}{2007}]{Tremonti2007} Tremonti C.~A., Moustakas J., Diamond-Stanic A.~M., 2007, ApJ, 663, L77 
\bibitem[\protect\citeauthoryear{Tripp \& Bowen}{2005}]{Tripp2005} Tripp T.~M., Bowen D.~V., 2005, pgqa.conf, 5 
\bibitem[\protect\citeauthoryear{Tytler \& Fan}{1992}]{Tytler1992} Tytler D., Fan X.-M., 1992, ApJS, 79, 1 
\bibitem[\protect\citeauthoryear{Tytler et al.}{2009}]{Tytler2009} Tytler D., et al., 2009, MNRAS, 392, 1539 
%
\bibitem[\protect\citeauthoryear{Vanden Berk et al.}{2008}]{Vandenberk2008} Vanden Berk D., et al., 2008, ApJ, 679, 239 
\bibitem[\protect\citeauthoryear{van de Voort \& Schaye}{2012}]{vandeVoort2012} van de Voort F., Schaye J., 2012, MNRAS, 423, 2991 
\bibitem[\protect\citeauthoryear{Vestergaard \& Wilkes}{2001}]{Vestergaard2001} Vestergaard M., Wilkes B.~J., 2001, ApJS, 134, 1 
\bibitem[\protect\citeauthoryear{Vestergaard}{2003}]{Vestergaard2003} Vestergaard M., 2003, ApJ, 599, 116 
\bibitem[\protect\citeauthoryear{Vestergaard \& Peterson}{2006}]{Vestergaard2006} Vestergaard M., Peterson B.~M., 2006, ApJ, 641, 689 
%
\bibitem[\protect\citeauthoryear{Werk et al.}{2013}]{Werk2013} Werk J.~K., Prochaska J.~X., Thom C., Tumlinson J., Tripp T.~M., O'Meara J.~M., Peeples M.~S., 2013, ApJS, 204, 17 
\bibitem[\protect\citeauthoryear{Weiner et al.}{2009}]{Weiner2009} Weiner B.~J., et al., 2009, ApJ, 692, 187
\bibitem[\protect\citeauthoryear{Wild et al.}{2008}]{Wild2008} Wild V., et al., 2008, MNRAS, 388, 227 
\bibitem[\protect\citeauthoryear{Wold et al.}{2001}]{Wold2001} Wold M., Lacy M., Lilje P.~B., Serjeant S., 2001, MNRAS, 323, 23
%
\bibitem[\protect\citeauthoryear{Zhang et al.}{2013}]{Zhang2013} Zhang S., Wang T., Wang H., Zhou H., 2013, ApJ, 773, 175
\bibitem[\protect\citeauthoryear{Zheng et al.}{2009}]{Zheng2009} Zheng Z., Zehavi I., Eisenstein D.~J., Weinberg D.~H., Jing Y.~P., 2009, ApJ, 707, 554
\bibitem[\protect\citeauthoryear{Zibetti et al.}{2007}]{Zibetti2007} Zibetti S., M{\'e}nard B., Nestor D.~B., Quider A.~M., Rao S.~M., Turnshek D.~A., 2007, ApJ, 658, 161 
\bibitem[\protect\citeauthoryear{Zwaan et al.}{2005}]{Zwaan2005} Zwaan M.~A., van der Hulst J.~M., Briggs F.~H., Verheijen M.~A.~W., Ryan-Weber E.~V., 2005, MNRAS, 364, 1467

\end{thebibliography}
\end{document}